\documentclass[a4paper,11pt]{article}

\usepackage{preamble}

\title{Dirac, Majorana, and Weyl Spinors in Arbitrary Dimension and Signature}

\author{Jan Hajer\email{jan.hajer@tecnico.ulisboa.pt}}

\affiliation{Departamento de Física, Instituto Superior Técnico (IST), Universidade de Lisboa, 1049-001 Lisboa, Portugal}

\begin{document}

\maketitle

\begin{abstract}
Fermionic field types in arbitrary signature are often inferred from complex Spin representations alone, which obscures distinctions associated with real structures, disconnected spacetime reflections, and action-level bilinear constraints.
We give a convention-explicit classification that treats the full real Clifford algebra and its even subalgebra separately, thereby distinguishing pinors from spinors and Pin-equivariant Majorana-type structures from structures that are only Spin-equivariant.
Adjoint, complex-conjugation, and transposition intertwiners are combined with elementary and collective reflection lifts into a unified sign calculus.
The resulting local Clifford classification is organised into thirty-two periodicity sectors determined by the Bott class and the total dimension modulo eight, without including theory-dependent gauge, anomaly, or global constraints.
For each sector, we identify the available Dirac, Weyl, Majorana, \sMlong, \MWlong, and \sMWlong fields.
The existence of an invariant pairing is distinguished from intertwiner compatibility, Grassmann and chiral selection rules, covariant index contraction, and Hermiticity of local bilinear operators.
\end{abstract}

\clearpage

\tableofcontents

\listoftables

\clearpage

\section{Introduction} \label{sec:introduction}

Spinor representations provide the kinematic description of fermions, from the \SM to supersymmetric and higher-dimensional theories.
In four-dimensional Lorentzian spacetime, the labels Dirac, Weyl, and Majorana have familiar operational meanings, and their combinations with symplectic reality or chirality are routinely used in supersymmetry and supergravity \cite{Wetterich:1982eh,Wetterich:1982ed,vanHolten:1982mx,VanProeyen:1999ni,Pal:2010ih,Dreiner:2008tw}.
Higher-dimensional model building, dimensional reduction, and Euclidean formulations require the same questions to be answered in other dimensions and signatures, where these labels no longer determine all representation-theoretic or action-level properties of a fermion field \cite{Gall:2021tiu}.

The underlying difficulty is that several distinct structures coincide in common four-dimensional conventions.
A complex representation of the connected Spin group determines neither its extension to disconnected spacetime reflections nor the real or quaternionic structure needed for a Majorana-type condition.
Likewise, the existence of a reality condition does not determine its compatibility with a single Weyl module, and neither statement guarantees a nonzero bilinear or, in Lorentzian signature, a Hermitian action \cite{DeAndrade:1994mb,DeAndrade:1999xa,DAuria:2000byu,Wetterich:2010ni,Stone:2020vva}.
Complexification can hide signature-dependent real structures, while restriction to the even Clifford algebra discards the odd elements that lift reflections.
Consequently, fields with the same complexified Spin content can differ as real Pin representations and can obey different bilinear selection rules.

These distinctions motivate a strict separation between the full Clifford algebra, its even subalgebra, and the action constructed from their modules.
The full algebra controls pinors and the lifts of spatial and temporal reflections, whereas the even algebra controls spinors of the connected group and, in even dimension, their Weyl decomposition.
An antilinear structure may therefore be Spin-equivariant without extending to the Pin action, and an odd reflection lift may exchange the two Weyl modules rather than act within either one.
The relation between Pin groups and disconnected orthogonal transformations has been studied extensively \cite{Dabrowski:1988kq,Berg:2000ne,Trautman:2005qx,Figueroa:2015}; here the reflection lifts and their signs are incorporated into the same convention system as spinor reality and fermion bilinears.

The classifications of real Clifford algebras, their modules, invariant bilinear forms, and their intertwiners are well established \cite{Atiyah:1964zz,Lawson:1998yr,Harvey:1990,Alekseevsky:1995,VanProeyen:1999ni,DAuria:2000byu,Wetterich:2010ni,Stone:2020vva,Park:2022pjv,Gil-Garcia:2025iqt}.
The purpose of this work is not to introduce new Clifford-algebra isomorphisms, but to assemble the information required for field-theory calculations in a single convention-explicit framework.
In particular, the elementary and collective reflection lifts are retained alongside the full and even module types, while the adjoint, transposition, and complex-conjugation intertwiners are assigned distinct basis-covariant roles.
This organisation makes every phase choice and sign relation traceable when translating a basis-independent statement into an explicit matrix representation of the Clifford algebra.

The calculation is organised around the intertwiners \(A^\pm\), \(\TransMat^\pm\), and \(\ConjMat^\pm\).
The adjoint intertwiners \(A^\pm\) define the candidate Dirac adjoints and control Hermiticity.
The conjugation intertwiners \(\ConjMat^\pm\) induce the real or quaternionic structures underlying Majorana and \sM fields, while the transposition intertwiners \(\TransMat^\pm\) define invariant complex-bilinear pairings and their Grassmann selection signs.
Their defining relations are basis independent, while fixed unphased ordered products provide a direct construction in an adapted unitary Clifford basis.
Compatibility phases between the three intertwiner families remain explicit, so conventional rephasings cannot be mistaken for invariant signs or for the absence of a relation.

Two independent periodic variables are needed to organise the module types and intertwiner signatures.
The Bott class \(\delta = s-t\bmod8\) fixes the real, complex, or quaternionic module type and the conjugation signatures, whereas the total dimension \(d = s+t\), considered modulo eight, supplies independent transposition and chirality information.
Because \(d\) and \(\delta\) have the same parity, their combination gives the thirty-two admissible local periodicity sectors collected in \cref{tab:master-32}.
These sectors retain information that is lost when only the eight matrix-size-independent real Clifford classes are specified, including the distinction between Pin- and Spin-equivariant reality structures and the signs relevant to scalar and kinetic bilinears.

The practical objective is to decide whether a candidate fermionic operator can occur in an action, rather than merely whether a suitable field representation exists.
An invariant spinor pairing and a compatible choice of intertwiners are required before Grassmann symmetry and, where applicable, Weyl projection can be tested.
Any remaining Clifford indices must then admit a covariant tensor contraction, and the internal gauge and flavour representations must contain the required invariant.
For a Lorentzian action, the coefficient must additionally make the integrated operator Hermitian after the necessary integrations by parts.
These logically independent tests are formulated in \cref{sec:bilinears-lagrangians}, where the signature-dependent signs are kept separate from theory-dependent internal contractions.
This separation prevents an available Majorana structure or a favourable transpose sign from being interpreted by itself as an allowed mass or interaction term.

The resulting classification is local and algebraic.
It determines the available Pin and Spin modules, their Majorana-type and Weyl restrictions, their candidate reflection and antilinear maps, and the necessary local conditions on fermion bilinears.
It does not classify gauge representations, flavour multiplicities, anomaly cancellation, global Spin or Pin structures, reflection positivity, or the dynamics of an interacting \QFT.
Moreover, a Clifford-module map represents a physical discrete symmetry only after its action on the complete theory, including internal quantum numbers and backgrounds, has been specified.

The same separation of Clifford-module structure from physical symmetry conditions also clarifies the relation to one-particle \QM and free-fermion condensed-matter systems.
In that setting the antilinear intertwiners become candidate antiunitary maps only after the appropriate Hamiltonian relation is imposed, while adjoining a normalised mass matrix gives the Clifford extension underlying the \AZlong tenfold way \cite{Schnyder:2008tya,Kitaev:2009mg,Ryu:2010zza,Kennedy:2014cia}.
The common Bott-periodic mechanism therefore provides a useful comparison without identifying a relativistic Majorana spinor with a Majorana zero mode or Weyl chirality with spectral symmetry.

The orthogonal groups and their Spin and Pin covers are reviewed in \cref{sec:orthogonal groups}, followed by the real and complex Clifford algebras in \cref{sec:clifford algebra}.
The Clifford realisations of the Pin and Spin groups and the convention-explicit intertwiner calculus are developed in \cref{sec:pin-spin-groups}, while the corresponding pinor and spinor fields, discrete module maps, and bilinear tests are presented in \cref{sec:spinor-pinor-classification}.
The thirty-two periodicity sectors, four-dimensional representatives, Lorentzian and Euclidean specialisations, and conformal embedding-space extension are assembled in \cref{sec:bott-class-field-content}.
The basis-independent \QM interpretation and its application to the condensed-matter tenfold way are given in \cref{sec:ABC-euclidean-QM,sec:condensed-matter-bott}.

\section{Orthogonal groups and their double covers} \label{sec:orthogonal groups}

Orthogonal groups describe transformations of spacetime vectors that preserve a non-degenerate metric.
Because a \(2\pi\) rotation acts trivially on vectors but can multiply a spinor by \(-1\), spinor fields transform under a double cover of the corresponding orthogonal group.
The Spin group covers the connected rotations, whereas including spacetime reflections requires a Pin group.
The discussion begins with the real Euclidean and indefinite cases, whose reflection lifts retain signature-dependent information that disappears after complexification.

\subsection{Euclidean orthogonal group} \label{sec:euclidean orthogonal group}

The \(d\)-dimensional Euclidean orthogonal group \(\O(d)\) consists of real \(d \times d\) matrices~\(\Mmat\) satisfying \(\Mmat^\trans \Mmat = \matident\).
The two values \(\det \Mmat = \pm 1\) label disconnected components related by the elementary reflection \(\p\), which is represented by a diagonal matrix \(\Pmat\) that flips one coordinate.
The orientation-reversing component therefore consists of transformations expressible as an odd number of reflections.
By contrast, the collective inversion \(\P x = -x\) reverses all coordinates and belongs to this component only for odd \(d\).
The two components are encoded by the zeroth homotopy group
\footnote{
The notation \(\pi_0(G)\) records the disconnected components of a group \(G\), so it is a compact way of saying which reflection components are present.
}
\begin{equation}\label{eq:orthogonal parity}
\pi_0(\O(d)) = \C^2_\p,
\end{equation}
where \(\C^2\) is the cyclic group of order two.
\footnote{
The cyclic group \(\C^n\) of order \(n\) is isomorphic to the group of integers modulo \(n\), \(\C^n \cong \mathbb Z_n \cong \mathbb Z/n\mathbb Z\).
}
The component containing the identity is the special orthogonal group \(\SO(d)\), which has \(\det \Mmat = 1\) and is path connected, so \(\pi_0(\SO(d)) = 0\).
The resulting quotient relation is
\begin{align} \label{eq:orthogonal parity 2}
\shortexact{\SO(d)}{\O(d)}{\C^2_\p}.
\end{align}
Connectedness does not imply simple connectedness: for \(d\geq2\), the first homotopy group of \(\SO(d)\) determines the required covering structure \(\cover\):
\begin{align} \label{eq:first homotopy}
\pi_1(\SO(d)) & =
\begin{cases}
0 & \text{for } d = 1,\\
\C_\cover & \text{for } d = 2,\\
\C^2_\cover & \text{for } d\geq3.
\end{cases}
\end{align}
Here \(\C\) denotes the infinite cyclic group.
\footnote{
The infinite cyclic group \(\C\) is isomorphic to the additive group of the integers \(\C \cong \mathbb Z\).
}
The universal cover of \(\SO(2)\) therefore has infinitely many sheets, whereas that of \(\SO(d)\) for \(d\geq3\) is a double cover.
Vector and tensor representations factor through \(\SO(d)\), but spinorial representations require this covering group.

\subsection{Euclidean spin and pin groups} \label{sec:euclidean spin group}

The spin group has the same Lie algebra as the special orthogonal group, \(\so(d) \cong \spin(d)\), and for \(d \geq 1\) double covers it:
\begin{align} \label{eq:Euclidean spin}
\shortexact{\C^2_\cover}{\Spin(d)}{\SO(d)}.
\end{align}
Its first homotopy group is therefore
\begin{align}
\pi_1(\Spin(d)) & = \begin{cases}
\C_\cover & \text{for } d = 2 , \\
0 & \text{otherwise} ,
\end{cases}
\end{align}
so for \(d\geq3\) it coincides with the universal cover of the special orthogonal group.
Its zeroth homotopy group is
\begin{equation}
\pi_0(\Spin(d)) = \begin{cases}
\C^2_\cover & \text{for } d = 1 ,\\
0 & \text{for } d\geq2 .
\end{cases}
\end{equation}

Because the Spin construction covers only the connected rotation group \(\SO(d)\), including orientation-reversing transformations also requires a lift \(\plift\) of the reflection \(\p\) to the pinor space \(S\).
The two possible lifts of a given reflection differ only by the central covering element \(-\ident\), whereas the invariant distinction between the two Euclidean Pin groups is the sign of the squared reflection lift, \(\plift^2 = \pm\ident\).
The pin group double covers the full orthogonal group,
\begin{align} \label{eq:Euclidean pin}
\shortexact{\C^2_\cover}{\Pin(d)}{\O(d)},
\end{align}
and its discrete extension is fixed by a double cover of the component group in \eqref{eq:orthogonal parity 2}:
\begin{align} \label{eq:Z^2 double covers}
\shortexact{\C^2_\cover}{H^4_\pliftsqr}{\C^2_\p} .
\end{align}
The two resulting groups of order four are distinguished by the square of the parity lift:
\begin{align} \label{eq:Z^2 double covers 2}
H^4_\pliftsqr & = \begin{cases}
\D^4_\four \cong \C^2_\cover \times \C^2_\plift & \text{for } \plift^2 = +\ident, \\
\Q^4_\four \cong \C^4_\four & \text{for } \plift^2 = -\ident .
\end{cases}
\end{align}
The dihedral group of order four \(\D^4\) is isomorphic to the product of two cyclic groups of order two and is also called the Klein four-group.
\footnote{The dihedral group of order \(2n\) is the group of symmetries of a regular \(n\)-polygon.}
The dicyclic group of order four~\(\Q^4\) is isomorphic to the cyclic group of order four.
\footnote{
The dicyclic group of order \(4n\) is the extension of the cyclic group of order two by a cyclic group of order \(2n\) and can be written as \(\shortexact*{\C^{2n}}{\Q^{4n}}{\C^2}\).
}
The quotient by the diagonal subgroup \(\C^2_\cover\) identifies the nontrivial central element of the discrete extension with the covering element \(-\ident\in\Spin(d)\).
Extending the fixed spin cover across the reflection component in \eqref{eq:orthogonal parity} then gives two pin groups, distinguished by the square of \(\plift\), as quotients of semidirect products \cite{Dabrowski:1988kq}:
\begin{align} \label{eq:Euclidean pin construction}
\shortexact{\C^2_\cover}{\Spin(d) \rtimes H^4_\pliftsqr}{\Pin_\pliftsqr(d)} .
\end{align}
The semidirect product records that the discrete lift acts on Spin transformations by conjugation,
\begin{align}
g &\mapsto \plift g \plift^{-1} , &
g &\in \Spin(d).
\end{align}
For \(d\geq2\), rotations involving the reflected direction change sign, whereas rotations within the unreflected hyperplane remain unchanged, so the chosen elementary-reflection presentation uses a nontrivial semidirect product.
For \(d = 1\), this conjugation action is trivial.
In odd dimensions the resulting Pin group can nevertheless admit a direct-product decomposition when its orientation-reversing component contains a central element of order two.
The two inequivalent pin groups are therefore
\begin{align} \label{eq:definite pin} &\shortexact{\C^2_\cover}{\Spin(d) \rtimes \D^4_\four}{\Pin_+(d)} , & &\shortexact{\C^2_\cover}{\Spin(d) \rtimes \Q^4_\four}{\Pin_-(d)} .
\end{align}

\subsection{Indefinite orthogonal group}

For a flat spacetime \(V^{s,t}\) whose diagonal metric \(\eta\) has \(s\) spatial \(+1\) entries and \(t\) temporal \(-1\) entries, the indefinite orthogonal group \(\O(s,t)\) consists of all matrices satisfying \(\Mmat^\trans\eta\Mmat = \eta\).
For \(s,t>0\), the four connected components distinguish whether spatial and temporal orientation are preserved or reversed.
The elementary reflections \(\p\) and \(\t\) reverse one spatial and one temporal coordinate and are represented by diagonal matrices \(\Pmat\) and \(\Tmat\), respectively.
Each matrix flips one coordinate and satisfies
\begin{align}
\Pmat^2 & = \Tmat^2 = \matident , &
\Pmat\Tmat & = \Tmat\Pmat , &
\det\Pmat & = \det\Tmat = -1 .
\end{align}
The collective inversions \(\P\) and \(\T\) act on \(x = (x_s,x_t)\), with \(x_s\in\mathbb R^s\) and \(x_t\in\mathbb R^t\), as
\begin{align}
\P x & = (-x_s,x_t) , &
\T x & = (x_s,-x_t) .
\end{align}
The two independent orientation signs give the Klein four-group of connected components,
\begin{align}
\pi_0(\O(s,t)) &\cong \D^4_\klein , &
\D^4_\klein &\cong \C^2_\p \times \C^2_\t .
\end{align}
Throughout, \(\properSO(s,t)\) denotes the connected identity component of \(\O(s,t)\), often written \(\SO_0(s,t)\) or \(\SO^+(s,t)\), whereas \(\O_\pt(s,t)\) denotes the full determinant-one subgroup for \(s,t>0\).
This convention differs from the common use of \(\SO(s,t)\) for the full determinant-one subgroup.
With this convention, the four disconnected components are
\begin{align}
\O(s,t) & = \properSO(s,t)
\sqcup \Pmat\properSO(s,t)
\sqcup \Tmat\properSO(s,t)
\sqcup \Pmat\Tmat\properSO(s,t) .
\end{align}
The determinant is positive on \(\properSO(s,t)\) and \(\Pmat\Tmat\properSO(s,t)\) and negative on the other two components.
The special orthogonal groups with exchanged spatial and temporal dimensions are isomorphic, and when either dimension vanishes they reduce to the definite special orthogonal group in \eqref{eq:orthogonal parity 2}:
\begin{align} \label{eq:SO symmetry}
\properSO(s,t) &\cong \properSO(t,s) , &
\properSO(d,0) &\cong \properSO(0,d) \cong \properSO(d) .
\end{align}
For \(s,t>0\), the component decomposition and \(\pi_0(\properSO(s,t)) = 0\) give
\begin{align} \label{eq:indefinite orthogonal}
\shortexact{\properSO(s,t)}{\O(s,t)}{\D^4_\klein} .
\end{align}
As in the Euclidean case, the connected special orthogonal group need not be simply connected.
Because its maximal compact subgroup is \(\SO(s) \times \SO(t)\), its first homotopy group factorises as \(\pi_1(\properSO(s, t)) = \pi_1(\SO(s)) \times \pi_1(\SO(t))\) \cite{Sati:2015ena}.
Using the Euclidean result \eqref{eq:first homotopy} then gives
\begin{align} \label{eq:indefinite homotopy}
\pi_1(\properSO(s, t)) & =
\begin{cases*}
\begin{array}{rccc} \toprule
s & \multicolumn3ct \\ \cmidrule{2-4}
& \leq1 & 2 & \geq 3 \\ \midrule
\leq1 & 0 & \C_\cover & \C^2_\cover \\
2 & \C_\cover & \C_\cover \times \C_\cover & \C_\cover \times \C^2_\cover \\
\geq 3 & \C^2_\cover & \C_\cover \times \C^2_\cover & \C^2_\cover \times \C^2_\cover \\
\bottomrule
\end{array}
\end{cases*}
.
\end{align}
Whenever this fundamental group is nontrivial, spinorial representations require a cover of the indefinite orthogonal group.

\subsection{Indefinite spin and pin groups}

The spin group \(\properSpin(s,t)\) is the double cover of the special orthogonal group \(\properSO(s,t)\)
\begin{align} \label{eq:indefinite spin}
\shortexact{\C^2_\cover}{\properSpin(s,t)}{\properSO(s,t)} .
\end{align}
For \(s+t\geq1\), its connected components are described by
\begin{equation}
\pi_0(\properSpin(s,t))
\cong
\begin{cases}
\C^2_\cover & \text{for } (s,t) \in
\set{(1,0),(0,1),(1,1)},\\
0 & \text{otherwise},
\end{cases}
\end{equation}
and its first homotopy group, based at the identity, is
\begin{align} \label{eq:indefinite homotopy 2}
\pi_1(\properSpin(s, t)) & =
\begin{cases*}
\begin{array}{rccc} \toprule
s & \multicolumn3ct \\ \cmidrule{2-4}
& \leq1 & 2 & \geq 3 \\ \midrule
\leq1 & 0 & \C_\cover & 0 \\
2 & \C_\cover & \C_\cover \times \C_\cover & \C_\cover \\
\geq 3 & 0 & \C_\cover & \C^2_\cover \\
\bottomrule
\end{array}
\end{cases*}
.
\end{align}
For \(d\geq3\), the proper spin group is the universal cover of the proper special orthogonal group when \(\min(s,t) \leq1\) and \(\max(s,t) \geq3\).
As for the special orthogonal groups in \eqref{eq:SO symmetry}, exchanging the spatial and temporal dimensions gives isomorphic spin groups.
When either dimension vanishes, the result can be identified with the Euclidean spin group in \eqref{eq:Euclidean spin}:
\begin{align}
\properSpin(s,t) &\cong \properSpin(t,s) , &
\properSpin(d,0) &\cong \properSpin(0,d) \cong \Spin(d) .
\end{align}

\begin{table}
\begin{tabular}{rr*4cl} \toprule
& \(H^8_\pin\) & \(i\) & \(\ptlift^2\) & \(\plift^2\) & \(\tlift^2\) & structure \\ \midrule
\multirow4*[-6pt]{\(\comm\plift\tlift = 0\)} & \(\D^2 \times \D^4\) & \(0\) & \(+\) & \(+\) & \(+\) & \(\C^2_\cover \times \C^2_\plift \times \C^2_\tlift\) \\ \cmidrule{2-7}
& & \multirow3*{\(2\)} & \multirow2*{\(-\)} & \(+\) & \(-\) & \(\C^4_{\cover,\tlift} \times \C^2_\plift\) \\
& \(\D^2 \times \Q^4\) & & & \(-\) & \(+\) & \(\C^4_{\cover,\plift} \times \C^2_\tlift\) \\ \cmidrule{5-7}
& & & \(+\) & \(-\) & \(-\) &
\(\C^4_{\cover,\plift}\times \C^2_\ptlift \cong \C^4_{\cover,\tlift}\times \C^2_\ptlift\) \\ \midrule
\multirow4*[-6pt]{\(\acomm\plift\tlift = 0\)} & & \multirow3*[-1ex]{\(1\)} & \(-\) & \(+\) & \(+\) & \(\D^8_{\cover,\plift,\tlift}\) \\ \cmidrule{5-7}
& \(\D^8\) & & \multirow2*{\(+\)} & \(+\) & \(-\) & generates \(\Pin(s,t)\) \\
& & & & \(-\) & \(+\) & generates \(\Pin(t,s)\) \\ \cmidrule{2-7}
& \(\Q^8\) & \(3\) & \(-\) & \(-\) & \(-\) & \(\Q^8_{\cover,\plift,\tlift}\) \\
\bottomrule \end{tabular}
\caption[Double covers of the Klein four-group]{
The Klein four-group has eight double covers belonging to four group isomorphism types.
Two product groups have commuting lifts \(\plift\) and \(\tlift\), while two further groups have anticommuting lifts \cite{Dummit:2003}.
The index \(i\) counts how many of the three nontrivial coset lifts square to minus the identity.
Only the two dihedral \(\D^8\) double covers with \(\ptlift^2 = +\ident\) are realised by the real Clifford constructions \(\Pin(s,t)\) and \(\Pin(t,s)\), for which the spatial and temporal reflection lifts anticommute and square with opposite signs.
} \label{tab:double covers}
\end{table}

For \(s,t>0\), the pin group constructed in the real Clifford algebra satisfies
\begin{align} \label{eq:indefinite pin}
\shortexact{\C^2_\cover}{\Pin(s,t)}{\O(s,t)} .
\end{align}
The possible double covers are determined by central extensions of the Klein four-group in \eqref{eq:indefinite orthogonal}:
\begin{align} \label{eq:Z^2_2 double covers}
\shortexact{\C^2_\cover}{H^8_\pin}{\D^4_\klein} .
\end{align}
For a chosen lift \(\ptlift\) of \(\pt\), the superscript signs of \(\Pin_{a,b,c}(s,t)\) are ordered according to
\begin{align} \label{eq:pin superscript convention}
\ptlift^2 & = a\ident , &
\plift^2 & = b\ident , &
\tlift^2 & = c\ident , &
a,b,c &\in\set{+1,-1} .
\end{align}
Although \(\p\) and \(\t\) commute, their lifts either commute or anticommute, and hence
\begin{align}
\ptlift & = \pm\tlift\plift , &
\ptlift^2 & = \pm\plift^2\tlift^2 .
\end{align}
The commuting cases give the abelian covers, while the anticommuting cases give the non-abelian covers.
The resulting eight covers and their structures are listed in \cref{tab:double covers} \cite{Trautman:2005qx}.
Each discrete cover extends the proper spin group to a double cover of the indefinite orthogonal group \cite{Dabrowski:1988kq}:
\begin{equation}\label{eq:indefinite pin 2}
\shortexact{\C^2_\cover}{\properSpin(s,t) \rtimes H^8_\pin}{\Pin_\pin(s,t)} .
\end{equation}
Only two of these eight double covers arise from the real Clifford algebra.
Their discrete groups are dihedral, with anticommuting generators whose squares have the signs of a spacelike and a timelike Clifford generator, respectively.
The product \(\ptlift\) is consequently an even Clifford element with \(\ptlift^2 = +\ident\) and represents a total-orientation-preserving spacetime transformation \cite{Trautman:2005qx,Berg:2000ne}.
With the sign ordering in \eqref{eq:pin superscript convention}, these covers are denoted by
\begin{align}
\Pin(s,t) & = \Pin_{+,+,-}(s,t) , &
\Pin(t,s) &\cong \Pin_{+,-,+}(s,t) .
\end{align}
The remaining six covers do not arise from this real Clifford-algebra construction, and their associated generalised Pin structures obey different global obstruction conditions \cite{Chamblin:1994crn}.
When either the temporal or spatial dimension vanishes, the connection with the definite case \eqref{eq:definite pin} is
\begin{align}
\Pin(d,0) & = \Pin_+(d) , &
\Pin(0,d) & = \Pin_-(d) .
\end{align}

\subsection{Improper orthogonal, spin, and pin groups}

\begin{table}
\begin{panels}2
\begin{tabular}{rr@{}lr@{}l} \toprule
& \multicolumn2c{\(\ident\)} & \multicolumn2c{\(\t\)} \\ \cmidrule{2-5}
\(\ident\) & \(\properSO\) & & \(\O\) & \({}_\t\) \\
\(\p\) & \(\O\) & \({}_\p\) & \(\O\) & \({}_\pt\) \\
\bottomrule \end{tabular}
\caption{Orthogonal subgroups.} \label{tab:orthogonal groups}
\panel
\begin{tabular}{rr@{}lr@{}l} \toprule
& \multicolumn2c{\(\ident\)} & \multicolumn2c{\(\tlift\)} \\ \cmidrule{2-5}
\(\ident\) & \(\properSpin\) & & \(\Pin\) & \({}_\tlift\) \\
\(\plift\) & \(\Pin\) & \({}_\plift\) & \(\Pin\) & \({}_\ptlift\) \\
\bottomrule \end{tabular}
\caption{Pin subgroups.} \label{tab:pin groups}
\end{panels}
\caption[Indefinite orthogonal and Pin subgroups]{
For \(s,t>0\), subgroups of the indefinite orthogonal group \(\O(s,t)\) and its double-covering group \(\Pin(s,t)\) are given in panels \subref{tab:orthogonal groups} and \subref{tab:pin groups}, respectively.
Spatial and temporal representatives adjoined to the identity component are arranged by row and column, using orthogonal matrices in panel \subref{tab:orthogonal groups} and their Clifford lifts in panel \subref{tab:pin groups}.
The subscripts label the reflection or reflection lift contained in each subgroup.
The displayed Pin groups are the inverse images of the orthogonal subgroups inside the Cliffordian group \(\Pin(s,t)\).
}
\end{table}

For \(s,t>0\), the determinant-one subgroup of the indefinite orthogonal group is the total-orientation-preserving improper special orthogonal group
\begin{align}
\O_\pt(s,t) & = \properSO(s,t)
\sqcup \Pmat\Tmat\properSO(s,t) .
\end{align}
Its two connected components form the component group \(\pi_0(\O_\pt(s,t))\cong\C^2_\pt\), which can be written as the quotient
\begin{align}
\shortexact{\properSO(s, t)}{\O_\pt(s, t)}{\C^2_\pt}.
\end{align}
Extending the fixed double cover \(\properSpin(s,t)\) across this \(\C^2_\pt\), in analogy with the definite case in \eqref{eq:Z^2 double covers}, gives two improper spin groups distinguished by \(\ptlift^2 = a\ident\) with \(a\in\set{+1,-1}\):
\begin{equation} \label{eq:spin double covers}
\shortexact{\C^2_\cover}{\properSpin(s,t) \rtimes H^4_a}{\Pin_{\ptlift}^a(s,t)}.
\end{equation}
The corresponding discrete extensions \(H^4_a\) are dihedral for \(a = +1\) and dicyclic for \(a = -1\), explicitly
\begin{align} \label{eq:spin double covers explicit} &\shortexact{\C^2_\cover}{\properSpin(s,t) \rtimes \D^4_{\cover,\ptlift}}{\Pin_{\ptlift+}(s,t)} , & &\shortexact{\C^2_\cover}{\properSpin(s,t) \rtimes \Q^4_{\cover,\ptlift}}{\Pin_{\ptlift-}(s,t)} .
\end{align}
Only the dihedral discrete extension has \(\ptlift^2 = +\ident\), so the improper Pin subgroup realised inside \(\Pin(s,t)\) is
\begin{align}
\Pin_\ptlift(s,t) &\cong \Pin_{\ptlift+}(s,t) .
\end{align}
It double covers the improper special orthogonal group through
\begin{align} \label{eq:improper spin cover}
\shortexact{\C^2_\cover}{\Pin_\ptlift(s,t)}{\O_\pt(s,t)} .
\end{align}
Exchanging the signs of the metric leaves the indefinite improper spin group isomorphic,
\begin{align}
\Pin_\ptlift(t,s) &\cong \Pin_\ptlift(s,t) .
\end{align}

Adjoining only one reflection component instead gives the subgroups
\begin{align}
\O_\p(s,t) & = \properSO(s,t)
\sqcup \Pmat\properSO(s,t) , &
\O_\t(s,t) & = \properSO(s,t)
\sqcup \Tmat\properSO(s,t) .
\end{align}
The groups \(\O_\p(s,t)\) and \(\O_\t(s,t)\) preserve temporal and spatial orientation, respectively.
Their component groups are the quotients
\begin{align} &\shortexact{\properSO(s, t)}{\O_\p(s, t)}{\C^2_\p} , & &\shortexact{\properSO(s, t)}{\O_\t(s, t)}{\C^2_\t}.
\end{align}
For the fixed double cover \(\properSpin(s,t)\), each subgroup likewise admits two extensions distinguished by the square of its reflection lift.
Inside \(\Pin(s,t)\), the Cliffordian cover of \(\O_\p(s,t)\) contains \(\plift\) with \(\plift^2 = +\ident\), whereas that of \(\O_\t(s,t)\) contains \(\tlift\) with \(\tlift^2 = -\ident\):
\begin{align} &\shortexact{\C^2_\cover}{\properSpin(s,t) \rtimes \D^4_{\cover,\plift}}{\Pin_\plift(s,t)} , & &\shortexact{\C^2_\cover}{\properSpin(s,t) \rtimes \Q^4_{\cover,\tlift}}{\Pin_\tlift(s,t)}.
\end{align}
The resulting Pin subgroups double cover their orthogonal counterparts:
\begin{align} \label{eq:parity double covers} &\shortexact{\C^2_\cover}{\Pin_\plift(s,t)}{\O_\p(s,t)} , & &\shortexact{\C^2_\cover}{\Pin_\tlift(s,t)}{\O_\t(s,t)}
.
\end{align}

\subsection{Complexification}

Complexification removes the distinction between real signatures because factors of~\(\i\) can absorb the metric signs.
All non-degenerate quadratic forms of dimension \(d\) are therefore equivalent over the complex numbers, and every \(\O(s,t)\) complexifies to \(\O(d,\mathbb C)\).

The determinant still takes the two values \(\det \Mmat = \pm1\).
The identity component is the complex special orthogonal group \(\SO(d,\mathbb C)\), while the second component is obtained by composing with a single reflection.
The corresponding Cliffordian double covers are the complex spin and pin groups
\begin{align} \label{eq:complex spin pin covers} &\shortexact{\C^2_\cover}{\Spin(d,\mathbb C)}{\SO(d,\mathbb C)} , & &\shortexact{\C^2_\cover}{\Pin(d,\mathbb C)}{\O(d,\mathbb C)} .
\end{align}
As in the Euclidean case, \(\O(d,\mathbb C)\) has a connected special-orthogonal component and a second component generated by one reflection, while passing from its Spin to its Pin cover enlarges the allowed representatives from even to general Clifford elements.
Unlike over the real numbers, however, multiplying a reflection generator by \(\i\) changes the sign of its square, so the two real Euclidean Pin groups become equivalent and the complex classification retains only the total dimension and determinant component.

\section{Clifford algebra} \label{sec:clifford algebra}

For a \(d = s+t\)-dimensional spacetime \(V^{s,t}\), the Clifford algebra \(\cl(s,t)\) is generated by elements \(\gamma_i\) satisfying
\begin{equation} \label{eq:Clifford algebra}
\frac{\acomm{\gamma_i}{\gamma_j}}2 = \eta_{ij} \ident ,
\end{equation}
where the metric \(\eta_{ij}\) is diagonal with \(s\) spatial entries \(+1\) and \(t\) temporal entries \(-1\).
As an associative algebra, \(\cl(s,t)\) is isomorphic to a matrix algebra over \(\mathbb R\), \(\mathbb C\), or \(\mathbb H\), or to a direct sum of two such matrix algebras.
\footnote{ \label{fn:quaternions}%
The quaternionic numbers \((\qu 1, \qu i, \qu j, \qu k) \in \mathbb H\) are defined by their algebra
\begin{equation*}
\begin{aligned}
\qu i^2 = \qu j^2 = \qu k^2 = \qu i \qu j \qu k & = - \qu 1 , &
\qu i \qu j & = \qu k , &
\qu j \qu k & = \qu i , &
\qu k \qu i & = \qu j .
\end{aligned}
\end{equation*}
A two-dimensional complex representation follows by identifying the quaternionic units with Pauli matrices \(\sigma_i \in M^2(\mathbb C)\):
\begin{equation*}
\begin{aligned}
\qu 1 & = \sigma_0 , &
\qu i & = - \i \sigma_1 , &
\qu j & = - \i \sigma_2 , &
\qu k & = - \i \sigma_3 .
\end{aligned}
\end{equation*}
Under this embedding, Hermitian conjugation of the complex representative implements quaternionic conjugation.
}
The matrix-algebra type identifies the available real or quaternionic module structure, while a direct sum signals two inequivalent irreducible Clifford modules.
\footnote{
For \(n\)-by-\(n\) matrices over \(\mathbb F = \mathbb R\), \(\mathbb C\), or \(\mathbb H\), we use interchangeably the notations \(\mathbb F^n = M^n(\mathbb F) \cong M(n, \mathbb F)\).
The direct sum of two such algebras is interchangeably labelled \(2\mathbb F^n = M^n(\mathbb F) \oplus M^n(\mathbb F) = 2M(n,\mathbb F)\).
}
As a real vector space the Clifford algebra has dimension \(2^d\), so each simple matrix block has size \(n = 2^{\floor{d/2}}\) in the real and complex cases and \(n = 2^{\floor{d/2-1}}\) in the quaternionic cases.
A faithful block-diagonal realisation of a split algebra \(2M^n(\mathbb F)\) has matrix size \(2n\).
Organising the Clifford basis by the number of generators gives
\begin{align} \label{eq:clifford basis}
 \basis & = \bigcup_{a = 0}^d \basis^a , & \basis^0 & = \set*{\ident} , & \basis^1 & = \set*{\gamma_i} , & \basis^2 & = \set*{\gamma_{[i}\gamma_{j]}} , & \basis^3 & = \set*{\gamma_{[i}\gamma_j\gamma_{k]}} ,
\end{align}
where the products are antisymmetrised.
The sectors \(\basis^0\), \(\basis^1\), and \(\basis^2\) are the scalar, vector, and antisymmetric-tensor Clifford structures, while the higher sectors supply their higher-rank analogues and, in even dimension, their Hodge duals.
A generic basis element containing \(a\) generators is denoted by
\begin{align} \label{eq:clifford basis elements}
 \element^a & = \gamma_{[i_1}\dots\gamma_{i_a]}
 \in \basis^a , & 1\leq i_1<\dots<i_a &\leq d ,
\end{align}
with \(\element^0 = \ident\).
The increasing order selects one representative for each independent antisymmetrised product.

\paragraph{Involutions}

\begin{table}
\begin{tabular}{l*4cl} \toprule
\(\eta\) & \multicolumn4c{\(a \bmod 4\)} \\ \cmidrule{2-5}
& \(0\) & \(1\) & \(2\) & \(3\) \\ \midrule
\(\grade\) & \(+\) & \(-\) & \(+\) & \(-\) & \((-1)^a\) \\
\(\rev\) & \(+\) & \(+\) & \(-\) & \(-\) & \((-1)^{\floor{a/2}} = (-1)^{a(a-1)/2}\) \\
\(\cliff\) & \(+\) & \(-\) & \(-\) & \(+\) & \((-1)^{\ceil{a/2}} = (-1)^{a(a+1)/2}\) \\
\bottomrule \end{tabular}
\caption[Clifford involution signatures]{
Signatures of the grade involution, reversion, and Clifford conjugation as functions of the order \(a\) of the homogeneous basis element \(\element^a\).
} \label{tab:involution}
\end{table}

Three involutions provide the elementary sign operations that enter Hermitian conjugation, Majorana structures, and fermion bilinears; applying any one of them twice returns the original Clifford element.
\begin{itemize}
\item The grade involution takes every generator to its negative, \((\gamma_i)_\grade = -\gamma_i\), and therefore acts on a rank-\(a\) basis element as
\begin{align} \label{eq:negation}
\element^a_\grade & = \eta_\grade^{}(a) \element^a , &
\eta_\grade^{}(a) & = (-1)^a .
\end{align}
\item The reversion involution reverses the order of the generators in a basis element:
\begin{align} \label{eq:reversion}
(\gamma_{[1} \dots \gamma_{a]})_\rev & = \gamma_{[a} \dots \gamma_{1]} = \eta_\rev^{}(a) \gamma_{[1} \dots \gamma_{a]} , &
\eta_\rev^{}(a) & = (-1)^{\floor{a/2}} = (-1)^{a(a-1)/2} .
\end{align}
\item Clifford conjugation is the combination of these involutions \(\element_\cliff = \element_{\rev\grade}\) and acts on a basis element as
\begin{align} \label{eq:cliff}
(\gamma_{[1} \dots \gamma_{a]})_\cliff & = (-1)^a \gamma_{[a} \dots \gamma_{1]} = \eta_\cliff^{}(a) \gamma_{[1} \dots \gamma_{a]} , &
\eta_\cliff^{}(a) = (-1)^{\ceil{a/2}} = (-1)^{a(a+1)/2} .
\end{align}
\end{itemize}
Their fourfold sign periodicity in the Clifford rank is summarised in \cref{tab:involution}.

The action of collective spatial and temporal inversions on the Clifford generators is \cite{Harvey:1990}
\begin{align} \label{eq:generator reflection}
(\gamma_i)_\P & = -\eta^{ii} \gamma_i , &
(\gamma_i)_\T & = \eta^{ii} \gamma_i , &
(\gamma_i)_\PT & = (\gamma_i)_\grade = -\gamma_i .
\end{align}
Hence collective spatial inversion reverses all spatial generators, collective temporal inversion reverses all temporal generators, and their combination implements the grade involution.
For a homogeneous basis element containing \(a^s\) spatial and \(a^\ft\) temporal generators, with \(a = a^s+a^\ft\), these inversions act as
\begin{align} \label{eq:reflection-signs-k-form}
\element^a_\P & = \eta_\grade^{}(a^s) \element^a , &
\element^a_\T & = \eta_\grade^{}(a^\ft) \element^a , &
\element^a_\PT & = \eta_\grade^{}(a) \element^a .
\end{align}
For (pseudo-)Euclidean signatures, these operations simplify to
\begin{align}
\element^a_\P & = \begin{cases}
\element^a_\grade & \text{for } t = 0,\\
\element^a & \text{for } s = 0,
\end{cases} &
\element^a_\T & = \begin{cases}
\element^a & \text{for } t = 0,\\
\element^a_\grade & \text{for } s = 0.
\end{cases}
\end{align}

The invertible elements of the Clifford algebra form the Clifford group \(\Cl(s,t)\) with the ordinary adjoint action
\begin{equation} \label{eq:adjoint action}
\begin{aligned}
\Ad_g(\gamma) & = g \gamma g^{-1} , &
\gamma &\in \cl(s,t) , &
g &\in \Cl(s,t).
\end{aligned}
\end{equation}
Although ordinary conjugation is the natural action on the algebra, an odd Clifford element does not produce the required orthogonal reflection on the embedded vector space.
For a non-null vector \(v\in V^{s,t}\), with inverse \(v^{-1} = v/v^2\), the ordinary adjoint fixes the line spanned by \(v\) and reverses its orthogonal complement, whereas a hyperplane reflection must reverse \(v\) and leave \(v_\perp\) fixed.
The required reflection is instead produced by the twisted adjoint action
\begin{equation} \label{eq:twisted adjoint action}
\begin{aligned}
\Ad^\PT_g(\gamma) & = g_\PT^{} \gamma g^{-1} , &
\gamma &\in \cl(s,t) , &
g &\in \Cl(s,t).
\end{aligned}
\end{equation}
For a non-null vector \(v\), one has \(v_\PT = -v\), and hence
\begin{equation} \label{eq:twisted vector reflection}
\Ad^\PT_v(u) = -v u v^{-1}.
\end{equation}
The twisted adjoint by \(v\) consequently acts on \(V^{s,t}\) as a hyperplane reflection, while products of vectors generate successive reflections in agreement with the \CD construction of the orthogonal group.
For even Clifford elements the twist is invisible, and the twisted adjoint therefore reduces to the ordinary adjoint on the Spin subgroup.

\paragraph{Even subalgebra}

The Clifford algebra elements that are even under the grade involution form the \(\PT\)-even subalgebra
\begin{equation} \label{eq:even Clifford algebra}
 \cl_\PT = \set{\element \in \cl \suchthat \element_\PT = \element} .
\end{equation}
Its vector-space basis consists of the sectors containing an even number of Clifford generators,
\begin{equation}
 \basis_\PT = \basis^0 \cup \basis^2 \cup \basis^4 \cup \dots .
\end{equation}
Infinitesimal Spin transformations are generated by the bivectors \(\gamma_{[i}\gamma_{j]}\in\basis^2\), while finite Spin transformations are represented by even products of unit vectors.
A single unit vector, represented in the Clifford algebra by \(v = v^i\gamma_i\), generates a reflection in the Pin group.
The distinction between even and odd Clifford elements explains why connected Lorentz transformations act within a Spin module, whereas including a reflection can require its extension to a Pin module.

\paragraph{Spin algebra}

The bivectors of the even Clifford algebra generate the spin algebra,
\begin{align}
\spin(s,t) & = \operatorname{span}_{\mathbb R}
\set*{
\gamma_{[i} \gamma_{j]}
\suchthat
i<j
}
\spininside
\left(
\cl_\PT(s,t),
{\comm{\cdot}{\cdot}}_{\cl}
\right) , &
{\comm{X}{Y}}_{\cl} & = XY-YX ,
\end{align}
and their commutators reproduce the \(\so(s,t)\) relations in Clifford form:
\begin{align}
\frac{{\comm{\gamma_{ij}}{\gamma_{kl}}_{\cl}}}2 = \eta_{jk}\gamma_{il}
-\eta_{ik}\gamma_{jl}
-\eta_{jl}\gamma_{ik}
+\eta_{il}\gamma_{jk}
.
\end{align}
Up to conventional normalisation, these bivectors are the Lorentz generators entering the spin connection and fermion covariant derivative, so the standard Lie-algebra action is recovered directly from the rank-two sector of the associative Clifford algebra.

\paragraph{Volume element}

The ordered product of all \(d\) generators defines the volume element
\begin{equation} \label{eq:volume element}
\omega = \element^d = \gamma_{[1} \gamma_{\vphantom[2} \dots \gamma_{\vphantom[d-1} \gamma_{d]} .
\end{equation}
\footnote{In four-dimensional Lorentzian spacetimes the volume element is closely related to \(\gamma_5\), with the crucial difference that for real Clifford algebras the volume element \(\omega\) cannot include factors of the imaginary unit \(\i\) and can therefore square to negative unity.}
By the anticommutation relation \eqref{eq:Clifford algebra}, its square is proportional to the identity, and the sign follows from reordering the generators and from the \(t\) negative metric entries:
\begin{align} \label{eq:volume element periodicity}
\omega^2 & = \eta_\rev^{}(\delta) \ident , &
\delta & = s - t .
\end{align}
\begin{itemize}
\item In \emph{odd dimensions} the volume element commutes with all generators and therefore lies in the centre of the Clifford algebra.
It acts as a scalar on each irreducible module, although these scalars need not agree between distinct simple summands.
When the algebra consists of complex matrices \(\mathbb C^n\), the volume element is proportional to the imaginary unit matrix \(\omega = \pm \i \matident^n\).
When the algebra consists of two copies, as in \(2\mathbb R^n\) and \(2\mathbb H^n\), the generators can be expressed with unified matrices of the form \(
\begin{psmallmatrix}1&0\\0&-1\end{psmallmatrix} \otimes \gamma_i\).
In these cases the volume element is proportional to \(\omega = \pm \begin{psmallmatrix}1&0\\0&-1\end{psmallmatrix} \otimes \matident^n\).

\item In \emph{even dimensions} the volume element anticommutes with all generators of the Clifford algebra and consequently commutes with the generators of the even subalgebra.
Furthermore, it serves as a negation intertwiner, since it can be used in a similarity transformation that takes all generators to their negative.
\begin{align} \label{eq:even negation intertwiner}
\omega \gamma_i \inv \omega & = - \gamma_i , &
\omega \element^a \inv \omega & = \eta_\grade^{}(a) \element^a .
\end{align}
\end{itemize}

Combining the even- and odd-dimensional cases, conjugation by the volume element acts on the generators as
\begin{equation}
\omega \gamma_i \inv \omega = - \eta_\grade^{}(d) \gamma_i ,
\end{equation}
and on a homogeneous element as
\begin{align} \label{eq:negation intertwiner}
\omega \element^a \inv \omega & = \eta_\grade^{}(a) \eta_\grade^{}(da) \element^a = [\xi_\grade^+(a) - \eta_\grade^{}(d) \xi_\grade^-(a)] \element^a , &
\xi_\grade^\pm(a) & = \frac{1\pm\eta_\grade^{}(a)}2
.
\end{align}

Adjoining \(\omega\) to an even-dimensional set of generators produces the Clifford relations of an odd \(d + 1\)-dimensional algebra on the original module, with the additional metric sign fixed by \eqref{eq:volume element periodicity}.
For even-dimensional metrics with \(\delta^+ \bmod 4 = 0,2\), the resulting odd-dimensional signature consequently satisfies
\begin{equation}\label{eq:omega as generator}
\delta^- =
\delta^+ + (\omega^+)^2
 = 1\bmod 4,
\end{equation}
where \((\omega^+)^2\) denotes the sign of the additional generator.
Because the odd-dimensional algebra is split and has twice the dimension of the original even-dimensional algebra, this construction represents only one simple summand rather than the full algebra faithfully.
Replacing \(\omega\) by \(-\omega\) selects the other summand; the arrows in \cref{tab:clifford algebras} display this relation between the even algebra and the two odd-dimensional representations.

\paragraph{Spinor structure map}

In even dimensions, the volume element supplies the negation intertwiner \eqref{eq:even negation intertwiner} for Clifford algebras over \(\mathbb R\) or \(\mathbb H\), whereas in odd dimensions it is central and cannot negate the generators.
Relating the two grade-related irreducible sectors in odd dimension therefore requires a structure map acting as
\begin{align} \label{eq:structure map action}
\structure \gamma_i \structure^{-1} & = -\gamma_i , &
\structure \element^a \structure^{-1} & = \eta_\grade^{}(a) \element^a , &
\structure \omega \structure^{-1} & = \eta_\grade^{}(d) \omega .
\end{align}
In these odd-dimensional cases, the structure map lies outside the represented Clifford algebra and maps between grade-related representations.
For two copies of real or quaternionic matrices, the generators can be chosen in the form \(\begin{psmallmatrix}a&0\\0&-a\end{psmallmatrix}\), so the structure map exchanges the two simple summands.
For the real and quaternionic cases, representative linear structure maps are
\begin{equation} \label{eq:structure map}
\structure =
\begin{cases}
\omega & \text{for } \mathbb R^n,\mathbb H^n, \\
\begin{psmallmatrix}0&\matident\\ \matident&0\end{psmallmatrix} & \text{for } 2\mathbb R^n,2\mathbb H^n .
\end{cases}
\end{equation}
For the complex matrix algebra \(\mathbb C^n\), the structure map is complex conjugation with respect to the central complex structure defined by the volume element.
It therefore fixes the even subalgebra, reverses the volume element, and implements the grade involution on the full Clifford algebra.

\subsection{Classification of Clifford algebras over the real numbers} \label{sec:real clifford algebras}

Since physical spacetime \(V^{s,t}\) is real, the natural starting point is the real Clifford algebra \(\cl(s,t)\): its full algebra controls Pin representations and reflections, while its even subalgebra generates Spin representations of the connected rotation group \(\properSO(s,t)\).
Classifying the associated fermion modules therefore reduces to identifying the matrix-algebra types of \(\cl(s,t)\) and \(\cl_\PT(s,t)\).
Signatures with \(s = t\) are referred to as split.
For \(d\geq2\), the cases \(t = 0\) and \(s = 0\) are called Euclidean and pseudo-Euclidean, respectively.
For \(d\geq3\), the cases \(t = 1\) and \(s = 1\) are called Lorentzian and pseudo-Lorentzian, respectively.
The conventions are first fixed in the lowest dimensions and then extended by an iterative construction that exposes the generic periodicity.

\subsubsection{Low-dimensional cases} \label{sec:low dim examples}

\begin{table}
\tikzset{every node/.style = {inner sep = 1ex}}
\newcommand{\symmetry}[1]{{\tikz[remember picture]\node(#1){};}}%
\newcommand{\volume}[2]{{\tikz[remember picture]\node[minimum width = 0pt,inner sep = 0pt, outer sep = 0pt,minimum height = 1.5ex](#1){#2};}}%
\newcolumntype{C}[1]{>{\centering\let\newline\\\arraybackslash}b{#1}}%
\NewDocumentCommand{\mathbbmat}{omm}{\IfValueT{#1}{#1}#3^{#2}}%
\begin{tabular}{@{ }l@{\hspace{1em}}*{19}{@{}C{2em}@{}}@{\hspace{.25em}}r@{}l@{ }}\toprule
\(d\) & & \multicolumn{17}c{\(\delta\)} & & \(\mathbb C\) \\ \cmidrule{3-19}
& & \(-8\) & \(-7\) & \(-6\) & \(-5\) & \(-4\) & \(-3\) & \(-2\) & \(-1\) & \(0\) & \(1\) & \(2\) & \(3\) & \(4\) & \(5\) & \(6\) & \(7\) & \(8\) \\ \midrule
\(0\) & & && && && && \volume{00}{\(\mathbb R\)} & \symmetry{01} & && && && && \(\mathbb C\) \\
\(1\) & & && && && & \(\mathbb C\) & & \volume{11}{\(2\mathbb R\)} & && && && && \(2\mathbb C\) \\
\(2\) & & && && & \symmetry{2-3} & \volume{2-2}{\(\mathbb H\)} & & \volume{20}{\(\mathbbmat2{\mathbb R}\)} & & \volume{22}{\(\mathbbmat2{\mathbb R}\)} & && && && & \(\mathbb C\) & \(^2\) \\
\(3\) & & && && & \volume{3-3}{\(2\mathbb H\)} & \symmetry{3-2} & \(\mathbbmat2{\mathbb C}\) & & \volume{31}{\(\mathbbmat[2]2{\mathbb R}\)} & & \(\mathbbmat2{\mathbb C}\) & && && && \(2\mathbb C\) & \(^2\) \\
\(4\) & & && && \volume{4-4}{\(\mathbbmat2{\mathbb H}\)} & & \volume{4-2}{\(\mathbbmat2{\mathbb H}\)} & & \volume{40}{\(\mathbbmat4{\mathbb R}\)} & & \volume{42}{\(\mathbbmat4{\mathbb R}\)} & & \volume{44}{\(\mathbbmat2{\mathbb H}\)} & \symmetry{45} & && && \(\mathbb C\) & \(^4\) \\
\(5\) & & && & \(\mathbbmat4{\mathbb C}\) & & \volume{5-3}{\(\mathbbmat[2]2{\mathbb H}\)} & & \(\mathbbmat4{\mathbb C}\) & & \volume{51}{\(\mathbbmat[2]4{\mathbb R}\)} & & \(\mathbbmat4{\mathbb C}\) & & \volume{55}{\(\mathbbmat[2]2{\mathbb H}\)} & && && \(2\mathbb C\) & \(^4\) \\
\(6\) & & & \symmetry{6-7} & \volume{6-6}{\(\mathbbmat8{\mathbb R}\)} & & \volume{6-4}{\(\mathbbmat4{\mathbb H}\)} & & \volume{6-2}{\(\mathbbmat4{\mathbb H}\)} & & \volume{60}{\(\mathbbmat8{\mathbb R}\)} & & \volume{62}{\(\mathbbmat8{\mathbb R}\)} & & \volume{64}{\(\mathbbmat4{\mathbb H}\)} & & \volume{66}{\(\mathbbmat4{\mathbb H}\)} & && & \(\mathbb C\) & \(^8\) \\
\(7\) & & & \volume{7-7}{\(\mathbbmat[2]8{\mathbb R}\)} & \symmetry{7-6} & \(\mathbbmat8{\mathbb C}\) & & \volume{7-3}{\(\mathbbmat[2]4{\mathbb H}\)} & & \(\mathbbmat8{\mathbb C}\) & & \volume{71}{\(\mathbbmat[2]8{\mathbb R}\)} & \symmetry{72} & \(\mathbbmat8{\mathbb C}\) & & \volume{75}{\(\mathbbmat[2]4{\mathbb H}\)} & & \(\mathbbmat8{\mathbb C}\) & && \(2\mathbb C\) & \(^8\) \\
\(8\) & \symmetry{8-8} & \(\mathbbmat{16}{\mathbb R}\) & \symmetry{8-7} & \(\mathbbmat{16}{\mathbb R}\) & \symmetry{8-5} & \(\mathbbmat8{\mathbb H}\) & & \(\mathbbmat8{\mathbb H}\) & & \(\mathbbmat{16}{\mathbb R}\) & \symmetry{81} & \(\mathbbmat{16}{\mathbb R}\) & \symmetry{83} & \(\mathbbmat8{\mathbb H}\) & & \(\mathbbmat8{\mathbb H}\) & & \(\mathbbmat{16}{\mathbb R}\) & \symmetry{89} & \(\mathbb C\) & \(^{16}\) \\ \midrule
\(\omega^2\) & & \(+\) & \(+\) & \(-\) & \(-\) & \(+\) & \(+\) & \(-\) & \(-\) & \(+\) & \(+\) & \(-\) & \(-\) & \(+\) & \(+\) & \(-\) & \(-\) & \(+\) & & \(+\) \\
\bottomrule \end{tabular}
\begin{tikzpicture}[remember picture,overlay]
\draw (01.center) -- (3-2.center);
\draw (01.center) -- (45.center);
\draw (2-3.center) -- (3-2.center);
\draw (2-3.center) -- (6-7.center);
\draw[dashed] (3-2.center) -- (72.center);
\draw[dashed] (3-2.center) -- (7-6.center);
\draw (45.center) -- (89.south east);
\draw (45.center) -- (72.center);
\draw (6-7.center) -- (8-8.south west);
\draw (6-7.center) -- (7-6.center);
\draw[dashed] (72.center) -- (81.south west);
\draw[densely dotted, semithick] (72.center) -- (83.south east);
\draw[dashed] (7-6.center) -- (8-5.south east);
\draw[densely dotted, semithick] (7-6.center) -- (8-7.south west);
\draw[->] (00.south east) -- (11.north west);
\draw[->] (2-2.south west) -- (3-3.north east);
\draw[->] (20.south east) -- (31.north west);
\draw[->] (22.south west) -- (31.north east);
\draw[->] (4-4.south east) -- (5-3.north west);
\draw[->] (4-2.south west) -- (5-3.north east);
\draw[->] (40.south east) -- (51.north west);
\draw[->] (42.south west) -- (51.north east);
\draw[->] (44.south east) -- (55.north west);
\draw[->] (6-6.south west) -- (7-7.north east);
\draw[->] (6-4.south east) -- (7-3.north west);
\draw[->] (6-2.south west) -- (7-3.north east);
\draw[->] (60.south east) -- (71.north west);
\draw[->] (62.south west) -- (71.north east);
\draw[->] (64.south east) -- (75.north west);
\draw[->] (66.south west) -- (75.north east);
\end{tikzpicture}
\caption[Real and complex Clifford algebras]{
Real Clifford algebras \(\cl(s,t)\) are displayed for dimensions up to \(d = 8\) and the indicated values of \(\delta = s-t\), with the corresponding complex algebras \(\cl(d,\mathbb C)\) in the final column.
The highlighted lines show the real-algebra symmetry around \(\delta = 1\bmod4\) described by \eqref{eq:clifford symmetry}.
The arrows connect an even-dimensional representation with the two simple-summand representations of the split odd-dimensional algebra obtained through \eqref{eq:omega as generator}.
For the real columns, the unphased volume-element square from \eqref{eq:volume element periodicity} appears in the final row; the complex entry instead refers to the phase-normalised element \(\chi^2 = \ident\) defined in \eqref{eq:complex volume element}.
} \label{tab:clifford algebras}
\end{table}

\begin{table}
\begin{tabular}{c*9cc} \toprule
\(d\) & \multicolumn9c{\(|\delta|\)} & \(\mathbb C\) \\ \cmidrule(lr){2-10}
& \(0\) & \(1\) & \(2\) & \(3\) & \(4\) & \(5\) & \(6\) & \(7\) & \(8\) \\ \midrule
\(1\) & & \(\mathbb R\) & & & & & & & & \(\mathbb C\) \\
\(2\) & \(2\mathbb R\) & & \(\mathbb C\) & & & & & & & \(2\mathbb C\) \\
\(3\) & & \(\mathbb R^2\) & & \(\mathbb H\) & & & & & & \(\mathbb C^2\) \\
\(4\) & \(2\mathbb R^2\) & & \(\mathbb C^2\) & & \(2\mathbb H\) & & & & & \(2\mathbb C^2\) \\
\(5\) & & \(\mathbb R^4\) & & \(\mathbb H^2\) & & \(\mathbb H^2\) & & & & \(\mathbb C^4\) \\
\(6\) & \(2\mathbb R^4\) & & \(\mathbb C^4\) & & \(2\mathbb H^2\) & & \(\mathbb C^4\) & & & \(2\mathbb C^4\) \\
\(7\) & & \(\mathbb R^8\) & & \(\mathbb H^4\) & & \(\mathbb H^4\) & & \(\mathbb R^8\) & & \(\mathbb C^8\) \\
\(8\) & \(2\mathbb R^8\) & & \(\mathbb C^8\) & & \(2\mathbb H^4\) & & \(\mathbb C^8\) & & \(2\mathbb R^8\) & \(2\mathbb C^8\) \\
\bottomrule \end{tabular}
\caption[Real and complex even Clifford subalgebras]{
Real even Clifford subalgebras \(\cl_\PT(s,t)\) are displayed by \(\abs\delta\) up to \(d = 8\), with the corresponding complex even subalgebras \(\cl_\PT(d,\mathbb C)\) in the final column.
} \label{tab:even clifford algebras}
\end{table}

\begin{table}
\begin{tabular}{cc*8{@{}c@{}}c} \toprule
\(d\) & \multicolumn9c{\(\abs \delta\)} & \(\mathbb C\) \\ \cmidrule(lr){2-10}
& \(0\) & \(1\) & \(2\) & \(3\) & \(4\) & \(5\) & \(6\) & \(7\) & \(8\) \\ \midrule
\(1\) & & \(0\) & & & & & & & & \(0\) \\
\(2\) & \(\gl(1,\mathbb R)\) & & \(\u(1)\) & & & & & & & \(\gl(1,\mathbb C)\) \\
\(3\) & & \(\sl(2,\mathbb R)\) & & \(\su(2)\) & & & & & & \(\sl(2,\mathbb C)\) \\
\(4\) & \(\sl(2,\mathbb R) \oplus\sl(2,\mathbb R)\) & & \(\sl(2,\mathbb C)\) & & \(\su(2) \oplus\su(2)\) & & & & & \(\sl(2,\mathbb C) \oplus\sl(2,\mathbb C)\) \\
\(5\) & & \(\sp(4,\mathbb R)\) & & \(\sp(1,1)\) & & \(\sp(2)\) & & & & \(\sp(4,\mathbb C)\) \\
\(6\) & \(\sl(4,\mathbb R)\) & & \(\su(2,2)\) & & \(\sl(2,\mathbb H)\) & & \(\su(4)\) & & & \(\sl(4,\mathbb C)\) \\
\(7\) & & \(\so(4,3)\) & & \(\so(5,2)\) & & \(\so(6,1)\) & & \(\so(7)\) & & \(\so(7,\mathbb C)\) \\
\(8\) & \(\so(4,4)\) & & \(\so(5,3)\) & & \(\so(6,2)\) & & \(\so(7,1)\) & & \(\so(8)\) & \(\so(8,\mathbb C)\) \\
\bottomrule \end{tabular}
\caption[Real and complex spin Lie algebras]{
Real spin Lie algebras extracted from \(\cl_\PT(s,t)\) are displayed by \(\abs\delta\) up to \(d = 8\), with the algebras extracted from \(\cl_\PT(d,\mathbb C)\) in the final column.
The identification of the even Clifford algebras in \cref{tab:even clifford algebras} with the spin algebras can be found \eg in \cite{Kugo:1982bn, Lee:2018}.
} \label{tab:spin algebras}
\end{table}

The six Clifford algebras in dimensions zero to two fix the conventions from which all higher-dimensional cases are constructed recursively.
Their dependence on \(\delta = s-t\bmod8\) is derived in \cref{sec:iterative construction} and summarised up to dimension eight in \cref{tab:clifford algebras}.
The corresponding even subalgebras depend on \(\abs\delta\bmod8\), as recorded in \cref{tab:even clifford algebras}, while the associated spin Lie algebras are identified without proof in \cref{tab:spin algebras}.
Before applying the recursion, the following low-dimensional calculations show explicitly how the generator squares determine these algebra types.

\paragraph{Zero dimensions}

The zero-dimensional split algebra \(\cl(0,0)\), with \(\delta = 0\), contains only real multiples of the unit and is therefore \(\cl(0,0) \cong \mathbb R\).

\paragraph{One dimension}

In one dimension, every Clifford element has the form \(a\ident+b\gamma\), and the structure map \eqref{eq:structure map} that negates the single generator lies outside the algebra.

\subparagraph{Euclidean}

For the Euclidean Clifford algebra \(\cl(1,0)\) with \(\delta \bmod 8 = 1\), the anticommutator relation \eqref{eq:Clifford algebra} requires \(\gamma^2 = \ident\).
A convenient real representation is \(\gamma = \sigma_3\), for which a generic algebra element becomes \(a \ident + b \gamma \cong \begin{psmallmatrix}a+b&0\\0&a-b\end{psmallmatrix}\).
The two diagonal entries are independent, which establishes \(\cl(1,0) \cong \mathbb R \oplus \mathbb R\).
The structure map \eqref{eq:structure map} corresponds in this case to \(\structure = \begin{psmallmatrix}0&1\\1&0\end{psmallmatrix}\).

\subparagraph{Pseudo-Euclidean}

For the pseudo-Euclidean Clifford algebra \(\cl(0,1)\) with \(\delta \bmod 8 = 7\), the generator squares to \(\gamma^2 = -\ident\) and therefore provides a complex structure.
\footnote{A complex structure on a real even-dimensional vector space identifies it with a complex vector space of half the real dimension.}
The correspondence \(a\ident+b\gamma\leftrightarrow a+b\i\) identifies this real Clifford algebra with the complex numbers, \(\cl(0,1)\cong\mathbb C\).
The structure map \eqref{eq:structure map} corresponds to complex conjugation.
On the vector space \(\mathbb C\), the Clifford generator acts by multiplication with the imaginary unit, \(\gamma z = \i z\), while the structure map is the real-linear map \(\structure z = z^\ast\).
It is antilinear with respect to the complex structure defined by \(\gamma\).
Consequently, for a generic Clifford element \(a\ident+b\gamma\), one has
\begin{equation*}
\structure (a\ident+b\gamma) \structure^{-1} = a\ident-b\gamma = (a\ident+b\gamma)_\grade .
\end{equation*}

\subparagraph{Even subalgebra}

In both cases the even subalgebra consists only of the identity element and can therefore be identified with the real numbers \(\cl_\PT(1,0) \cong \cl_\PT(0,1) \cong \cl(0,0) \cong \mathbb R\).

\paragraph{Two dimensions}

The two-dimensional Clifford algebras have two generators \(\basis^1 = \set{\gamma_1, \gamma_2}\).
The even subalgebra \(\cl_\PT\) is spanned by the identity and the volume element \(\basis^2 = \set{\omega = \gamma_{[1} \gamma_{2]}}\).
The structure map \eqref{eq:structure map} coincides with the volume element \eqref{eq:volume element}, \(\structure = \omega\).

\subparagraph{Euclidean}

The Euclidean Clifford algebra \(\cl(2, 0)\) has \(\delta \bmod 8 = 2\), and the anticommutator relation \eqref{eq:Clifford algebra} requires \(\gamma_1^2 = \gamma_2^2 = \ident\).
A convenient real representation is
\begin{align} \label{eq:algebra two Euclidean}
\gamma_1 & = \sigma_1 = \begin{pmatrix} 0 & 1 \\ 1 & 0\end{pmatrix} , &
\gamma_2 & = \sigma_3 = \begin{pmatrix} 1 & 0 \\ 0 & -1\end{pmatrix} , &
\omega & = - \i \sigma_2 = \begin{pmatrix} 0 & -1\\ 1& 0\end{pmatrix} .
\end{align}
Together with the identity, these matrices span \(M^2(\mathbb R)\), establishing \(\cl(2,0) \cong M^2(\mathbb R)\).
The volume element \eqref{eq:volume element} squares to \(\omega^2 = - \matident\).

\subparagraph{Pseudo-Euclidean}

In the pseudo-Euclidean Clifford algebra \(\cl(0,2)\) with \(\delta \bmod 8 = 6\), the generators square to \(\gamma_1^2 = \gamma_2^2 = - \ident\), as realised by
\begin{align} \label{eq:algebra two pseudo Euclidean}
\gamma_1 & = - \i \sigma_1 = -\begin{pmatrix}0&\i\\\i&0\end{pmatrix} , &
\gamma_2 & = - \i \sigma_2 = \begin{pmatrix}0&1\\-1&0\end{pmatrix} , &
\omega & = -\i\sigma_3 = \begin{pmatrix}-\i&0\\0&\i \end{pmatrix} .
\end{align}
The three nontrivial basis elements square to minus the identity and satisfy the quaternionic multiplication rules.
The identification \((\qu 1,\qu i,\qu j,\qu k) = (\matident, \gamma_1, \gamma_2, \omega)\) therefore establishes \(\cl(0,2) \cong \mathbb H\).

\subparagraph{Even Euclidean subalgebra}

In both definite signatures, \(\omega^2 = -\matident\), so the map \(\omega\leftrightarrow\i\) equips the even subalgebra \(a\matident+b\omega\) with a complex structure.
Both even subalgebras are consequently complex, \(\cl_\PT(2,0) \cong \cl_\PT(0,2) \cong \cl(0,1) \cong \mathbb C\).
The corresponding spin algebra is \(\cl_\PT(2,0) \cong \cl_\PT(0,2) \spincontains \spin(2) \cong \u(1)\).

\subparagraph{Split}

The split Clifford algebra \(\cl(1,1)\) with \(\delta = 0\) is defined by \(\gamma_1^2 = - \gamma_2^2 = \ident\), as realised by
\begin{align}
\gamma_1 & = \sigma_1 = \begin{pmatrix}0&1\\1&0\end{pmatrix} , &
\gamma_2 & = - \i \sigma_2 = \begin{pmatrix}0&-1\\1&0\end{pmatrix} , &
\omega & = \sigma_3 = \begin{pmatrix}1&0\\0&-1\end{pmatrix} .
\end{align}
These matrices again span the two-by-two real matrices, \(\cl(1,1) \cong M^2(\mathbb R)\).
In contrast to the definite cases, \(\omega^2 = \matident\), so the even subalgebra consists of real diagonal matrices and splits as \(\cl_\PT(1,1) \cong \cl(1,0) \cong \mathbb R \oplus \mathbb R\).
The corresponding spin algebra is \(\cl_\PT(1,1) \spincontains \spin(1,1) \cong \gl(1, \mathbb R)\).

\subsubsection{Iterative construction and periodicity} \label{sec:iterative construction}

Higher-dimensional Clifford representations follow recursively by combining a \(d\)-dimensional representation with one of the two-dimensional building blocks:
\begin{equation} \label{eq:iterative}
\gamma_i^{{(d+2)}} =
\begin{cases}
\gamma_i^{(d)} \otimes \omega^{(2)} & \text{for } 1 \leq i \leq d , \\
\matident^{(d)} \otimes \gamma_{i-d}^{(2)} & \text{for } 1 \leq i-d \leq 2
\end{cases}.
\end{equation}
The signature of the enlarged representation follows from the tensor-product squares and the three two-dimensional building blocks:
\begin{align}
(\gamma_i^{{(d+2)}})^2 & =
\begin{cases}
(\gamma_i^{(d)})^2 \otimes (\omega^{(2)})^2 & \text{for } 1 \leq i \leq d , \\
(\matident^{(d)})^2 \otimes (\gamma_{i-d}^{(2)})^2 & \text{for } 1 \leq i-d \leq 2 ,
\end{cases}& &\begin{array}{rlccc}\toprule
& & \gamma_1^2 & \gamma_2^2 & \omega^2 \\ \midrule
\cl(2,0) & \mathbb R^2 & + & + & - \\
\cl(1,1) & \mathbb R^2 & + & - & + \\
\cl(0,2) & \mathbb H & - & - & - \\
\bottomrule
\end{array}
\end{align}
The two two-dimensional (pseudo-)Euclidean algebras imply the relations \cite{Floerchinger:2019oeo}
\begin{align} \label{eq:recursive construction 1}
\cl(s+2,t) &\cong \cl(t,s) \otimes \cl(2,0) , &
\cl(s,t+2) &\cong \cl(t,s) \otimes \cl(0,2) ,
\end{align}
while the two-dimensional split algebra implies
\begin{align} \label{eq:recursive construction 2}
\cl(s+1,t+1) &\cong \cl(s,t) \otimes \cl(1,1) .
\end{align}
An eightfold periodicity in \(\delta\) at fixed total dimension follows by combining the two Euclidean relations in \eqref{eq:recursive construction 1}:
\begin{equation} \label{eq:period 8}
\cl(s + 4,t) \cong \cl(s,t) \otimes \cl(0,2) \otimes \cl(2,0) \cong \cl(s,t+4) .
\end{equation}
Using the split relation \eqref{eq:recursive construction 2}, this periodicity can be extended to varying \(d\):
\begin{equation} \label{eq:period 8 in delta}
\begin{rcases}
\cl(s+8,t) \\ \cl(s,t+8)
\end{rcases}
\cong \cl(s+4,t+4) \cong \cl(s,t) \otimes \cl^4(1,1) \cong M^{2^4}(\cl(s,t)) .
\end{equation}
The tensor factors change the matrix size but not the underlying real, complex, quaternionic, or split type, leaving \(\delta\bmod8\) as the periodic classification variable used in \cref{tab:clifford real}.

\begin{table}
\begin{tabular}{ccll} \toprule
\(\delta \bmod 8\) & \(\omega^2\) & \header{\(\cl(s,t)\)} & \header{\(\cl_\PT(s,t)\)} \\ \midrule
\(0\) & \(+\) & \(\phantom2M(2^{\frac d2},\mathbb R)\) & \(2M(2^{\frac{d-2}2},\mathbb R)\) \\
\(1\) & \(+\) & \(2M(2^{\frac{d-1}2},\mathbb R)\) & \(\phantom2M(2^{\frac{d-1}2},\mathbb R)\) \\
\(2\) & \(-\) & \(\phantom2M(2^{\frac d2},\mathbb R)\) & \(\phantom2M(2^{\frac{d-2}2},\mathbb C)\) \\
\(3\) & \(-\) & \(\phantom2M(2^{\frac{d-1}2},\mathbb C)\) & \(\phantom2M(2^{\frac{d-3}2},\mathbb H)\) \\
\(4\) & \(+\) & \(\phantom2M(2^{\frac{d-2}2},\mathbb H)\) & \(2M(2^{\frac{d-4}2},\mathbb H)\) \\
\(5\) & \(+\) & \(2M(2^{\frac{d-3}2},\mathbb H)\) & \(\phantom2M(2^{\frac{d-3}2},\mathbb H)\) \\
\(6\) & \(-\) & \(\phantom2M(2^{\frac{d-2}2},\mathbb H)\) & \(\phantom2M(2^{\frac{d-2}2},\mathbb C)\) \\
\(7\) & \(-\) & \(\phantom2M(2^{\frac{d-1}2},\mathbb C)\) & \(\phantom2M(2^{\frac{d-1}2},\mathbb R)\) \\
\bottomrule \end{tabular}
\caption[Periodicity of real Clifford algebras and even subalgebras]{
Real Clifford algebras \(\cl(s,t)\) and their even subalgebras \(\cl_\PT(s,t)\) as functions of \(\delta = s-t\bmod8\) for \(d>0\).
The zero-dimensional case is given separately in \cref{tab:clifford algebras}.
The column \(\omega^2\) indicates whether the volume element of the full Clifford algebra squares to the positive or negative identity.
} \label{tab:clifford real}
\end{table}

\paragraph{Even subalgebra and symmetry}

Multiplying every remaining generator by one fixed temporal or spatial generator produces a generating set for the even subalgebra \(\cl_\PT(s,t)\).
The Clifford relation \eqref{eq:Clifford algebra} identifies these sets with the generators of \(\cl(s,t-1)\) or \(\cl(t,s-1)\), respectively, and hence
\begin{equation}
\cl_\PT(s,t) \cong
\begin{cases}
\cl(s,t-1) & \text{for } t \geq 1 \\
\cl(t,s-1) & \text{for } s \geq 1
\end{cases}
.
\end{equation}
Whenever all displayed signatures are non-negative, it follows for Clifford algebras and their even subalgebras that
\begin{align} \label{eq:symmetry}
\cl(s,t) &\cong \cl(t+1,s-1) , &
\cl_\PT(s,t) &\cong \cl_\PT(t,s) .
\end{align}
The even subalgebras with exchanged spatial and temporal dimensions are therefore isomorphic and exhibit the \(\abs\delta\bmod8\) periodicity used in \cref{tab:even clifford algebras,tab:spin algebras}.
Correspondingly, the Spin groups generated by these even subalgebras are isomorphic and double cover the respective connected rotation groups.
Combining the Clifford algebra property of \eqref{eq:symmetry} with the periodicity \eqref{eq:period 8} shows that the Clifford algebras exhibit a symmetry around signatures \(i = 1 \bmod 4\):
\begin{equation} \label{eq:clifford symmetry}
\cl(s,t) \cong \cl(t+i,s-i) \text{ for } i = 1 \bmod 4 .
\end{equation}
The resulting reflection symmetry across the table of signatures is visible in \cref{tab:clifford algebras}.

\subsubsection{Examples of relevance to physical spacetimes} \label{sec:physically relevant spacetimes}

Together with the low-dimensional cases, the classes \(\delta = 3,4,5\bmod8\) display every algebra type in one Bott period.
Explicit representatives in three and four dimensions then connect the abstract classification to Clifford bases commonly used in relativistic field theory.

\paragraph{Three dimensions}

A three-dimensional Clifford algebra has the generators \(\basis^1 = \set{\gamma_1, \gamma_2, \gamma_3}\).
Its even subalgebra is spanned by the identity and the three bivectors \(\basis^2 = \set{\gamma_{[1} \gamma_{2]}, \gamma_{[2} \gamma_{3]}, \gamma_{[1} \gamma_{3]}}\).
The volume element \(\basis^3 = \set{\omega = \gamma_{[1} \gamma_{\vphantom[2} \gamma_{3]}}\) is odd, and the structure map lies outside the Clifford algebra.

\subparagraph{Euclidean}

The Euclidean case with \(\delta \bmod 8 = 3\) is defined by \(\gamma_1^2 = \gamma_2^2 = \gamma_3^2 = \ident\).
The recursive construction \eqref{eq:recursive construction 1} gives
\begin{equation}
\cl(3,0) = \cl(0,1) \otimes \cl(2,0) \cong \mathbb C \otimes M^2(\mathbb R) \cong M^2(\mathbb C) .
\end{equation}
The Euclidean Clifford algebra is therefore the algebra of two-by-two complex matrices, represented conveniently by the Pauli matrices:
\begin{align} \label{eq:Pauli matrices}
\gamma_1 & = \sigma_1 = \begin{pmatrix}0&1\\1&0\end{pmatrix} , &
\gamma_2 & = \sigma_3 = \begin{pmatrix}1&0\\0&-1\end{pmatrix} , &
\gamma_3 & = \sigma_2 = \begin{pmatrix}0&-\i\\\i&0\end{pmatrix} .
\end{align}
The volume element is \(\omega = - \i \sigma_0\), squares to \(\omega^2 = - \matident\), and provides the complex structure in this representation.
The generators of the even subalgebra are
\begin{align} \label{cl+(3 0)}
\gamma_{[2} \gamma_{3]} & = \omega \gamma_1 = - \i \sigma_1 , &
\gamma_{[1} \gamma_{3]} & = -\omega \gamma_2 = \i \sigma_3 , &
\gamma_{[1} \gamma_{2]} & = \omega \gamma_3 = - \i \sigma_2 .
\end{align}
Since all three of them square to minus the identity, the basis of the even subalgebra can be identified with the quaternionic units via \((\qu 1,\qu i,\qu j,\qu k) = (\matident, \gamma_{[2} \gamma_{3]}, \gamma_{[1} \gamma_{2]}, -\gamma_{[1} \gamma_{3]})\).
The structure map \eqref{eq:structure map} can be represented by \(\i \sigma_2\) followed by componentwise complex conjugation,
\begin{align*}
\structure&:S\to S , &
\structure\psi & = \i\sigma_2\psi^\ast , &
\psi&\in S ,
\end{align*}
where \(\psi^\ast\) denotes componentwise complex conjugation.

\subparagraph{Pseudo-Euclidean}

In the pseudo-Euclidean case \(\cl(0,3)\), with \(\delta = 5\bmod8\), the generators obey \(\gamma_1^2 = \gamma_2^2 = \gamma_3^2 = - \ident\), and the recursive construction \eqref{eq:recursive construction 1} gives
\begin{equation} \label{eq:cl(0 3)}
\cl(0,3) = \cl(1,0) \otimes \cl(0,2) \cong (\mathbb R \oplus \mathbb R) \otimes \mathbb H = \mathbb H \oplus \mathbb H .
\end{equation}
The algebra therefore consists of two quaternionic blocks, which can be represented by two-by-two diagonal quaternionic matrices:
\begin{align}
\gamma_1 & = \begin{pmatrix}\qu i&0\\0&-\qu i\end{pmatrix} , &
\gamma_2 & = \begin{pmatrix}\qu j&0\\0&-\qu j\end{pmatrix} , &
\gamma_3 & = \begin{pmatrix}\qu k&0\\0&-\qu k\end{pmatrix} , &
\omega & = \begin{pmatrix}- \qu 1&0\\0&\qu 1\end{pmatrix}
.
\end{align}
A complex representative in terms of the Pauli matrices is given in \cref{fn:quaternions}.
The volume element squares to \(\omega^2 = \matident\) and the generators of the even Clifford algebra are
\begin{align} \label{cl+(0 3)}
\gamma_{[2} \gamma_{3]} & = - \omega \gamma_1
 = \begin{pmatrix}\qu i&0\\0&\qu i\end{pmatrix} , &
\gamma_{[1} \gamma_{3]} & = \omega \gamma_2
 = -\begin{pmatrix}\qu j&0\\0&\qu j\end{pmatrix} , &
\gamma_{[1} \gamma_{2]} & = - \omega \gamma_3
 = \begin{pmatrix}\qu k&0\\0&\qu k\end{pmatrix}
.
\end{align}
The generators of the even subalgebra can be identified with the quaternionic units via \((\qu 1,\qu i,\qu j,\qu k) = (\matident, \gamma_{[2} \gamma_{3]}, -\gamma_{[1} \gamma_{3]}, \gamma_{[1} \gamma_{2]})\).
The structure map corresponds to \(\structure = \begin{psmallmatrix}0&\qu 1\\\qu 1&0\end{psmallmatrix}\).

\subparagraph{Even Euclidean subalgebra}

The three nontrivial basis elements of each even Euclidean subalgebra in \eqref{cl+(3 0),cl+(0 3)} are anticommuting complex structures and therefore span
\begin{equation}
\begin{rcases}
\cl_\PT(3,0) \\ \cl_\PT(0,3)
\end{rcases}
\cong \cl(0,2)
\cong \mathbb H
\spincontains \spin(3) \cong \sp(1) \cong \su(2) ,
\end{equation}
where the last isomorphisms realise the compact real form of the accidental rank-one Dynkin isomorphism \(A_1\cong B_1\cong C_1\).

\subparagraph{Lorentzian}

The three-dimensional Lorentzian signatures have \(\delta = 1\) and \(7\bmod8\), so their algebra types follow from the corresponding one-dimensional classes after the recursive matrix-size extension.
The Lorentzian algebra can be identified with two copies of the algebra of two-by-two real matrices, \(\cl(2,1) \cong 2 M^2(\mathbb R)\).
The generators can be written as
\begin{align}
\gamma_1 & = \begin{pmatrix}\sigma_1&0\\0&-\sigma_1\end{pmatrix} , &
\gamma_2 & = \begin{pmatrix}\sigma_3&0\\0&-\sigma_3\end{pmatrix} , &
\gamma_3 & = \i \begin{pmatrix}\sigma_2&0\\0&-\sigma_2\end{pmatrix} .
\end{align}
These choices give the generators of the even subalgebra as
\begin{align} \label{cl(21) even}
\gamma_{[2} \gamma_{3]} & = \omega \gamma_1 = \begin{pmatrix}\sigma_1&0\\0&\sigma_1\end{pmatrix} , &
\gamma_{[1} \gamma_{3]} & = -\omega \gamma_2 = -\begin{pmatrix}\sigma_3&0\\0&\sigma_3\end{pmatrix} , &
\gamma_{[1} \gamma_{2]} & = -\omega \gamma_3 = -\i \begin{pmatrix}\sigma_2&0\\0&\sigma_2\end{pmatrix} ,
\end{align}
where the volume element is given by \(\omega = \begin{psmallmatrix}\sigma_0&0\\0&-\sigma_0\end{psmallmatrix}\) and squares to the identity.
The structure map is given by \(\structure = \begin{psmallmatrix}0&\sigma_0\\\sigma_0&0\end{psmallmatrix}\).

\subparagraph{Pseudo-Lorentzian}

The pseudo-Lorentzian algebra is isomorphic to the algebra of two-by-two complex matrices, \(\cl(1,2) \cong M^2(\mathbb C)\).
\footnote{\label{fn:isomorphisms}%
Throughout, we use the additional isomorphisms
\begin{align*}
\mathbb R \otimes_{\mathbb R} \mathbb R &\cong \mathbb R , &
\mathbb R \otimes_{\mathbb R} \mathbb C &\cong \mathbb C , &
\mathbb R \otimes_{\mathbb R} \mathbb H &\cong \mathbb H , &
\mathbb C \otimes_{\mathbb R} \mathbb C &\cong 2\mathbb C , &
\mathbb H \otimes_{\mathbb R} \mathbb C &\cong \mathbb C^2 , &
\mathbb H \otimes_{\mathbb R} \mathbb H &\cong \mathbb R^4 .
\end{align*}
}
Its generators can be written as
\begin{align}
\gamma_1 & = \sigma_2 = \i \begin{pmatrix}0&-1\\1&0\end{pmatrix} , &
\gamma_2 & = \i \sigma_1 = \i \begin{pmatrix}0&1\\1&0\end{pmatrix} , &
\gamma_3 & = \i \sigma_3 = \i \begin{pmatrix}1&0\\0&-1\end{pmatrix} .
\end{align}
The ordered volume element is \(\omega = \i\sigma_0\) and squares to minus the identity.
Their products give the even-subalgebra generators
\begin{align} \label{cl(12) even}
\gamma_{[2} \gamma_{3]} & = \omega \gamma_1 = \begin{pmatrix}0&1\\-1&0\end{pmatrix} , &
\gamma_{[1} \gamma_{3]} & = \omega \gamma_2 = \begin{pmatrix}0&-1\\-1&0\end{pmatrix} , &
\gamma_{[1} \gamma_{2]} & = -\omega \gamma_3 = \begin{pmatrix}1&0\\0&-1\end{pmatrix} .
\end{align}
The structure map acts by componentwise complex conjugation, \(\structure \psi = \psi^\ast\).

\subparagraph{Even Lorentzian subalgebra}

Both even subalgebras \eqref{cl(12) even,cl(21) even} are isomorphic to the algebra of two-by-two real matrices
\begin{equation}
\begin{rcases}
\cl_\PT(2,1) \\ \cl_\PT(1,2)
\end{rcases}
\cong \cl(2,0)
\cong \cl(1,1)
\cong M^2(\mathbb R)
\spincontains \spin(2,1)
\cong \sp(2,\mathbb R)
\cong \sl(2, \mathbb R)
\cong \su(1,1) ,
\end{equation}
where the last isomorphisms realise the split real form of the accidental rank-one Dynkin isomorphism \(A_1\cong B_1\cong C_1\).

\paragraph{Four dimensions}

In all four-dimensional signatures, the Clifford generators are labelled \(\gamma_1,\gamma_2,\gamma_3,\gamma_4\), with the spatial generators ordered before the temporal generators, contrary to the convention often used in Lorentzian applications.
Besides the identity element, the even subalgebra contains the six products \(\gamma_{[i}\gamma_{j]}\) and the volume element \(\omega\), which is the ordered product of all four generators.
The structure map corresponds to the volume element \(\structure = \omega\).

\subparagraph{(Pseudo) Euclidean}

Both definite four-dimensional Clifford algebras have \(\delta = 4\bmod8\) and are therefore isomorphic, as follows directly from the recursive construction \eqref{eq:recursive construction 1}:
\begin{equation}
\begin{rcases}
\cl(0,4) = \cl(2,0) \otimes \cl(0,2) \\
\cl(4,0) = \cl(0,2) \otimes \cl(2,0)
\end{rcases}
\cong M^2(\mathbb R) \otimes \mathbb H = M^2(\mathbb H) .
\end{equation}
Thus both algebras are realised by two-by-two quaternionic matrices.
In both signatures, the quaternionic units \((\qu i,\qu k,\qu j)\) are assigned to \((\gamma_2,\gamma_3,\gamma_4)\).

\subparagraph{Euclidean}

In the Euclidean case the generators satisfy \(\gamma_1^2 = \gamma_2^2 = \gamma_3^2 = \gamma_4^2 = \matident\) and they can be represented by
\begin{align} \label{eq:gamma-40-quaternionic}
\gamma_1 & = \begin{pmatrix}0 & \qu 1 \\ \qu 1 & 0\end{pmatrix} , &
\gamma_2 & = \begin{pmatrix}0 & \qu i \\ - \qu i& 0\end{pmatrix} , &
\gamma_3 & = \begin{pmatrix}0 & \qu k\\ -\qu k& 0\end{pmatrix} , &
\gamma_4 & = \begin{pmatrix}0 & \qu j\\ -\qu j& 0\end{pmatrix} .
\end{align}
The volume element \(\omega = \begin{psmallmatrix}\qu 1&0\\0&-\qu 1\end{psmallmatrix}\) squares to \(\omega^2 = \matident\) and the generators of the even Clifford algebra are
\begin{subequations}\label{eq:4d Euclidean subalgebra}
\begin{align}
\gamma_{[1} \gamma_{2]} & = - \omega \qu i = \begin{pmatrix}-\qu i&0\\0& \qu i\end{pmatrix} , &
\gamma_{[1} \gamma_{3]} & = - \omega \qu k = \begin{pmatrix}-\qu k&0\\0& \qu k\end{pmatrix} , &
\gamma_{[1} \gamma_{4]} & = - \omega \qu j = \begin{pmatrix}-\qu j&0\\0& \qu j\end{pmatrix} , \\
\gamma_{[3} \gamma_{4]} & = \matident \qu i = \begin{pmatrix}\qu i&0\\0& \qu i\end{pmatrix} , &
\gamma_{[2} \gamma_{4]} & = - \matident \qu k = -\begin{pmatrix}\qu k&0\\0& \qu k\end{pmatrix} , &
\gamma_{[2} \gamma_{3]} & = \matident \qu j = \begin{pmatrix}\qu j&0\\0& \qu j\end{pmatrix} .
\end{align}
\end{subequations}

\subparagraph{Pseudo-Euclidean}

In the pseudo-Euclidean case with \(\gamma_1^2 = \gamma_2^2 = \gamma_3^2 = \gamma_4^2 = - \matident\), the generators can be chosen to be
\begin{align} \label{eq:gamma-04-quaternionic}
\gamma_1 & = \begin{pmatrix}0 & \qu 1 \\ - \qu 1 & 0\end{pmatrix} , &
\gamma_2 & = \begin{pmatrix}0 & \qu i \\ \qu i& 0\end{pmatrix} , &
\gamma_3 & = \begin{pmatrix}0 & \qu k\\ \qu k& 0\end{pmatrix} , &
\gamma_4 & = \begin{pmatrix}0 & \qu j\\ \qu j& 0\end{pmatrix} ,
\end{align}
and the volume element \(\omega = \begin{psmallmatrix}\qu 1&0\\0&-\qu 1\end{psmallmatrix}\) squares to \(\omega^2 = \matident\).
The generators of the even subalgebra are
\begin{subequations}\label{eq:4d pseudo Euclidean subalgebra}
\begin{align}
\gamma_{[1} \gamma_{2]} & = \omega \qu i = \begin{pmatrix}\qu i&0\\0& -\qu i\end{pmatrix} , &
\gamma_{[1} \gamma_{3]} & = \omega \qu k = \begin{pmatrix}\qu k&0\\0& -\qu k\end{pmatrix} , &
\gamma_{[1} \gamma_{4]} & = \omega \qu j = \begin{pmatrix}\qu j&0\\0& -\qu j\end{pmatrix} , \\
\gamma_{[3} \gamma_{4]} & = - \matident \qu i = -\begin{pmatrix}\qu i&0\\0& \qu i\end{pmatrix} , &
\gamma_{[2} \gamma_{4]} & = \matident \qu k = \begin{pmatrix}\qu k&0\\0& \qu k\end{pmatrix} , &
\gamma_{[2} \gamma_{3]} & = - \matident \qu j = -\begin{pmatrix}\qu j&0\\0& \qu j\end{pmatrix} .
\end{align}
\end{subequations}

\subparagraph{Even Euclidean subalgebra}

In both Euclidean cases the generators of the even subalgebras \eqref{eq:4d Euclidean subalgebra,eq:4d pseudo Euclidean subalgebra} are diagonal matrices of quaternions.
The even subalgebra consequently splits into two quaternionic blocks,
\begin{equation}
\begin{rcases}
\cl_\PT(4,0) \\
\cl_\PT(0,4)
\end{rcases}
\cong \cl(0,3)
\cong \mathbb H \oplus \mathbb H
\spincontains \spin(4)
\cong \sp(1) \oplus \sp(1)
\cong \su(2) \oplus \su(2) ,
\end{equation}
where the last isomorphism reflects the accidental Dynkin-diagram isomorphism \(D_2\cong A_1\oplus A_1\).

\subparagraph{(Pseudo) Lorentzian}

The signatures \((3,1)\) and \((1,3)\) describe the same connected Lorentz Lie algebra but lead to inequivalent full real Clifford algebras, so their Pin-level properties must be distinguished.

\subparagraph{Lorentzian}

The Lorentzian Clifford algebra defined by \(\gamma_1^2 = \gamma_2^2 = \gamma_3^2 = -\gamma_4^2 = \ident\) belongs to the real class \(\delta = 2\bmod8\), as made explicit by the recursive construction \eqref{eq:recursive construction 1}:
\begin{equation}
\cl(3,1) = \cl(1,1) \otimes \cl(2,0) \cong M^2(\mathbb R) \otimes M^2(\mathbb R) \cong M^4(\mathbb R) .
\end{equation}
For the mostly-plus metric \(\eta = \diag(1,1,1,-1)\) with the Clifford algebra definition \eqref{eq:Clifford algebra}, the algebra admits the real four-by-four representation:
\begin{align} \label{eq:gamma-31}
\gamma_1 & = \begin{pmatrix}-\sigma_1 & 0 \\ 0 & \sigma_1 \end{pmatrix} , &
\gamma_2 & = \begin{pmatrix}0 & \sigma_0 \\ \sigma_0 & 0 \end{pmatrix} , &
\gamma_3 & = \begin{pmatrix}-\sigma_3&0\\0&\sigma_3 \end{pmatrix} , &
\gamma_4 & = \i \begin{pmatrix}-\sigma_2 & 0 \\ 0 & \sigma_2 \end{pmatrix} .
\end{align}
This is one possible Majorana basis \cite{Majorana:1937vz}.
The volume element is
\(\omega = \begin{psmallmatrix}0&-\sigma_0\\\sigma_0&0 \end{psmallmatrix}\) and squares to \(\omega^2 = - \matident\).
Because \(\omega^2 = -\matident\), it provides a commuting complex structure on the even subalgebra generated by
\begin{subequations} \label{eq:lorentzian spin algebra}
\begin{align}
\gamma_{[1} \gamma_{4]} & = \begin{pmatrix}-\sigma_3&0\\0&-\sigma_3 \end{pmatrix} , &
\gamma_{[2} \gamma_{3]} & = \begin{pmatrix}0 & \sigma_3\\-\sigma_3&0\end{pmatrix} = \omega \gamma_{[1} \gamma_{4]} , \\
\gamma_{[2} \gamma_{4]} & = \i \begin{pmatrix}0 & \sigma_2\\-\sigma_2&0\end{pmatrix} , &
\gamma_{[1} \gamma_{3]} & = \i \begin{pmatrix} \sigma_2&0\\0&\sigma_2 \end{pmatrix} = -\omega \gamma_{[2} \gamma_{4]} , \\
\gamma_{[3} \gamma_{4]} & = \begin{pmatrix}\sigma_1&0\\0& -\sigma_1 \end{pmatrix} , &
\gamma_{[1} \gamma_{2]} & = \begin{pmatrix}0 & -\sigma_1\\\sigma_1&0\end{pmatrix} = \omega \gamma_{[3} \gamma_{4]} .
\end{align}
\end{subequations}

\subparagraph{Pseudo-Lorentzian}

The pseudo-Lorentzian Clifford algebra is defined by \(\gamma_1^2 = -\gamma_2^2 = -\gamma_3^2 = -\gamma_4^2 = \ident\).
Its class \(\delta = 6\bmod8\) is quaternionic, as follows from the recursive construction \eqref{eq:recursive construction 1}:
\begin{equation}
\cl(1,3) = \cl(1,1) \otimes \cl(0,2) \cong M^2(\mathbb R) \otimes \mathbb H = M^2(\mathbb H) \ncong M^4(\mathbb R) = \cl(3,1) ,
\end{equation}
which makes the difference from the real algebra \(\cl(3,1)\) explicit.
For the mostly-minus metric \(\eta = \diag(1,-1,-1,-1)\), an irreducible two-by-two quaternionic representation is
\begin{align} \label{eq:gamma-13-quaternionic}
\gamma_1 & = \begin{pmatrix}0 & \qu 1 \\ \qu 1 & 0\end{pmatrix} , &
\gamma_2 & = \begin{pmatrix}\qu i & 0 \\ 0 & -\qu i\end{pmatrix} , &
\gamma_3 & = \begin{pmatrix}\qu j & 0 \\ 0 & -\qu j\end{pmatrix} , &
\gamma_4 & = \begin{pmatrix}\qu k & 0 \\ 0 & -\qu k\end{pmatrix} .
\end{align}
This is closely related to the Dirac basis \cite{Dirac:1928hu}.
The volume element is then \(\omega = \begin{psmallmatrix}0&\qu 1\\-\qu 1&0\end{psmallmatrix}\), squares to \(\omega^2 = -\matident\), and provides a complex structure on the even subalgebra, which is generated by
\begin{subequations} \label{eq:pseudo lorentzian spin algebra}
\begin{align}
\gamma_{[1} \gamma_{2]} & = \begin{pmatrix}0 & -\qu i\\\qu i&0\end{pmatrix} = - \qu i \omega , &
\gamma_{[1} \gamma_{3]} & = \begin{pmatrix}0 & -\qu j\\\qu j&0\end{pmatrix} = - \qu j \omega , &
\gamma_{[1} \gamma_{4]} & = \begin{pmatrix}0 & - \qu k\\\qu k&0\end{pmatrix} = - \qu k \omega , \\
\gamma_{[3} \gamma_{4]} & = \begin{pmatrix}\qu i&0\\0& \qu i\end{pmatrix} = \qu i \matident , &
\gamma_{[2} \gamma_{4]} & = -\begin{pmatrix}\qu j&0\\0& \qu j\end{pmatrix} = -\qu j \matident , &
\gamma_{[2} \gamma_{3]} & = \begin{pmatrix}\qu k&0\\0& \qu k\end{pmatrix} = \qu k \matident .
\end{align}
\end{subequations}
Here the boosts are proportional to the volume element, while the rotations are proportional to the identity.

\subparagraph{Even Lorentzian subalgebra}

Although the full real algebras differ, their even subalgebras in \eqref{eq:pseudo lorentzian spin algebra,eq:lorentzian spin algebra} share the complex structure supplied by the volume element and are both
\begin{equation}
\begin{rcases}
\cl_\PT(1,3) \\
\cl_\PT(3,1)
\end{rcases}
\cong \cl(3,0)
\cong \cl(1,2)
\cong M^2(\mathbb C)
\spincontains \spin(3,1)
\cong \sl(2, \mathbb C)_{\mathbb R}
\cong \sp(2, \mathbb C)_{\mathbb R} .
\end{equation}
Here \(\sl(2,\mathbb C)_{\mathbb R}\) and \(\sp(2,\mathbb C)_{\mathbb R}\) denote the corresponding complex Lie algebras regarded as real Lie algebras.
Their complexifications are isomorphic to \(\spin(3,1)_{\mathbb C}\cong \sl(2,\mathbb C)\oplus\sl(2,\mathbb C)\), in accordance with the accidental rank-two Dynkin isomorphism \(D_2\cong A_1\oplus A_1\), while the last isomorphism reflects the accidental rank-one isomorphism \(A_1\cong C_1\).
A suitable basis for the associative algebra \(M^2(\mathbb C)\) is given by the identity together with the three Pauli matrices in \eqref{eq:Pauli matrices}.

\subparagraph{Split}

The four-dimensional split Clifford algebra with \(\gamma_1^2 = \gamma_2^2 = - \gamma_3^2 = - \gamma_4^2 = \ident\) has \(\delta \bmod 8 = 0\) and can be regarded as a larger version of the two-dimensional split algebra.
It is isomorphic to the four-by-four real matrices \(\cl(2,2) \cong M^4(\mathbb R)\), while its even subalgebra satisfies \(\cl_\PT(2,2) \cong \cl(2,1) \cong M^2(\mathbb R)\oplus M^2(\mathbb R)\) and contains the spin algebra \(\sl(2,\mathbb R)\oplus\sl(2,\mathbb R)\).
A real matrix representation adapted to two spatial followed by two temporal generators is
\begin{align} \label{eq:gamma-22}
\gamma_1 & = \sigma_1\otimes\sigma_3 , &
\gamma_2 & = \sigma_0\otimes\sigma_1 , &
\gamma_3 & = -\i\sigma_2\otimes\sigma_3 , &
\gamma_4 & = -\i\sigma_0\otimes\sigma_2 .
\end{align}
The first two matrices square to \(\matident\), the last two square to \(-\matident\), and all distinct pairs anticommute.

\subsection{Classification of Clifford algebras over the complex numbers} \label{sec:complex clifford algebras}

\begin{table}
\begin{tabular}{cll} \toprule
\(d\bmod 2\) & \header{\(\cl(d,\mathbb C)\)} & \header{\(\cl_\PT(d,\mathbb C)\)} \\ \midrule
\(0\) & \(\phantom2M(2^{\frac d2},\mathbb C)\) & \(2M(2^{\frac{d-2}2},\mathbb C)\) \\
\(1\) & \(2M(2^{\frac{d-1}2},\mathbb C)\) & \(\phantom2M(2^{\frac{d-1}2},\mathbb C)\) \\
\bottomrule \end{tabular}
\caption[Periodicity of complex Clifford algebras and even subalgebras]{
Complex Clifford algebras \(\cl(d, \mathbb C)\) and even subalgebras \(\cl_\PT(d, \mathbb C)\) as functions of \(d\bmod2\) for \(d>0\).
The zero-dimensional case is given separately in \cref{tab:clifford algebras}.
} \label{tab:clifford complex}
\end{table}

Complex Clifford representations depend only on the total dimension because
\begin{equation} \label{eq:Clifford algebra complexification}
\cl(s,t) \otimes_{\mathbb R}\mathbb C
\cong
\cl(d,\mathbb C) ,
\end{equation}
and factors of \(\i\) remove the distinction between real signatures.
The real eightfold periodicity in \(\delta\) consequently reduces to a twofold periodicity in \(d\).
Although this simplification suffices for unconstrained complex Dirac pinors, it erases the real or quaternionic structures needed to decide whether Majorana or \sMlong conditions exist.

The available complex phases also allow the volume element to be normalised:
\footnote{
In four-dimensional Lorentzian spacetime, the resulting complexified volume element \(\chi\) is the usual \(\gamma_5\).
}
\begin{align} \label{eq:complex volume element}
\chi & = \i^{(d^2 -\delta)/2} \gamma_1 \dots \gamma_d , &
\chi^2 & = \ident .
\end{align}
In even dimensions, this phase-normalised volume element determines chirality: Spin transformations preserve its eigenspaces, whereas a one-coordinate Pin reflection exchanges them.
Consequently, a single Weyl field naturally carries a Spin representation rather than a representation of the full Pin group.
An internal action of the full Pin group requires both chiralities, whereas a reflection may instead map a chiral theory to a distinct theory of opposite chirality, in which case the original theory remains only Spin-covariant.

The low-dimensional examples below apply the isomorphisms collected in \cref{fn:isomorphisms}, while the general results are summarised in \cref{tab:clifford algebras,tab:even clifford algebras,tab:spin algebras,tab:clifford complex}.

\paragraph{Zero dimensions}

In zero dimensions the complexification \(\cl(0, \mathbb C) \cong \cl^{\mathbb C}(0, 0) \cong \mathbb R \otimes_{\mathbb R} \mathbb C \cong \mathbb C\) leads to the algebra of the complex numbers.

\paragraph{One dimension}

The complexifications of the one-dimensional Clifford algebras are
\begin{equation}
\cl(1, \mathbb C) \cong
\left\{\begin{alignedat}2
\cl^{\mathbb C}(1, 0) &\cong{}&{} 2\mathbb R \otimes_{\mathbb R} \mathbb C & \\
\cl^{\mathbb C}(0, 1) &\cong{}&{} \mathbb C \otimes_{\mathbb R} \mathbb C &
\end{alignedat} \right\}
\cong 2 \mathbb C .
\end{equation}
Thus \(\cl(1,\mathbb C)\) contains two copies of the complex numbers and has a faithful realisation by two-by-two complex diagonal matrices.
The complexification of the even subalgebras results in the algebra of the complex numbers \(\cl_\PT(1,\mathbb C) \cong \mathbb R \otimes_{\mathbb R} \mathbb C = \mathbb C\).

\paragraph{Two dimensions}

The complexifications of the two-dimensional Clifford algebras are
\begin{equation}
\cl(2, \mathbb C) \cong
\left\{\begin{alignedat}2
\cl^{\mathbb C}(2, 0) \cong \cl^{\mathbb C}(1, 1) &\cong \mathbb R^2 &{} \otimes_{\mathbb R} \mathbb C \\
\cl^{\mathbb C}(0, 2) &\cong \mathbb H &{} \otimes_{\mathbb R} \mathbb C
\end{alignedat} \right\}
\cong \mathbb C^2 .
\end{equation}
Thus \(\cl(2,\mathbb C)\) is the algebra of two-by-two complex matrices, with basis given by the identity and the three Pauli matrices in \eqref{eq:Pauli matrices}.
The two-dimensional even Clifford algebra is
\begin{equation}
\cl_\PT(2, \mathbb C) \cong
\left\{\begin{alignedat}2
\cl_\PT^{\mathbb C}(2, 0) &\cong {}& \mathbb C \otimes_{\mathbb R} \mathbb C & \\
\cl_\PT^{\mathbb C}(1, 1) &\cong {}& 2 \mathbb R \otimes_{\mathbb R} \mathbb C &
\end{alignedat} \right\}
\cong 2 \mathbb C \spincontains \spin(2, \mathbb C) \cong \gl(1, \mathbb C) .
\end{equation}

\paragraph{Three dimensions}

The complexifications of the three-dimensional Clifford algebras are
\begin{equation}
\cl(3, \mathbb C) \cong
\left\{\begin{alignedat}4
\cl^{\mathbb C}(3, 0) \cong \cl^{\mathbb C}(1, 2) &\cong{}& \mathbb C&^2&{} \otimes_{\mathbb R} \mathbb C \\
\cl^{\mathbb C}(2, 1) &\cong{}& 2 \mathbb R&^2&{} \otimes_{\mathbb R} \mathbb C \\
\cl^{\mathbb C}(0, 3) &\cong{}& 2 \mathbb H& &{} \otimes_{\mathbb R} \mathbb C
\end{alignedat} \right\}
\cong 2\mathbb C^2 .
\end{equation}
Thus \(\cl(3,\mathbb C)\) consists of two copies of the two-by-two complex matrices, with basis \(\diag(\sigma_i,0)\) and \(\diag(0,\sigma_i)\) for \(i = 0,1,2,3\).
The three-dimensional even Clifford algebra is
\begin{equation}
\cl_\PT(3, \mathbb C) \cong
\left\{\begin{alignedat}2
\cl_\PT^{\mathbb C}(3, 0) &\cong \mathbb H &{} \otimes_{\mathbb R} \mathbb C \\
\cl_\PT^{\mathbb C}(2, 1) &\cong \mathbb R^2 &{} \otimes_{\mathbb R} \mathbb C
\end{alignedat} \right\}
\cong \mathbb C^2
\spincontains \spin(3, \mathbb C) \cong \sl(2, \mathbb C)
\cong \sp(2, \mathbb C) .
\end{equation}
The last isomorphisms are due to the accidental isomorphism of rank-one Dynkin diagrams \(A_1\cong B_1 \cong C_1\).
The identity together with the three Pauli matrices in \eqref{eq:Pauli matrices} gives a basis for the associative algebra.

\paragraph{Four dimensions}

The complexifications of the four-dimensional Clifford algebras are
\begin{equation}
\cl(4, \mathbb C) \cong
\left\{\begin{alignedat}2
\cl^{\mathbb C}(4, 0) \cong \cl^{\mathbb C}(1, 3) \cong \cl^{\mathbb C}(0, 4) &\cong \mathbb H^2 &{}\otimes_{\mathbb R} \mathbb C \\
\cl^{\mathbb C}(3, 1) \cong \cl^{\mathbb C}(2, 2) &\cong \mathbb R^4 &{}\otimes_{\mathbb R} \mathbb C
\end{alignedat} \right\}
\cong \mathbb C^4 .
\end{equation}
Thus \(\cl(4,\mathbb C)\) is the algebra of four-by-four complex matrices, with one convenient set of Clifford generators given by
\begin{align}
\gamma_i & = \begin{pmatrix}0 & \sigma_i \\ \sigma_i & 0\end{pmatrix} , &
i & = 1,2,3 , \\
\gamma_4 & = \i \begin{pmatrix}0 & -\sigma_0 \\ \sigma_0 &  0\end{pmatrix} , &
\omega & = \gamma_1 \gamma_2 \gamma_3 \gamma_4 = \begin{pmatrix}-\sigma_0 & 0 \\ 0 & \sigma_0 \end{pmatrix} ,
\end{align}
which is closely related to the Weyl basis \cite{Weyl:1929fm}.
\footnote{The Weyl basis is adapted to Lorentzian applications, although the same matrices generate the complexified Clifford algebra independently of the original real signature.}
The even subalgebra is then
\begin{equation}
\cl_\PT(4, \mathbb C) \cong
\left\{\begin{alignedat}4
\cl_\PT^{\mathbb C}(4, 0) &\cong{}& 2 \mathbb H& &{}\otimes_{\mathbb R} \mathbb C & \\
\cl_\PT^{\mathbb C}(3, 1) &\cong{}& \mathbb C&^2 &{}\otimes_{\mathbb R} \mathbb C & \\
\cl_\PT^{\mathbb C}(2, 2) &\cong{}& 2 \mathbb R&^2 &{}\otimes_{\mathbb R} \mathbb C &
\end{alignedat} \right\}
\cong 2\mathbb C^2 \spincontains \spin(4, \mathbb C) \cong \sl(2, \mathbb C) \oplus \sl(2, \mathbb C) .
\end{equation}
The last isomorphism is due to the accidental isomorphism of rank-two Dynkin diagrams \(D_2 \cong A_1 \oplus A_1\).
A basis is given by two independent copies of the identity and the three Pauli matrices in \eqref{eq:Pauli matrices}, whose traceless parts are usually labelled left- and right-handed.

\section{Pin and spin groups} \label{sec:pin-spin-groups}

Spinor transformation matrices arise from Clifford elements whose action on the generators reproduces an orthogonal transformation of spacetime.
Because conjugation by a general invertible Clifford element can mix a vector generator with higher-rank elements, this property selects a subgroup of the invertible Clifford algebra.
The \CL group therefore consists of the invertible Clifford elements whose twisted adjoint action \eqref{eq:twisted adjoint action} preserves the embedded vector subspace:
\begin{equation} \label{eq:Clifford-Lipschitz group}
\cL(s,t) = \set{g \in \Cl(s,t) \suchthat g_\PT^{} V^{s,t} g^{-1} = V^{s,t}} .
\end{equation}
As the normaliser of the spacetime-vector subspace, \(\cL(s,t)\) is precisely the group whose action can be projected onto \(\O(s,t)\).
Restricting the twisted adjoint to \(V^{s,t}\) defines the corresponding orthogonal matrix \(\Mmat\) through
\begin{align} \label{eq:Clifford orthogonal action}
g_\PT^{}\gamma_i g^{-1} & = \gamma_j\Mmat^j{}_i , &
g &\in \cL(s,t) , &
\gamma_i &\in V^{s,t}\subset\cl(s,t) , &
\Mmat &\in \O(s,t) .
\end{align}
Imposing the same requirements on the even Clifford subalgebra gives the even \CL group
\begin{equation}
\cL_\PT(s,t) = \cL(s,t) \cap \cl_\PT(s,t) .
\end{equation}
The Clifford norm of a group element \(g\) is defined by the Clifford conjugation
\begin{equation} \label{eq:Clifford norm}
N(g) = g_\cliff^{} g .
\end{equation}
For the real non-degenerate quadratic spaces considered here, the Clifford-group theorem implies both that \(N(g)\in\mathbb R^\times\) for every \(g\in\cL(s,t)\) and that every such element is a nonzero real scalar multiple of a product of non-null vectors \cite{Lawson:1998yr,Harvey:1990}.
For a homogeneous Pin element, the sign of the Clifford norm records spatial orientation, while combining it with the element's \(\PT\) sign records temporal orientation.
After normalising the vector factors, the pin group has the equivalent definitions \cite{Lawson:1998yr,Harvey:1990,Floerchinger:2019oeo}
\begin{equation} \label{eq:pin definition}
\Pin(s,t) = \left\langle v\in V^{s,t}\suchthat v^2 = \pm\ident\right\rangle = \set{g\in\cL(s,t) \suchthat N(g) = \pm\ident} .
\end{equation}
Every product of unit vectors has Clifford norm \(\pm\ident\), and the only Pin elements acting trivially on spacetime are
\begin{equation} \label{eq:pin kernel}
\set{
g\in\Pin(s,t)
\suchthat
\eval{\Ad^\PT_g}_{V^{s,t}} = \ident
} = \set{\pm\ident} .
\end{equation}
The kernel \eqref{eq:pin kernel} establishes that \(\Pin(s,t)\) double covers \(\O(s,t)\), as expressed for definite and indefinite signatures in \eqref{eq:definite pin,eq:indefinite pin}.
For \(s,t>0\), the three two-component orthogonal subgroups lift to the following Clifford subgroups.
The time-orientation-preserving pin group contains the spatial-reflection lift and has the Clifford realisation
\begin{equation}
\Pin_\plift(s,t) = \set{
g\in\Pin(s,t)
\suchthat
g_\PT^{} = N(g)g
} .
\end{equation}
The space-orientation-preserving pin group contains the temporal-reflection lift and has the Clifford realisation
\begin{equation}
\Pin_\tlift(s,t) = \set{
g\in\Pin(s,t)
\suchthat
N(g) = \ident
} .
\end{equation}
Both groups double cover their orthogonal counterparts in \eqref{eq:parity double covers}.
The total-orientation-preserving improper pin group is the even subgroup of the pin group,
\begin{equation}
\Pin_\ptlift(s,t) = \Pin(s,t) \cap\cL_\PT(s,t) = \set{
g\in\Pin(s,t)
\suchthat
g_\PT^{} = g
} .
\end{equation}
It double covers the improper special orthogonal group in \eqref{eq:improper spin cover}.
In a definite signature, the absent orientation character is trivial, so these intermediate inverse images reduce to either the full pin group or the proper spin group rather than defining three distinct subgroups.
For every signature, imposing both spatial- and temporal-orientation preservation gives the proper spin group, whose Clifford realisation corresponds to the definite and indefinite covers in \cref{eq:Euclidean spin,eq:indefinite spin}:
\begin{equation}
\properSpin(s,t) = \set{
g\in\Pin(s,t)
\suchthat
g_\PT^{} = g,
N(g) = \ident
} .
\end{equation}

\subsection{Representations}

Restricting the twisted adjoint action \eqref{eq:twisted adjoint action} to \(V^{s,t}\) associates each Pin element with the orthogonal transformation in \eqref{eq:Clifford orthogonal action} \cite{Atiyah:1964zz}.
For \(s,t>0\), a spatial index \(i_s\) and a temporal index \(i_t\) specify the coordinates reflected by \(\Pmat\) and \(\Tmat\), and the corresponding elementary Clifford lifts are
\begin{align} \label{eq:pt lifts}
\plift & = \gamma_{i_s} , &
\tlift & = \gamma_{i_t} .
\end{align}
Their twisted adjoint actions reproduce the corresponding reflection matrices:
\begin{align}
\eval{\Ad^\PT_{\plift}}_{V^{s,t}} & = \Pmat , &
\eval{\Ad^\PT_{\tlift}}_{V^{s,t}} & = \Tmat .
\end{align}
While each elementary lift has odd Clifford degree, their product \(\ptlift\) has even degree.
In an adapted ordering with all spatial generators followed by all temporal generators, representatives of the dimension-dependent lifts of the collective inversions \(\P\) and \(\T\) can be chosen as the products
\begin{align} \label{eq:collective-reflection-lifts}
\Plift & = \gamma_1\cdots\gamma_s , &
\Tlift & = \gamma_{s+1}\cdots\gamma_{s+t} , &
\PTlift & = \omega .
\end{align}
Their squares and relative commutation sign are
\begin{align} \label{eq:collective-reflection-signs}
\Plift^2 & = \eta_\rev^{}(s) \ident , &
\Tlift^2 & = \eta_\cliff^{}(t) \ident , &
(\PTlift)^2 & = \eta_\rev^{}(\delta) \ident , &
\PTlift & = \eta_\grade^{}(s t) \Tlift\Plift .
\end{align}
The elementary relations \(\plift^2 = +\ident\), \(\tlift^2 = -\ident\), and \(\plift\tlift = -\tlift\plift\) determine the Cliffordian Pin extension, whereas the dimension-dependent signs in \eqref{eq:collective-reflection-signs} describe the explicitly ordered collective lifts in \eqref{eq:collective-reflection-lifts}.
In particular, the square of the full spacetime-inversion lift is the volume-element square in \eqref{eq:volume element periodicity}.

By a similarity transformation, the Clifford generators can be brought to a unitary basis \cite{VanNieuwenhuizen:1981ab, VanNieuwenhuizen:1981ae, VanNieuwenhuizen:1985be}:
\begin{align} \label{eq:similarity transformation}
U^{-1} \gamma_i U & = \gamma_i^\prime , &
(\gamma_i^\prime)^\dagger\gamma_i^\prime & = \matident , &
U\in\GL(S^{\mathbb C})
.
\end{align}
Dropping the primes, the Clifford relation \eqref{eq:Clifford algebra} consequently implies
\begin{equation} \label{eq:gamma hermiticity}
\gamma_i^\dagger = \eta_{ii}\gamma_i .
\end{equation}
Thus spatial generators are Hermitian, while temporal generators are anti-Hermitian.
For a homogeneous Clifford basis element \(\element^a\) of degree \(a = a^s+a^t\), with \(a^s\) spatial and \(a^t\) temporal indices, one obtains
\begin{equation} \label{eq:homogeneous-hermiticity}
(\element^a)^\dagger = \eta_\rev^{}(a) \eta_\grade^{}(a^t) \element^a .
\end{equation}

\subsection{Intertwiners}

Hermitian adjoints, antilinear reality structures, and transpose pairings require three pairs of Clifford intertwiners, defined by
\begin{align} \label{eq:intertwiner}
A^\pm \gamma_i (A^\pm)^{-1} & = \pm (\gamma_i)^\dagger , &
\TransMat^\pm \gamma_i (\TransMat^\pm)^{-1} & = \pm (\gamma_i)^\trans , &
(\ConjMat^\pm)^{-1} \gamma_i \ConjMat^\pm & = \pm (\gamma_i)^\ast .
\end{align}
Under an arbitrary change of the Clifford basis
\begin{align}
(\gamma_i)^\prime &= U^{-1} \gamma_i U, &
U &\in \GL(n,\mathbb C)
\end{align}
the intertwiners transform as
\begin{align} \label{eq:intertwiner basis change}
(A^\pm)^\prime & = U^\dagger A^\pm U , &
(\TransMat^\pm)^\prime & = U^\trans \TransMat^\pm U , &
(\ConjMat^\pm)^\prime & = U^{-1} \ConjMat^\pm U^\ast .
\end{align}
Accordingly, the \(A^\pm\) are sesquilinear forms used in Dirac adjoints, the \(\TransMat^\pm\) are complex-bilinear forms used in Majorana self-contractions, and the \(\ConjMat^\pm\) map the complex-conjugate spinor space back to the original module \cite{Park:2022pjv,Gil-Garcia:2025iqt}.
The corresponding adjoint, transposed, and charge-conjugate (s)pinors are
\begin{align}
\widebar\psi^\pm &= \psi^\dagger A^\pm , &
\psi^{\pm\ft} &= \psi^\trans\TransMat^\pm , &
\psi^{\pm\charge} &= \ConjMat^\pm \psi^\ast .
\end{align}

\begin{table}
\begin{tabular}{r*8c} \toprule
& \multicolumn8c{\(\delta\bmod8\)} \\ \cmidrule{2-9}
& \(0\) & \(1\) & \(2\) & \(3\) & \(4\) & \(5\) & \(6\) & \(7\) \\ \midrule
\(\cl\) & \(\mathbb R\) & \(2\mathbb R\) & \(\mathbb R\) & \(\mathbb C\) & \(\mathbb H\) & \(2\mathbb H\) & \(\mathbb H\) & \(\mathbb C\) \\
\(\xi_\bott\) & \(+\) & \(+\) & \(+\) & \(0\) & \(-\) & \(-\) & \(-\) & \(0\) \\
\bottomrule \end{tabular}
\caption[Bott signature]{
Bott signature derived from the classification of Clifford algebras.
When this periodic sign is used for an intertwiner, a zero indicates that the corresponding intertwiner does not exist on the chosen irreducible complex Clifford module; it is not a vanishing square.
} \label{tab:bott signature}
\end{table}

\begin{table}
\begin{panels}2
\begin{tabular}{lcc} \toprule
\(d \bmod 4\) & \multicolumn2c{\(\delta\bmod4\)} \\ \cmidrule{2-3}
& \(1\) & \(3\) \\ \midrule
\(1\) & \(A^+\), \(\TransMat^+\), \(\ConjMat^+\) & \(A^-\), \(\TransMat^+\), \(\ConjMat^-\) \\
\(3\) & \(A^-\), \(\TransMat^-\), \(\ConjMat^+\) & \(A^+\), \(\TransMat^-\), \(\ConjMat^-\) \\
\bottomrule \end{tabular}
\caption{
Existing intertwiners.
} \label{tab:existence}
\panel
\begin{tabular}{lcc} \toprule
\(d \bmod 4\) & \multicolumn2c{\(\delta\bmod4\)} \\ \cmidrule{2-3}
& \(1\) & \(3\) \\ \midrule
\(1\) & \(\hphantom-\gamma^\dagger\), \(\hphantom-\gamma^\trans\), \(\gamma^\ast\) &
\(-\gamma^\dagger\), \(\hphantom-\gamma^\trans\), \(-\gamma^\ast\) \\
\(3\) & \(-\gamma^\dagger\), \(-\gamma^\trans\), \(\gamma^\ast\) &
\(\hphantom-\gamma^\dagger\), \(-\gamma^\trans\), \(-\gamma^\ast\) \\
\bottomrule \end{tabular}
\caption{
Equivalence classes of transformed generators.
} \label{tab:equivalence classes}
\end{panels}
\caption[Odd-dimensional intertwiner and generator equivalence classes]{
For a fixed irreducible complex representation in odd dimensions, the realised member of each intertwiner pair \eqref{eq:intertwiner} appears in panel \subref{tab:existence} for every pair of congruence classes \(d\bmod4\) and \(\delta\bmod4\).
The corresponding equivalence classes of transformed generators appear in panel \subref{tab:equivalence classes}.
} \label{tab:odd dimensions}
\end{table}

In a Clifford basis with definite Hermiticity, reality, and transpose symmetry, six ordered products provide explicit unitary candidates for the intertwiners:
\begin{align} \label{eq:intertwiner candidates}
A^\sym & = \prod_{\mathclap{\text{Hermitian}}} \gamma_i , &
A^\asym & = \prod_{\mathclap{\text{anti-Hermitian}}} \gamma_i , &
\TransMat^\sym & = \prod_{\mathclap{\text{symmetric}}} \gamma_i , &
\TransMat^\asym & = \prod_{\mathclap{\text{antisymmetric}}} \gamma_i , &
\ConjMat^\sym & = \prod_{\mathclap{\text{real}}} \gamma_i , &
\ConjMat^\asym & = \prod_{\mathclap{\text{imaginary}}} \gamma_i ,
\end{align}
where an empty product equals the identity.
The labels \(\sym\) and \(\asym\) describe the selected Clifford generators, while the actual signs in their intertwining relations are
\begin{align} \label{eq:intertwiner candidate actions}
A^{\flatfrac\asym\sym} \gamma_i (A^{\flatfrac\asym\sym})^{-1} & = \eta_A^{\flatfrac\asym\sym} (\gamma_i)^\dagger , &
\TransMat^{\flatfrac\asym\sym} \gamma_i (\TransMat^{\flatfrac\asym\sym})^{-1} & = \eta_\TransMat^{\flatfrac\asym\sym} (\gamma_i)^\trans , &
(\ConjMat^{\flatfrac\asym\sym})^{-1} \gamma_i \ConjMat^{\flatfrac\asym\sym} & = \eta_\ConjMat^{\flatfrac\asym\sym} (\gamma_i)^\ast ,
\end{align}
where the adjoint-intertwiner sign is given by
\begin{align} \label{eq:intertwiner candidate signs}
\eta_A^{\flatfrac\asym\sym}(s,t) & = \eta^{\flatfrac\asym\sym}\begin{cases}
\eta_\grade^{}(t) & \text{for } \asym , \\
\eta_\grade^{}(s)& \text{for } \sym ,
\end{cases} &
\eta^{\flatfrac\asym\sym} & = \begin{cases}
1 & \text{for } \asym , \\
-1& \text{for } \sym ,
\end{cases}
\end{align}
while the transposed- and conjugate-intertwiner signs are
\begin{align} \label{eq:intertwiner candidate signs 2}
\eta_\ConjMat^{\flatfrac\asym\sym}(\delta) & = \eta^{\flatfrac\asym\sym}
\xi_\bott(\delta) \xi_\grade^+(\delta) + \eta_\grade^{}\left(\frac{\delta-1}2\right) \xi_\grade^-(\delta) &
\eta_\TransMat^{\flatfrac\asym\sym}(s,t) & = \eta_\grade^{}(t)
\eta_\ConjMat^{\flatfrac\asym\sym}(s-t).
\end{align}
The signs are conveniently expressed through the eight-periodic Bott signature
\begin{equation} \label{eq:bott sign}
\xi_\bott(n)
 = \sgn\left[\cos\left(\frac{n-1}4\pi\right) \right]
 = \eta_\rev^{} \left(\floor*{\flatfrac n2}\right) [1-\xi_\grade^-(n) \xi_\grade^-(\floor{\flatfrac n2})]
\end{equation}
whose values follow directly from the Clifford-algebra classification in \cref{sec:real clifford algebras} and are displayed in \cref{tab:bott signature}.
In odd dimensions, the antisymmetric and symmetric signs \eqref{eq:intertwiner candidate signs,eq:intertwiner candidate signs 2} coincide:
\begin{align}
\eta_A^\asym(s,t) & = \eta_A^\sym(s,t) , &
\eta_\TransMat^\asym(s,t) & = \eta_\TransMat^\sym(s,t) , &
\eta_\ConjMat^\asym(\delta) & = \eta_\ConjMat^\sym(\delta) , &
\delta = 1\bmod2 .
\end{align}
Hence, in odd dimensions, one member of each intertwiner pair in \eqref{eq:intertwiner} fails to exist on a fixed irreducible complex Clifford module, with the periodic pattern displayed in \cref{tab:odd dimensions} \cite{Gil-Garcia:2025iqt}.
The obstruction is the central volume element, which acts as a scalar in an irreducible complex representation and therefore cannot provide the negation intertwiner \eqref{eq:negation intertwiner}.
In even dimensions, by contrast, all six intertwiners \eqref{eq:intertwiner} are realised, and multiplication by the volume element \eqref{eq:volume element} relates the two members of each pair.

\begin{table}
\begin{panels}2
\begin{tabular}{cc*4c} \toprule
\(a\bmod4\) & \(s\bmod4\) & \multicolumn4c{\(t\bmod4\)} \\ \cmidrule{3-6}
& & \(0\) & \(1\) & \(2\) & \(3\) \\ \midrule
\multirow4*{\(0,1\)} & \(0\) & \(+\) & \(0\) & \(-\) & \(0\) \\
& \(1\) & \(+\) & \(+\) & \(-\) & \(+\) \\
& \(2\) & \(+\) & \(0\) & \(-\) & \(0\) \\
& \(3\) & \(+\) & \(-\) & \(-\) & \(-\) \\ \cmidrule{2-6}
\multirow4*{\(2,3\)} & \(0\) & \(-\) & \(0\) & \(+\) & \(0\) \\
& \(1\) & \(-\) & \(-\) & \(+\) & \(-\) \\
& \(2\) & \(-\) & \(0\) & \(+\) & \(0\) \\
& \(3\) & \(-\) & \(+\) & \(+\) & \(+\) \\
\bottomrule \end{tabular}
\caption{\(\xi_A^+(s,t,a)\)} \label{tab:xi A plus compact}
\panel
\begin{tabular}{cc*4c} \toprule
\(a\bmod4\) & \(s\bmod4\) & \multicolumn4c{\(t\bmod4\)} \\ \cmidrule{3-6}
& & \(0\) & \(1\) & \(2\) & \(3\) \\ \midrule
\multirow4*{\(0,3\)} & \(0\) & \(+\) & \(+\) & \(+\) & \(+\) \\
& \(1\) & \(0\) & \(-\) & \(0\) & \(+\) \\
& \(2\) & \(-\) & \(-\) & \(-\) & \(-\) \\
& \(3\) & \(0\) & \(-\) & \(0\) & \(+\) \\ \cmidrule{2-6}
\multirow4*{\(1,2\)} & \(0\) & \(-\) & \(-\) & \(-\) & \(-\) \\
& \(1\) & \(0\) & \(+\) & \(0\) & \(-\) \\
& \(2\) & \(+\) & \(+\) & \(+\) & \(+\) \\
& \(3\) & \(0\) & \(+\) & \(0\) & \(-\) \\
\bottomrule \end{tabular}
\caption{\(\xi_A^-(s,t,a)\)} \label{tab:xi A minus compact}
\end{panels}
\caption[Adjoint-intertwiner Hermiticity signatures]{
Periodicity of the rank-dependent adjoint-intertwiner Hermiticity signatures in the spatial and temporal dimensions and in the rank of the homogeneous Clifford element.
The base cases are \(\xi_A^\pm(s,t) = \xi_A^\pm(s,t,0)\).
Each zero entry indicates that the corresponding adjoint intertwiner is absent on the fixed irreducible complex representation.
} \label{tab:symmetry A}
\end{table}

\begin{table}
\begin{panels}[t]2
\begin{tabular}{lc*8c} \toprule
\multicolumn2r{\(a\bmod 2\)} & \multicolumn8c{\(d\bmod8\)} \\ \cmidrule{3-10}
& & \(0\) & \(1\) & \(2\) & \(3\) & \(4\) & \(5\) & \(6\) & \(7\) \\ \midrule
\multirow2*{\(\xi_\TransMat^+\)}& \(0,1\) & \(+\) & \(+\) & \(+\) & \(0\) & \(-\) & \(-\) & \(-\) & \(0\) \\
& \(2,3\) & \(-\) & \(-\) & \(-\) & \(0\) & \(+\) & \(+\) & \(+\) & \(0\) \\ \cmidrule{3-10}
\multirow2*{\(\xi_\TransMat^-\)}& \(0,3\) & \(+\) & \(0\) & \(-\) & \(-\) & \(-\) & \(0\) & \(+\) & \(+\) \\
& \(1,2\) & \(-\) & \(0\) & \(+\) & \(+\) & \(+\) & \(0\) & \(-\) & \(-\) \\
\bottomrule \end{tabular}
\caption{Majorana selection signature \(\xi_\TransMat^\pm(d, a)\)} \label{tab:symmetry trans}
\panel
\begin{tabular}{lc*8c} \toprule
\multicolumn2r{\(a\bmod 2\)} & \multicolumn8c{\(\delta\bmod8\)} \\ \cmidrule{3-10}
& & \(0\) & \(1\) & \(2\) & \(3\) & \(4\) & \(5\) & \(6\) & \(7\) \\ \midrule
\(\xi_\ConjMat^+\) & & \(+\) & \(+\) & \(+\) & \(0\) & \(-\) & \(-\) & \(-\) & \(0\) \\ \cmidrule{3-10}
\multirow2*{\(\xi_\ConjMat^-\)}& \(0\) & \(+\) & \(0\) & \(-\) & \(-\) & \(-\) & \(0\) & \(+\) & \(+\) \\
& \(1\) & \(-\) & \(0\) & \(+\) & \(+\) & \(+\) & \(0\) & \(-\) & \(-\) \\
\bottomrule \end{tabular}
\caption{Majorana structure signature \(\xi_\ConjMat^\pm(\delta, a)\)} \label{tab:symmetry conj}
\end{panels}
\caption[Majorana structure and selection signatures]{
Periodicity of the Majorana selection and structure signatures in panels \subref{tab:symmetry trans} and \subref{tab:symmetry conj}, respectively.
The base cases are \(\xi_\TransMat^\pm(d) = \xi_\TransMat^\pm(d, 0)\) and \(\xi_\ConjMat^\pm(\delta) = \xi_\ConjMat^\pm(\delta, 0)\).
Each zero entry indicates that the corresponding conjugation or transposition intertwiner is absent on the fixed irreducible complex representation.
} \label{tab:symmetry trans conj}
\end{table}

Applying Hermitian conjugation, transposition, or complex conjugation to the corresponding intertwiner relation \eqref{eq:intertwiner} shows that \((A^\pm)^{-1}(A^\pm)^\dagger\), \((\TransMat^\pm)^{-1}(\TransMat^\pm)^\trans\), and \(\ConjMat^\pm(\ConjMat^\pm)^\ast\) commute with all Clifford generators.
On an irreducible complex Clifford module, Schur's lemma therefore restricts each of these combinations to a scalar multiple of the identity.
With the normalisations adopted here, these scalars reduce to the signatures
\begin{align} \label{eq:intertwiner relation}
(A^\pm)^\dagger & = \xi_A^\pm(s,t) A^\pm , &
(\TransMat^\pm)^\trans & = \xi_\TransMat^\pm(d) \TransMat^\pm , &
\ConjMat^\pm (\ConjMat^\pm)^\ast & = \xi_\ConjMat^\pm(\delta) \matident .
\end{align}
The numerical value of an adjoint-intertwiner signature depends on the fixed phase of \(A^\pm\).
For the unphased ordered products in \eqref{eq:intertwiner candidates}, these signatures are
\begin{subequations} \label{eq:signature A}
\begin{align}
\xi_A^+(s,t) & = \xi_\grade^+(t) \eta_\cliff^{}(t)
+
\xi_\grade^-(t) \xi_\grade^-(s) \eta_\rev^{}(s) , &
\xi_A^-(s,t) & = \xi_\grade^+(s) \eta_\rev^{}(s)
+
\xi_\grade^-(s) \xi_\grade^-(t) \eta_\cliff^{}(t).
\end{align}
\end{subequations}
The resulting signatures are listed explicitly in \cref{tab:symmetry A}.
The signatures of the conjugation and transposition intertwiners are
\begin{align} \label{eq:signature trans conj}
\xi_\TransMat^\pm(d) & = \begin{cases}
\xi_\bott(d) & \text{for } + , \\
\xi_\bott(d+2) & \text{for } - ,
\end{cases} &
\xi_\ConjMat^\pm(\delta) & = \begin{cases}
\xi_\bott(\delta) & \text{for } + , \\
\xi_\bott(\delta+2) & \text{for } {-} .
\end{cases}
\end{align}
The corresponding periodicity patterns are collected in \cref{tab:symmetry trans conj}.
For a unitary basis, the conjugation-intertwiner relation \eqref{eq:intertwiner relation} is equivalent to \((\ConjMat^\pm)^\trans = \xi_\ConjMat^\pm(\delta) \ConjMat^\pm\).

In even dimensions, the unphased volume element \(\omega\) \eqref{eq:volume element} implements the negation of the Clifford generators \eqref{eq:negation intertwiner}, while chirality is defined by the phase-normalised volume element \(\chi\) in \eqref{eq:complex volume element}.
The products \(A^+\omega\), \(\TransMat^+\omega\), and \(\omega \ConjMat^+\) satisfy the defining relations of \(A^-\), \(\TransMat^-\), and \(\ConjMat^-\), respectively.
For an irreducible complex Clifford representation, uniqueness up to complex scalars therefore gives
\begin{align} \label{eq:intertwiner negation}
A^- & = \xi_A^+(s,t) A^+ \omega , &
\TransMat^- & = \zeta_\TransMat \TransMat^+ \omega , &
\ConjMat^- & = \zeta_\ConjMat \omega \ConjMat^+  , &
\zeta_\TransMat,\zeta_\ConjMat &\in \U(1) .
\end{align}
Although adjoining the volume element fixes each relative phase, rephasing the representatives can remove \(\zeta_\TransMat\) and \(\zeta_\ConjMat\) without changing the invariant transpose or conjugation signatures.
The intertwiners act on the volume element according to
\begin{equation} \label{eq:intertwiner candidates-on-volume-element}
\begin{aligned}
A^\pm \omega (A^\pm)^{-1} & = \eta_\grade^\pm(d) \eta_\rev^{}(d) \omega^\dagger , \\
\TransMat^\pm \omega (\TransMat^\pm)^{-1} & = \eta_\grade^\pm(d) \eta_\rev^{}(d) \omega^\trans , &
(\ConjMat^\pm)^{-1}\omega \ConjMat^\pm & = \eta_\grade^\pm(d) \omega^\ast ,
\end{aligned}
\end{equation}
where we have introduced
\begin{equation}
\eta_\grade^\pm(d)
\equiv
\begin{cases}
1 & \text{for } + , \\
\eta_\grade^{}(d) & \text{for } {-} .
\end{cases}
\end{equation}

For each existing intertwiner, the intertwiner relations \eqref{eq:intertwiner} extend to a homogeneous Clifford basis element \(\element^a\) as
\begin{subequations}
\begin{align}
A^\pm \element^a (A^\pm)^{-1} & = \eta_\grade^\pm(a) \eta_\rev^{}(a) (\element^a)^\dagger , \\
\TransMat^\pm \element^a (\TransMat^\pm)^{-1} & = \eta_\grade^\pm(a) \eta_\rev^{}(a) (\element^a)^\trans , &
(\ConjMat^\pm)^{-1} \element^a \ConjMat^\pm & = \eta_\grade^\pm(a) (\element^a)^\ast .
\end{align}
\end{subequations}
Together with the signatures in \eqref{eq:signature A,eq:signature trans conj}, these relations determine the rank-dependent signs controlling the Hermiticity, Grassmann symmetry, and reality of fermion bilinears.
For the adjoint intertwiner, this gives
\begin{align} \label{eq:Aa sym}
(A^\pm\element^a)^\dagger & = \xi_A^\pm(s,t,a)
A^\pm\element^a ,&
\xi_A^\pm(s,t,a) & = \xi_A^\pm(s,t) \eta_\grade^\pm(a) \eta_\rev^{}(a) ,
\end{align}
as collected in \cref{tab:symmetry A}, while the transposition intertwiner gives
\begin{align} \label{eq:Ca sym}
(\TransMat^\pm\element^a)^\trans & = \xi_\TransMat^\pm(d,a)
\TransMat^\pm\element^a
,&
\xi_\TransMat^\pm(d,a) & = \xi_\TransMat^\pm(d)
\eta_\grade^\pm(a) \eta_\rev^{}(a)
,
\end{align}
as collected in \cref{tab:symmetry trans}.
The corresponding conjugation-intertwiner relation is
\begin{align}\label{eq:Ba sym}
(\element^a \ConjMat^\pm)(\element^a \ConjMat^\pm)^\ast & = \xi_\ConjMat^\pm(\delta,a) (\element^a)^2 , &
\xi_\ConjMat^\pm(\delta,a) & = \xi_\ConjMat^\pm(\delta)
\eta_\grade^\pm(a) ,
\end{align}
as collected in \cref{tab:symmetry conj}.

\section{Pinors and spinors} \label{sec:spinor-pinor-classification}

\begin{table}
\begin{tabular}{c*3{rl}} \toprule
\(\delta\bmod8\) & \multicolumn2c{Pin \(J_\Ch^+\)} & \multicolumn2c{Spin restriction} & \multicolumn2c{Spin \(J_\Ch^-\)} \\ \cmidrule(r){2-3} \cmidrule(lr){4-5} \cmidrule(l){6-7}
& \(\cl(s,t)\) & constrained field & \(\cl_\PT\) & minimal field & \(\cl(t,s)\) & constrained field \\
\midrule
\(0\) & \(\mathbb R\) & Majorana & \(2\mathbb R\) & $\MW$ & \(\mathbb R\) & $\MW$ \\
\(1\) & \(2\mathbb R\) & two Majorana classes & \(\mathbb R\) & Majorana & \(\mathbb C\) & none \\
\(2\) & \(\mathbb R\) & Majorana & \(\mathbb C\) & Weyl & \(\mathbb H\) & $\sM$ \\
\(3\) & \(\mathbb C\) & none & \(\mathbb H\) & $\sM$ & \(2\mathbb H\) & $\sM$ \\
\(4\) & \(\mathbb H\) & $\sM$ & \(2\mathbb H\) & $\sMW$ & \(\mathbb H\) & $\sMW$ \\
\(5\) & \(2\mathbb H\) & two \sM classes & \(\mathbb H\) & $\sM$ & \(\mathbb C\) & none \\
\(6\) & \(\mathbb H\) & $\sM$ & \(\mathbb C\) & Weyl & \(\mathbb R\) & Majorana \\
\(7\) & \(\mathbb C\) & none & \(\mathbb R\) & Majorana & \(2\mathbb R\) & Majorana \\
\bottomrule
\end{tabular}
\caption[Pinor and spinor field content]{
Periodic Pin and Spin field content as a function of the Bott class \(\delta = s-t\bmod8\).
The first pair of columns specifies the full Clifford algebra and the Pin-equivariant Majorana-type condition induced by \(J_\Ch^+\), while the middle pair specifies the even Clifford algebra and the minimal irreducible Spin field, including its Weyl type in even dimensions.
The final pair contains the signature-reversed algebra and the Spin-equivariant Majorana-type condition induced by \(J_\Ch^-\).
For \(\delta = 0,4\bmod8\), both \(J_\Ch^+\) and \(J_\Ch^-\) preserve either Weyl space, whereas for \(\delta = 2,6\bmod8\) both maps exchange the two Weyl spaces and a corresponding condition requires their direct sum.
An unconstrained complex Dirac field can be formed in every class, and every \sM condition requires the even multiplet in \eqref{eq:quaternionic structure}.
An entry labelled ``none'' means that the indicated \(J_\Ch^\pm\) is absent on the chosen irreducible complex module.
It does not exclude a Spin-level condition supplied by the other intertwiner or a condition on enlarged field content after adjoining the conjugate module.
A prefactor of two denotes two simple summands: for \(\cl(s,t)\) in the classes \(\delta = 1,5\bmod8\) these give two inequivalent Pin modules, while for \(\cl_\PT(s,t)\) in the classes \(\delta = 0,4\bmod8\) they give the two Weyl modules.
The two signature-reversed Pin modules for \(\delta = 3,7\bmod8\) restrict to the same Spin module.
} \label{tab:spinor-pinor-classes}
\end{table}

In general signature, covariance under the connected Spin group must be distinguished from its extension to the full Pin group, even though the distinction is often hidden in four-dimensional vectorlike theories.
A spinor transforms under \(\properSpin(s,t)\), the double cover of the identity component \(\properSO(s,t)\), and is therefore the object usually called a Lorentz spinor.
A pinor additionally carries an action of \(\Pin(s,t)\), as required when parity, time reflection, or a Pin structure is part of the physical problem.
This additional structure affects Weyl fields, reality conditions, and the possible reflection assignments.

Algebraically, a pinor space \(S\) carries a representation of the full real Clifford algebra, whose complexification supplies the corresponding complex field space:
\begin{align} \label{eq:pinor}
\cl(s,t) &\to \End_{\mathbb R}(S) , &
S^{\mathbb C} & = S\otimes_{\mathbb R}\mathbb C
.
\end{align}
The matrices \(\gamma_i\) representing the Clifford generators on \(S\) satisfy \eqref{eq:Clifford algebra}, and an unconstrained complex pinor takes values in an irreducible complex constituent of \(S^{\mathbb C}\).
Restricting this representation to \(\Pin(s,t)\) gives the action of rotations and spacetime reflections on the field.

By contrast, a spinor space \(S_\PT\) need only represent the even Clifford algebra:
\begin{align} \label{eq:spinor}
 \cl_\PT(s,t) &\to \End_{\mathbb R}(S_\PT) ,&
 S_\PT^\mathbb C & = S_\PT\otimes_{\mathbb R}\mathbb C .
\end{align}
A complex spinor field may then take values in a single irreducible complex constituent of \(S_\PT^\mathbb C\).
Additional constituents are required only when an antilinear structure or reflection relates them, or when the field content is nonchiral; they must then be combined before a Majorana-type condition can select an invariant real subspace.

Pinor classes consequently follow from the single-block or split matrix-algebra type of \(\cl(s,t)\), whereas spinor classes follow from the corresponding type of \(\cl_\PT(s,t)\).
The matrix-algebra type determines whether the relevant module is real, complex, or quaternionic, while the conjugation intertwiners determine which reality conditions can be imposed, as summarised in \cref{tab:spinor-pinor-classes}.

In even dimensions the volume element \eqref{eq:volume element} anticommutes with every Clifford generator and therefore commutes with the even Clifford algebra.
After phase normalisation \eqref{eq:complex volume element}, it consequently defines a \(\C^2_\chi\) grading of the complex spinor representation into the eigenspaces.
The corresponding chiral projectors are
\begin{align} \label{eq:chiral projectors}
\psi^\lambda &\in S_\PT^\lambda , &
S_\PT^\lambda & = \Pi^\lambda S_\PT^\mathbb C , &
\Pi^\lambda & = \frac{\matident+\lambda\chi}2 , &
\lambda & = \pm1 .
\end{align}
Each eigenspace is invariant under the Spin action and therefore defines a Weyl module in which a Weyl spinor field can take its values.
An antilinear reality structure can be imposed on a single Weyl module only when it preserves the corresponding eigenspace; otherwise it exchanges the two Weyl modules and requires their direct sum.
This compatibility is demonstrated in the Spin columns of \cref{tab:spinor-pinor-classes}.

On the complexified modules, the complex-conjugation intertwiners induce the antilinear maps
\begin{align} \label{eq:Majorana structure}
J_\Ch^\pm \psi & = \ConjMat^\pm \psi^\ast , &
(J_\Ch^\pm)^2 & = \xi_\ConjMat^\pm(\delta) \ident .
\end{align}
Its dependence on \(\delta\) follows from \eqref{eq:Ba sym} and is listed in \cref{tab:symmetry conj}.
An antilinear map is called Spin- or Pin-equivariant when it commutes with the corresponding group action.
The defining relations for \(\ConjMat^\pm\) imply that \(J_\Ch^\pm\) commute with the even Clifford action, whereas \(J_\Ch^+\) commutes and \(J_\Ch^-\) anticommutes with every odd Clifford generator.
Both maps can therefore define Spin-equivariant reality conditions, but only \(J_\Ch^+\) is automatically equivariant under the full Pin action.
A fixed-point condition \(J_\Ch^\pm\psi = \psi\) is consistent only when the antilinear map squares to \(+\ident\), since applying a negative-square map twice would instead give \(\psi = -\psi\).
A Majorana condition can be imposed when the corresponding conjugation is a real structure,
\begin{align} \label{eq:real structure}
\psi & = \ConjMat^\pm \psi^\ast , &
(J_\Ch^\pm)^2 = \xi_\ConjMat^\pm(\delta) & = +1 .
\end{align}
When the relevant equivariant map instead squares to \(-\ident\), the second minus sign supplied by an antisymmetric internal pairing makes a \sMfirst condition possible on an even multiplet:
\begin{align} \label{eq:quaternionic structure}
\psi_A & = \Omega_{AB} \ConjMat^\pm \psi_B^\ast , &
(J_\Ch^\pm)^2 = \xi_\ConjMat^\pm(\delta) & = -1 .
\end{align}
Here \(A = 1,\ldots,2N\), and the symplectic matrix \(\Omega_{AB}\) is
\begin{align} \label{eq:symplectic matrix}
\Omega & = \matident^N\otimes\begin{psmallmatrix}0&1\\-1&0\end{psmallmatrix} , &
\Omega^\trans & = -\Omega , &
\Omega\Omega^\ast & = -\matident^{2N} .
\end{align}
The case \(N = 1\) is the minimal doublet, while for every \(N\) the combined antilinear map in \eqref{eq:quaternionic structure} squares to \(\ident\).
In even dimensions a reality condition is compatible with a Weyl condition precisely when its antilinear map preserves each chiral subspace; if the map exchanges them, the condition instead requires \(S_\PT^+\oplus S_\PT^-\).

The resulting fields are
\begin{description}[style = nextline]
\item[Dirac (s)pinor] An unconstrained complex pinor or spinor, containing both Weyl modules in even dimension.
\item[Weyl spinor] A chiral complex spinor in even dimension.
\item[Majorana (s)pinor] A real field satisfying \(\psi = \ConjMat^\pm \psi^\ast\).
\item[\sMlong (s)pinor] A quaternionic even field multiplet satisfying \(\psi = \Omega \ConjMat^\pm \psi^\ast\).
\item[\MWlong spinor] A real chiral spinor satisfying \(\psi = \ConjMat^\pm \psi^\ast\).
\item[\sMWlong spinor] A quaternionic chiral even field multiplet satisfying \(\psi = \Omega \ConjMat^\pm \psi^\ast\).
\end{description}

\subsection{Discrete transformations}

\begin{table}
\begin{tabular}{lr@{}lr@{}lc} \toprule
&\multicolumn2c{map} & \multicolumn2c{spinor part} & chirality \\ \midrule
\multirow3*{linear}
&\(\P\) & & \(A^- \psi\) & & \(\eta_\grade^{}(s)\) \\
&\(\T\) & & \(A^+\psi\) & & \(\eta_\grade^{}(t)\)  \\
&\(\PT\) & & \(\omega\psi\) & & \(1\) \\ \cmidrule{2-6}
\multirow5*{antilinear}
&\(\Ch\) & \({}^\pm\) & \(\ConjMat^\pm\psi\) & \({}^\ast\) & \(\eta_\rev^{}(\delta)\) \\
&\(\PC\) & \({}^\pm\) & \(A^-\ConjMat^\pm\psi\) & \({}^\ast\) & \(\eta_\rev^{}(d)\) \\
&\(\TC\) & \({}^\pm\) & \(A^+\ConjMat^\pm\psi\) & \({}^\ast\) & \(\eta_\rev^{}(d)\) \\
&\(\PTC\) & \({}^\pm\) & \(\omega\ConjMat^\pm\psi\) & \({}^\ast\) & \(\eta_\rev^{}(\delta)\) \\
&\(\alT\) & \({}^\pm\) & \((\TransMat^\pm)^{-1}\psi\) & \({}^\ast\) & \(\eta_\rev^{}(d)\) \\
\bottomrule
\end{tabular}
\caption[Chirality signatures of candidate Clifford-module maps]{
Chirality signatures of candidate Clifford-module maps associated with collective inversions and antilinear transformations in even dimensions.
Positive chirality signature means that the map preserves each Weyl representation, whereas a negative signature means that the map exchanges the Weyl spaces \(S_\PT^+\) and \(S_\PT^-\).
} \label{tab:weyl-discrete-transformations}
\end{table}

\begin{table}
\begin{tabular}{cccccc}
\toprule
\(d\bmod4\) & \(\delta\bmod4\) & \(\P,\T\) & \(\PT\) & \(\Ch^\pm,\PTC^\pm\) & \(\alT^\pm,\PC^\pm,\TC^\pm\) \\
\midrule
\multirow2*{\(0\)} & \(0\) & \(+\) & \(+\) & \(+\) & \(+\) \\
& \(2\) & \(-\) & \(+\) & \(-\) & \(+\) \\
\midrule
\multirow2*{\(2\)} & \(0\) & \(-\) & \(+\) & \(+\) & \(-\) \\
& \(2\) & \(+\) & \(+\) & \(-\) & \(-\) \\
\bottomrule \end{tabular}
\caption[Signature dependence of candidate Clifford-module maps]{
Signature dependence of candidate Clifford-module maps acting on Weyl spinor spaces in even dimension.
The signs indicate whether the corresponding map sends \(S_\PT^\lambda\) to itself or to \(S_\PT^{-\lambda}\).
\(\PT\) is always even and never exchanges the Weyl spinor spaces.
} \label{tab:weyl-transformations-by-signature}
\end{table}

Discrete transformations can act on the spacetime coordinates, the spinor components, or both, with coordinate reflections determining the transformation of \(x\) and Clifford-module maps acting on the spinor indices.
The intertwiners \(A^\pm\), \(\ConjMat^\pm\), and \(\TransMat^\pm\), together with the structure map \(\structure\), supply the relevant local maps and determine whether they preserve or exchange chirality.
The resulting candidates \(\Ch^\pm\), \(\alT^\pm\), \(\PC^\pm\), \(\TC^\pm\), and \(\PTC^\pm\) do not by themselves define symmetries of the \QFT.
A physical symmetry also requires a compatible action on the coordinates, internal representations, and complete field content, together with invariance of the operator algebra and action; time reversal must moreover be represented antiunitarily on the Hilbert space.
Although the overall phases of the module maps are conventional, their existence and action on chirality are fixed by the Clifford-module structure.
The elementary spatial and temporal reflections \(\p\) and \(\t\), by contrast, are represented by the one-generator Clifford lifts \eqref{eq:pt lifts}; in even dimensions each therefore exchanges the two Weyl spaces, while their product \(\ptlift\) preserves them.
The chirality actions of all candidate module maps in even dimension are given in \cref{tab:weyl-discrete-transformations,tab:weyl-transformations-by-signature}.

\subsubsection{Spatial and temporal inversions}

The collective spatial and temporal inversions \eqref{eq:generator reflection}, together with their product, act on the spinor field as
\begin{align} \label{eq:field reflections}
\psi_\P(x) & = U_\P \psi(\P x) , &
\psi_\T(x) & = U_\T \psi(\T x) , &
\psi_\PT(x) & = U_\PT \psi(\PT x).
\end{align}
On a spinor field space admitting the required Clifford-module maps, the corresponding spinor transformations can be represented by
\begin{align}
 U_\P
 & =
 \zeta_\P
 \left[
 \xi_\grade^+(s)A^-
 +
 \xi_\grade^-(s)\structure A^+
 \right] , &
 U_\PT
 & =
 \zeta_\PT\structure
 \\
 U_\T
 & =
 \zeta_\T
 \left[
 \xi_\grade^+(t)A^+
 +
 \xi_\grade^-(t)\structure A^-
 \right] , &
 \zeta_\P,\zeta_\T,\zeta_\PT &\in \U(1)
 .
\end{align}
The adjoint intertwiners are defined in \eqref{eq:intertwiner}, while \(\zeta_\P\), \(\zeta_\T\), and \(\zeta_\PT\) parameterise the phase freedom of their spinor-space implementations.
The matrices \(U_\P\), \(U_\T\), and \(U_\PT\) implement the corresponding automorphisms of the represented Clifford generators by ordinary conjugation, whereas \(\Plift\), \(\Tlift\), and \(\PTlift\) in \eqref{eq:collective-reflection-lifts} are canonical \CL-group lifts under the twisted adjoint.
Because ordinary and twisted adjoint actions differ on odd elements, the two sets of representatives need not coincide and their squares must not be identified; the canonical Pin-lift signs are fixed by \eqref{eq:collective-reflection-signs}.
In odd dimensions, any factor of \(\structure\) may exchange the two grade-related irreducible Pin sectors, so the corresponding formula defines an internal operator only when the field space contains their direct sum.

In even dimensions the volume element determines chirality, and the volume-element transformation rules \eqref{eq:intertwiner candidates-on-volume-element} imply
\begin{align}
U_\P \chi(U_\P)^{-1} & = \eta_\grade^{}(s) \chi,&
U_\T \chi(U_\T)^{-1} & = \eta_\grade^{}(t) \chi,&
U_\PT \chi(U_\PT)^{-1} & = \eta_\grade^{}(d) \chi,
\end{align}
as shown in \cref{tab:weyl-discrete-transformations}.
The lifted reflections \eqref{eq:field reflections} consequently act on the chiral projectors as
\begin{align}
U_X \Pi^\lambda (U_X)^{-1} & = \Pi^{\lambda^\prime} , &
U_X S_\PT^\lambda &\subset
S_\PT^{\lambda^\prime} , &
\lambda^\prime & = \eta_X\lambda
,
\end{align}
where \(X\in\set{\P,\T,\PT}\).
Equivalently, a reflection preserves Weyl representations if it commutes with the chirality operator and exchanges them if it anticommutes with it.
Since \(d\) is even in the Weyl case, \(\PT\) always preserves chirality, while \(\P\) and \(\T\) preserve or exchange chirality according to the parity of \(s\) and \(t\), respectively, as shown in \cref{tab:weyl-transformations-by-signature}.
When a lifted reflection exchanges the Weyl spaces, it acts internally only on field content containing both chiralities.
For purely chiral field content, it instead maps the theory to a distinct theory of opposite chirality, so the original theory remains only Spin-covariant.
In odd dimensions the volume element is central in an irreducible representation and no Weyl splitting is defined.

\subsubsection{Antilinear transformations}

The complex-conjugation intertwiners \eqref{eq:intertwiner} define two candidate charge-conjugation maps on the spinor module,
\begin{align} \label{eq:field charge conjugated}
\psi_\Ch^\pm(x) & = \zeta_\Ch^\pm \ConjMat^\pm\psi^\ast(x).
\end{align}
In even dimensions, the negation intertwiner \eqref{eq:intertwiner negation} relates the two definitions.
After the fixed phase relating \(\omega\) to \(\chi\) is absorbed into the conventional phase \(\zeta_\Ch^-\), \eqref{eq:intertwiner negation} gives
\begin{align}
\psi_\Ch^-(x) & = \zeta_\Ch \chi \psi_\Ch^+(x) , &
\psi_\Ch^{\lambda-}(x) & = \lambda \zeta_\Ch \psi_\Ch^{\lambda+}(x) , &
\zeta_\Ch &= \frac{\zeta_\Ch^-\zeta_\ConjMat}{\zeta_\Ch^+} .
\end{align}
Here \(\psi_\Ch^{\lambda\pm}\equiv\Pi^\lambda\psi_\Ch^\pm\) denotes the chirality-\(\lambda\) component of the transformed field.
Hence \(\Ch^-\) differs from \(\Ch^+\) only by composition with the chirality operator and a conventional phase.
The chirality of this candidate charge-conjugated spinor is determined by the antilinear action on the complex volume element.
For the Majorana structure \eqref{eq:Majorana structure}, one has
\begin{align}
J_\Ch^\pm \chi (J_\Ch^\pm)^{-1} & = \eta_\rev^{}(\delta) \chi , &
\Ch^\pm \Pi^\lambda (\Ch^\pm)^{-1} & = \Pi^{\lambda^\prime} , &
\lambda^\prime & = \eta_\rev^{}(\delta) \lambda ,
\end{align}
see also \cref{tab:weyl-discrete-transformations}.
The candidate charge-conjugation maps therefore preserve Weyl representations for \(\delta = 0,4\bmod8\) and exchange them for \(\delta = 2,6\bmod8\), as summarised in \cref{tab:weyl-transformations-by-signature}.
Consequently, \(\delta = 0\) permits a \MW condition and \(\delta = 4\) permits a \sMW condition on an even multiplet, whereas the corresponding Majorana or \sM condition pairs opposite chiralities for \(\delta = 2,6\bmod8\).
In odd dimensions the volume element cannot act as a negation intertwiner \eqref{eq:negation intertwiner} and the two intertwiners \(\ConjMat^\pm\) belong to different irreducible equivalence classes, as indicated in \cref{tab:odd dimensions}.

In quantum theory, time reversal is represented by an antiunitary operator rather than by a purely linear action on the Hilbert space.
The linear Pin lift implementing a coordinate reflection \eqref{eq:field reflections} must therefore be distinguished from an antilinear map on spinor components.
In the unitary Clifford bases fixed by \eqref{eq:gamma hermiticity}, the transposition intertwiners \eqref{eq:intertwiner} define the candidate antilinear module maps
\begin{align}
\psi_\alT^\pm(x) & = \zeta_\alT^\pm (\TransMat^\pm)^{-1} \psi^\ast(\T x) , &
K_\alT^\pm\psi & = (\TransMat^\pm)^{-1}\psi^\ast .
\end{align}
Indeed, under a unitary change of Clifford basis, \(((\TransMat^\pm)^\prime)^{-1} = U^{-1}(\TransMat^\pm)^{-1}U^\ast\), which is the transformation law required for the matrix part of an antilinear module map.
For each existing \(\TransMat^\pm\), the transpose signature in \eqref{eq:intertwiner relation} fixes the square of this map:
\begin{equation}
(K_\alT^\pm)^2 = \xi_\TransMat^\pm(d) \ident .
\end{equation}
Either map can represent the spinor part of physical time reversal only when combined with the required coordinate reflection and when the complete theory is invariant.
In even dimensions the two definitions are related by the negation intertwiner.
Inverting the basis-covariant ordering in \eqref{eq:intertwiner negation} and absorbing the fixed phase relating \(\omega^{-1}\) to \(\chi\) into a redefinition of the conventional phase \(\zeta_\alT^-\) gives
\begin{align}
\psi_\alT^-(x) & = \zeta_\alT \chi \psi_\alT^+(x) , &
\psi_\alT^{\lambda-}(x) & = \lambda \zeta_\alT \psi_\alT^{\lambda+}(x) , &
\zeta_\alT & = \frac{\zeta_\alT^-}{\zeta_\alT^+\zeta_\TransMat} .
\end{align}
The two candidate maps consequently differ only by composition with the chirality operator and a conventional phase.
\footnote{
In four-dimensional Lorentzian signature these two choices give the Wigner and Schwinger time-reversal operations \cite{Wigner:1932,Schwinger:1951xk}.
The Wigner time reversal is the antiunitary time reversal from \QM, whose square controls Kramers-type degeneracies.
The Schwinger time reversal is natural in covariant field-theoretic transformation laws.
}
Since \(K_\alT^\pm\) is antilinear, it complex-conjugates the phase in the complex volume element, and therefore
\begin{align}
K_\alT^\pm \chi (K_\alT^\pm)^{-1} & = \eta_\rev^{}(d) \chi , &
K_\alT^\pm \Pi^\lambda (K_\alT^\pm)^{-1} & = \Pi^{\lambda^\prime} , &
\alT^\pm:S_\PT^\lambda &\to
S_\PT^{\lambda^\prime} , &
\lambda^\prime & = \eta_\rev^{}(d) \lambda ,
\end{align}
see also \cref{tab:weyl-discrete-transformations}.
The candidate antilinear maps \(K_\alT^\pm\) therefore preserve Weyl representations for \(d = 0,4\bmod8\) and exchange them for \(d = 2,6\bmod8\), as shown in \cref{tab:weyl-transformations-by-signature}.

\subsubsection{Composite discrete transformations}

Combining the candidate charge-conjugation map \eqref{eq:field charge conjugated} with collective spatial inversion \eqref{eq:field reflections} gives
\begin{align}
 \psi_\PC^\pm(x)
 & =
 \zeta_\Ch^\pm U_\P\ConjMat^\pm\psi^\ast(\P x) ,
&
 \PC^\pm
 & =
 \P\circ\Ch^\pm .
\end{align}
In even dimensions the two candidate parity-charge maps and the chirality-\(\lambda\) components are related by
\begin{align}
 \psi_\PC^-(x) &= \eta_\grade^{}(s) \zeta_\Ch \chi \psi_\PC^+(x) , &
 \psi_\PC^{\lambda-}(x) &= \lambda \eta_\grade^{}(s) \zeta_\Ch \psi_\PC^{\lambda+}(x).
\end{align}
Their action on chirality in even dimensions combines the chirality signatures of spatial inversion and charge conjugation,
\begin{align}
 \PC^\pm \Pi^\lambda (\PC^\pm)^{-1}
 & = \Pi^{\lambda^\prime} , &
 \lambda^\prime
 & = \eta_\rev^{}(d)\lambda .
\end{align}
Independently of the signature, the candidate parity-charge maps therefore preserve the two Weyl representations for \(d=0,4\bmod8\) and exchange them for \(d=2,6\bmod8\), as recorded in \cref{tab:weyl-discrete-transformations,tab:weyl-transformations-by-signature}.

Replacing spatial by temporal inversion gives the candidate temporal-charge maps
\begin{align}
 \psi_\TC^\pm(x)
 & =
 \zeta_\Ch^\pm U_\T\ConjMat^\pm\psi^\ast(\T x) ,
&
 \TC^\pm
 & =
 \T\circ\Ch^\pm .
\end{align}
In even dimensions the two candidate temporal-charge maps and the chirality-\(\lambda\) components are related by
\begin{align}
 \psi_\TC^-(x) &= \eta_\grade^{}(t) \zeta_\Ch \chi \psi_\TC^+(x) , &
 \psi_\TC^{\lambda-}(x) &= \lambda \eta_\grade^{}(t) \zeta_\Ch \psi_\TC^{\lambda+}(x).
\end{align}
Their action on chirality in even dimensions combines the chirality signatures of temporal inversion and charge conjugation,
\begin{align}
 \TC^\pm \Pi^\lambda (\TC^\pm)^{-1}
 & = \Pi^{\lambda^\prime} , &
 \lambda^\prime
 & = \eta_\rev^{}(d)\lambda .
\end{align}
Independently of the signature, the candidate temporal-charge maps likewise preserve the two Weyl representations for \(d=0,4\bmod8\) and exchange them for \(d=2,6\bmod8\), as recorded in \cref{tab:weyl-discrete-transformations,tab:weyl-transformations-by-signature}.

Combining charge conjugation with the inversion of all spatial and temporal coordinates gives the candidate parity-time-charge maps
\begin{align}
 \psi_\PTC^\pm(x)
 & =
 U_\PT\left[\zeta_\Ch^\pm\ConjMat^\pm\psi^\ast(\PT x)\right] , &
 \PTC^\pm
 & =
 \PT\circ\Ch^\pm .
\end{align}
In even dimensions, the lift of the full spacetime inversion is proportional to the volume element, \(U_\PT = \zeta_\PT\omega\), and therefore
\begin{equation}
 \psi_\PTC^\pm(x) = \zeta_\PT\zeta_\Ch^\pm\omega\ConjMat^\pm\psi^\ast(\PT x).
\end{equation}
The two candidate parity-time-charge maps and their chirality-\(\lambda\) components are then related by
\begin{align}
 \psi_\PTC^-(x) &= \zeta_\Ch\chi\psi_\PTC^+(x) , &
 \psi_\PTC^{\lambda-}(x) &= \lambda\zeta_\Ch\psi_\PTC^{\lambda+}(x).
\end{align}
Their action on chirality combines the chirality signatures of full spacetime inversion and charge conjugation,
\begin{align}
 \PTC^\pm \Pi^\lambda (\PTC^\pm)^{-1}
 & = \Pi^{\lambda^\prime} , &
 \lambda^\prime
 & = \eta_\rev^{}(\delta)\lambda .
\end{align}
The candidate parity-time-charge maps therefore preserve the two Weyl representations for \(\delta=0,4\bmod8\) and exchange them for \(\delta=2,6\bmod8\), as recorded in \cref{tab:weyl-discrete-transformations,tab:weyl-transformations-by-signature}.

The Clifford-algebraic statement concerns only the existence and intertwining behaviour of these candidate module maps \cite{Wetterich:2010ni,Stone:2020vva}.
It neither establishes that a displayed candidate acts as a symmetry of a given theory nor replaces the assumptions or conclusion of the Lorentzian \(\CPT\) theorem.

\subsection{Bilinears and Lagrangians} \label{sec:bilinears-lagrangians}

Whether a fermion bilinear can enter a Lagrangian depends jointly on \(\properSpin(s,t)\) covariance, Hermiticity, Majorana and Weyl constraints, Grassmann symmetry, and the internal representation content \cite{DeAndrade:1994mb,DeAndrade:1999xa,Wetterich:2010ni,Stone:2020vva}.
For example, \(\psi^\trans\TransMat^\beta\element^a\psi\) vanishes identically for anticommuting fields unless \(\TransMat^\beta\element^a\) has the required transpose symmetry.
A candidate term must pass six independent tests:
\begin{inlinelist}
\item its spinor indices must admit an invariant pairing
\item its intertwiners must be mutually compatible
\item it must survive the Grassmann and chiral-projector conditions
\item its free Clifford indices must be contracted covariantly
\item its internal indices must admit invariant gauge and flavour contractions compatible with any imposed reality condition
\item its coefficients must make the integrated action Hermitian after any required integration by parts
\end{inlinelist}
Failure of any one test excludes the term, although the physical interpretation of the algebraic conditions depends on the setting.
In Lorentzian signature with one time direction, the adjoint identities and coefficient relations below directly determine the Hermiticity of the action.
In a Euclidean functional integral, \(\psi\) and \(\widebar\psi\) are normally independent Grassmann variables, so reflection positivity replaces naive Hermiticity, while signatures with several time directions likewise admit no canonical positive-energy Hilbert-space interpretation.
In a conformal field theory the same local Clifford-module conditions apply, supplemented by conformal covariance of the complete operator, including its scaling dimension and tensor structure.

For \(g\in\properSpin(s,t)\) with \(N(g)=+\ident\), the defining intertwiner relations give the invariant-pairing identities
\begin{align}
g^\dagger A^\alpha g & = A^\alpha , &
\alpha&\in\{+,-\} , &
g^\trans \TransMat^\beta g & = \TransMat^\beta , &
\beta&\in\{+,-\} .
\end{align}
Every existing \(A^\pm\) or \(\TransMat^\pm\) consequently supplies a \(\properSpin(s,t)\)-invariant pairing \cite{Alekseevsky:1995,Gil-Garcia:2025iqt}.
The signature- and Clifford-module-dependent signs controlling intertwiner existence, Grassmann nonvanishing, and coefficient phases are encoded in \cref{tab:symmetry A,tab:symmetry trans conj}, while compatibility with the field content, tensor covariance, and internal invariant contractions remain separate conditions.
A Majorana-type constraint is available only when the tensor product of the spinor and internal representations admits an equivariant antilinear involution.
A real spinor structure can combine with a real internal structure, while a quaternionic spinor structure requires a pseudoreal internal pairing \(\Omega\) \cite{Gall:2021tiu}.
A genuinely complex internal representation instead requires the conjugate representation in the field content.
Every coefficient \(c_{i_1\cdots i_a}\) below accordingly denotes an antisymmetric covariant tensor field or background that contracts the contravariant Clifford indices; a constant non-scalar coefficient preserves \(\properSpin(s,t)\) only when it is an invariant tensor of the theory.

\subsubsection{Dirac fields}

For an anticommuting Dirac spinor or pinor, every existing \(A^\alpha\) defines a Dirac adjoint
\begin{align} \label{eq:Dirac adjoint}
\widebar\psi^\alpha(x) & = \psi^\dagger(x)A^\alpha , &
\alpha &\in\{+,-\} .
\end{align}
In even dimensions, when both adjoint intertwiners exist, the volume element relates them through \eqref{eq:intertwiner negation}.
Up to the conventional phase and rank-dependent signs, inserting the volume element therefore maps bilinears constructed with \(\widebar\psi^+\) to those constructed with \(\widebar\psi^-\), so the two choices form Hodge-dual families.
Under \(\properSpin(s,t)\), both rank-zero families are invariant, while their designation as scalar or pseudoscalar, and correspondingly as vector or axial vector at rank one, depends on the chosen action of the disconnected reflections.

The corresponding candidate bilinear density is
\begin{align} \label{eq:Dirac Lagrangian}
\mathcal L_D^\alpha & = \kappa^\alpha\widebar\psi^\alpha\gamma_\mu \partial^\mu \psi - \mu^\alpha\widebar\psi^\alpha\psi - \sum_{a = 1}^{d} c^\alpha_{i_1\cdots i_a}\widebar\psi^\alpha\gamma^{i_1\cdots i_a}\psi .
\end{align}
The adjoint bilinear signatures are defined in \eqref{eq:Aa sym} and listed in \cref{tab:symmetry A}.
If \(\xi_A^\alpha(s,t,a)=0\), the corresponding adjoint structure is absent on the module.
Individual Hermiticity of an existing non-derivative rank-\(a\) Dirac bilinear requires
\begin{equation} \label{eq:Hermiticity condition}
(c^\alpha_{i_1\cdots i_a})^\ast = \xi_A^\alpha(s,t,a)c^\alpha_{i_1\cdots i_a} .
\end{equation}
The scalar coefficient obeys the same condition at \(a=0\),
\begin{equation}
(\mu^\alpha)^\ast = \xi_A^\alpha(s,t,0)\mu^\alpha .
\end{equation}
A coefficient that fails the appropriate condition instead requires the operator to be accompanied by its Hermitian conjugate with the conjugate coefficient.
For a spacetime-independent kinetic coefficient, Hermitian conjugation followed by integration by parts introduces an additional minus sign:
\begin{equation} \label{eq:kinetic-Hermiticity}
(\kappa^\alpha)^\ast = -\xi_A^\alpha(s,t,1)\kappa^\alpha .
\end{equation}
An explicit factor \(\i\) is therefore appropriate when \(\xi_A^\alpha(s,t,1)=+1\), whereas a negative sign requires a real kinetic coefficient.

Although both adjoint families are retained in the bilinear classification, the signature-adapted Clifford representation distinguishes the ordered product \(A^\asym\) of the anti-Hermitian generators.
With the conventions of \eqref{eq:intertwiner candidates}, this is
\begin{align}
A^\asym &= A^\alpha, &
\alpha &= \eta_\grade^{}(t) .
\end{align}
This relation selects \(A^+\) for even \(t\) and \(A^-\) for odd \(t\).
This choice reproduces the standard Dirac adjoint in Lorentzian signature, reduces to the ordinary Hermitian pairing in Euclidean signature, and is precisely the adjoint that survives on each fixed irreducible complex Clifford module in odd dimensions.

For either choice, the associated rank-one bilinear is
\begin{equation}
j_i^\pm = \widebar\psi^\pm\gamma_i\psi = \psi^\dagger A^\pm\gamma_i\psi .
\end{equation}
In even-dimensional Lorentzian signature with \(t=1\), the unique temporal generator is \(\gamma_d\).
The unphased ordered products in \eqref{eq:intertwiner candidates} give \(A^\asym=A^-=\gamma_d\), whereas \(A^+\) is the product of all spatial generators.
Since \(\gamma_d^2=-\matident\), raising the temporal index changes the sign and gives
\begin{equation}
(j^-)^d = -j_d^- = \psi^\dagger\psi .
\end{equation}
The matrix defining this component is positive definite on the complex spinor module and therefore reproduces the standard positive density of a one-particle Dirac wavefunction.
By contrast, the temporal component associated with \(A^+\) satisfies, up to the fixed phase relating the volume element to \(\chi\),
\begin{equation}
(j^+)^d = -j_d^+ \propto \psi^\dagger\chi\psi = (\psi^+)^\dagger\psi^+ - (\psi^-)^\dagger\psi^- .
\end{equation}
The resulting chirality-weighted form is indefinite on a Dirac module containing both Weyl sectors.
For a spatial direction \(i_s\), the matrices \(A^-\gamma_{i_s}\) and \(A^+\gamma_{i_s}\) have Clifford degrees \(2\) and \(d-2\), respectively, and hence have the same degree in four dimensions.
In four-dimensional Lorentzian spacetime, the positive temporal component rather than Clifford degree therefore distinguishes the conventional Dirac adjoint.

In Euclidean signature, \(t=0\), the same prescription gives \(A^\asym=A^+=\matident\), since the ordered product over anti-Hermitian generators is empty.
The associated rank-one bilinear is
\begin{equation}
j_i^+ = \psi^\dagger\gamma_i\psi .
\end{equation}
Its matrix contains a single Hermitian Clifford generator.
When \(d>0\) is even, the complementary \(A^-\) is the product of all \(d\) generators, so \(A^-\gamma_i\) has Clifford degree \(d-1\).
The signature-adapted choice therefore also selects the lower-degree representative for \(d>2\), while the two degrees coincide for \(d=2\).

The Lorentzian and Euclidean cases therefore select the same signature-adapted adjoint \(A^\asym=A^{\eta_\grade^{}(t)}\), namely the ordered product over the anti-Hermitian Clifford generators.
In odd dimensions, the existence pattern in \cref{tab:existence} leaves precisely this adjoint on each fixed irreducible complex Clifford module.
The opposite sign can be realised only after the two grade-related sectors are combined and the intertwiner is allowed to exchange them.
For \(t>1\), the same prescription remains the natural algebraic continuation of the signature-adapted construction, although it no longer defines a preferred positive density because a spacetime with several time directions has no canonical positive-energy Hilbert-space interpretation.
Accordingly, an unlabelled Dirac adjoint denotes
\begin{align} \label{eq:preferred Dirac adjoint}
\widebar\psi(x) & = \psi^\dagger(x)A^\alpha , &
\alpha & = \eta_\grade^{}(t) .
\end{align}

\subsubsection{Majorana fields}

Because \(\ConjMat^\varsigma\) defines the Majorana condition, \(\TransMat^\beta\) pairs a Majorana self-bilinear, and \(A^\alpha\) tests its Hermiticity, these three independently defined intertwiners must form a compatible triple.
For an anticommuting Majorana (s)pinor satisfying \eqref{eq:real structure}, each existing \(\TransMat^\beta\) defines a candidate Majorana adjoint
\begin{align}
\psi^{\beta\ft}(x) & = \psi^\trans(x) \TransMat^\beta , &
\beta&\in\{+,-\} .
\end{align}
For the Majorana structure
\begin{align}
J_\Ch^\varsigma\psi &= \ConjMat^\varsigma\psi^\ast , &
\varsigma&\in\{+,-\} ,
\end{align}
the transpose and Dirac constructions represent the same adjoint precisely when the three intertwiners form a compatible triple,
\begin{align} \label{eq:compatible-ABC}
\psi^\trans \TransMat^\beta & = \zeta_M^{\alpha\varsigma}\psi^\dagger A^\alpha , &
\TransMat^\beta & = \zeta_M^{\alpha\varsigma} (\ConjMat^\varsigma)^\dagger A^\alpha , &
\zeta_M^{\alpha\varsigma} &\in\U(1) , &
\beta & = \alpha\varsigma .
\end{align}
The compatibility phase \(\zeta_M^{\alpha\varsigma}\) records the relative phase between the independently fixed intertwiner representatives without affecting their defining relations or signatures.
The candidate Majorana density is
\begin{align}
\mathcal L_M^\beta & = \frac{\kappa^\beta}2 \psi^{\beta\ft} \gamma_\mu \partial^\mu \psi
-\frac{\mu^\beta}2 \psi^{\beta\ft} \psi
- \frac12\sum_{a = 1}^{d} c^\beta_{i_1\cdots i_a}
\psi^{\beta\ft}
\gamma^{i_1\cdots i_a}
\psi .
\end{align}
The factor \(\flatfrac12\) avoids double counting for a real field.
The transpose symmetry in \eqref{eq:Ca sym} determines whether a self-bilinear vanishes; for a non-derivative term, the Grassmann selection rule requires
\begin{align} \label{eq:Grassmann condition}
\xi_\TransMat^\beta(d,a) & = -1 ,
\end{align}
so the matrix in the bilinear must be antisymmetric.
For a derivative term, integration by parts reverses this rule, so a nonzero kinetic term requires
\begin{align}
\xi_\TransMat^\beta(d,1) & = +1 .
\end{align}
Equivalently, the matrix multiplying the derivative must be symmetric.
The sign is defined in \eqref{eq:Ca sym}, with its periodicity displayed in \cref{tab:symmetry trans}.
Using \eqref{eq:compatible-ABC}, Hermiticity of the integrated Majorana action independently requires
\begin{align} \label{eq:Majorana coefficient Hermiticity}
(\kappa^\beta)^\ast & = -\xi_A^\alpha(s,t,1) (\zeta_M^{\alpha\varsigma})^2 \kappa^\beta , &
(\mu^\beta)^\ast & = \xi_A^\alpha(s,t,0) (\zeta_M^{\alpha\varsigma})^2 \mu^\beta , &
(c^\beta_{i_1\cdots i_a})^\ast & = \xi_A^\alpha(s,t,a) (\zeta_M^{\alpha\varsigma})^2 c^\beta_{i_1\cdots i_a} .
\end{align}
The factors \((\zeta_M^{\alpha\varsigma})^2\) follow from \((\zeta_M^{\alpha\varsigma})^\ast=(\zeta_M^{\alpha\varsigma})^{-1}\) and do not enter the independent Grassmann conditions above.

\subsubsection{\sMlong fields}

For an anticommuting \sMfirst (s)pinor satisfying \eqref{eq:quaternionic structure}, each existing \(\TransMat^\beta\) defines a candidate \sM adjoint
\begin{align}
\psi_A^{\beta\ft} & = \psi_A^\trans \TransMat^\beta , &
\beta&\in\{+,-\} .
\end{align}
The compatible adjoint obeys \eqref{eq:compatible-ABC} for the \(\ConjMat^\varsigma\) entering the \sM condition.
The candidate \sM density is
\begin{align}
\mathcal L_\text{\sM}^\beta & = \frac{\kappa^\beta}2 \psi_A^{\beta\ft}
\gamma_\mu
\Omega_{AB}
\partial^\mu \psi_B
-\frac{\mu^\beta}2 \psi_A^{\beta\ft}
\Omega_{AB}
\psi_B
- \frac12\sum_{a = 1}^{d} c^\beta_{i_1\cdots i_a}
\psi_A^{\beta\ft}
\gamma^{i_1\cdots i_a}
\Omega_{AB}
\psi_B .
\end{align}
Because the symplectic matrix in \eqref{eq:symplectic matrix} is antisymmetric, it reverses the Majorana selection sign, so a non-derivative \sM self-bilinear can be nonzero only if
\begin{align} \label{eq:sM constraint}
\xi_\TransMat^\beta(d,a) & = +1 ,
\end{align}
so the spinorial matrix must be symmetric.
For the kinetic term, integration by parts reverses the derivative selection rule, so a nonzero \sM kinetic term requires
\begin{equation}
\xi_\TransMat^\beta(d,1) = -1 .
\end{equation}
The coefficients obey the Hermiticity conditions \eqref{eq:Majorana coefficient Hermiticity} for the compatible triple with \(\alpha = \beta\varsigma\), while the antisymmetric internal form changes only the independent Grassmann conditions above.

\subsubsection{Weyl spinors}

A Weyl spinor \eqref{eq:chiral projectors} exists only in even dimensions, and its projection can eliminate an otherwise available Dirac bilinear by placing the two spinors in orthogonal chiral subspaces.
The chirality projection is
\begin{align}
\psi^\lambda & = \Pi^\lambda\psi , &
\lambda = \pm1 .
\end{align}
The corresponding Dirac adjoint is
\begin{align}
\widebar\psi^{\lambda\alpha} & = \widebar\psi^\alpha \Pi^{\lambda^\prime} , &
\Pi^{\lambda^\prime} & = (A^\alpha)^{-1} \Pi^\lambda A^\alpha , &
\lambda^\prime & = \lambda \begin{cases}
\eta_\grade^{}(t) & \text{for } \alpha = +,\\
\eta_\grade^{}(s) & \text{for } \alpha = - .
\end{cases}
\end{align}
Projecting the Dirac density \eqref{eq:Dirac Lagrangian} with \eqref{eq:chiral projectors} gives the candidate single-Weyl density
\begin{align} \label{eq:Weyl Lagrangian}
\mathcal L_W^{\lambda\alpha} & = \kappa^{\lambda\alpha}
\widebar\psi^{\lambda\alpha}
\gamma_\mu \partial^\mu \psi^\lambda
-
\mu^{\lambda\alpha}\widebar\psi^{\lambda\alpha} \psi^\lambda
-
\sum_{a = 1}^{d}
c^{\lambda\alpha}_{i_1\cdots i_a}
\widebar\psi^{\lambda\alpha}
\gamma^{i_1\cdots i_a}
\psi^\lambda .
\end{align}
A homogeneous Clifford element of degree \(a\) maps chirality as
\begin{align}
\element^a \Pi^\lambda & = \Pi^{\lambda^{\prime\prime}}\element^a , &
\lambda^{\prime\prime} & = \eta_\grade^{}(a) \lambda .
\end{align}
A rank-\(a\) Clifford bilinear therefore survives only when the chirality selected by the adjoint agrees with that produced by the Clifford element:
\begin{align} \label{eq:Weyl constraint}
\lambda^\prime & = \lambda^{\prime\prime} , &
\eta_\grade^{}(a) & = \begin{cases}
\eta_\grade^{}(t) & \text{for } \alpha = +,\\
\eta_\grade^{}(s) & \text{for } \alpha = - .
\end{cases}
\end{align}
The resulting periodicity is displayed in \cref{tab:weyl-adjoint-projector-terms}.
The Weyl kinetic term survives the projector only if
\begin{align}
\gamma_\mu \Pi^\lambda & = \Pi^{-\lambda}\gamma_\mu , &
-1 & = \begin{cases}
\eta_\grade^{}(t) & \text{for } \alpha = +,\\
\eta_\grade^{}(s) & \text{for } \alpha = - .
\end{cases}
\end{align}
The mass term for a single Weyl spinor survives the projector only if
\begin{align}
1 & = \begin{cases}
\eta_\grade^{}(t) & \text{for } \alpha = +,\\
\eta_\grade^{}(s) & \text{for } \alpha = - .
\end{cases}
\end{align}
For a fixed adjoint intertwiner, the single-Weyl kinetic and mass terms therefore require opposite signatures.

\subsubsection{Weyl spinor pairs}

\begin{table}
\begin{tabular}{cccc}
\toprule
\(d\bmod4\) & \(\delta\bmod4\) & \multicolumn2c{required \(\eta_\grade^{}(a)\)} \\ \cmidrule{3-4}
& & diagonal & off-diagonal \\
\midrule
\multirow2*{\(0\)} & \(0\) & \(+\) & \(-\) \\
& \(2\) & \(-\) & \(+\) \\ \cmidrule{2-4}
\multirow2*{\(2\)} & \(0\) & \(-\) & \(+\) \\
& \(2\) & \(+\) & \(-\) \\
\bottomrule \end{tabular}
\caption[Projector conditions for Dirac-adjoint Weyl bilinears]{
Projector conditions for diagonal and off-diagonal Dirac-adjoint Weyl bilinears.
The required value of \(\eta_\grade^{}(a)\) follows from the two Weyl constraints \cref{eq:Weyl constraint,eq:Weyl pair constraint}.
The diagonal column refers to single-Weyl terms, while the off-diagonal column refers to Weyl-pair terms.
The mass term has \(a = 0\), and the kinetic term has \(a = 1\).
Only the projector test is encoded in the table.
} \label{tab:weyl-adjoint-projector-terms}
\end{table}

For a pair of opposite Weyl modules, applying both chiral projectors to the Dirac Lagrangian \eqref{eq:Dirac Lagrangian} retains both terms diagonal in chirality and off-diagonal terms coupling the two modules.
The resulting density is
\begin{multline}
\mathcal L_{2W}^{\alpha} = \mathcal L_W^{+\alpha}
+
\mathcal L_W^{-\alpha}
+
\kappa^\pm\widebar\psi^{+\alpha}\gamma_\mu \partial^\mu \psi^-
-
\mu^\pm\widebar\psi^{+\alpha}\psi^-
+
\kappa^\mp\widebar\psi^{-\alpha}\gamma_\mu \partial^\mu \psi^+
-
\mu^\mp\widebar\psi^{-\alpha}\psi^+ \\
-
\sum_{a = 1}^{d}
\left(
c^\pm_{i_1\cdots i_a}
\widebar\psi^{+\alpha}
\gamma^{i_1\cdots i_a}
\psi^-
+
c^\mp_{i_1\cdots i_a}
\widebar\psi^{-\alpha}
\gamma^{i_1\cdots i_a}
\psi^+
\right).
\end{multline}
For the off-diagonal rank-\(a\) bilinear \(\widebar\psi^{\lambda\alpha}\element^a\psi^{-\lambda}\), the adjoint selects
\begin{equation}
\lambda^\prime = \lambda \begin{cases}
\eta_\grade^{}(t) & \text{for } \alpha = +,\\
\eta_\grade^{}(s) & \text{for } \alpha = - .
\end{cases}
\end{equation}
Acting on \(\psi^{-\lambda}\), the Clifford element instead produces
\begin{align}
\element^a \Pi^{-\lambda} & = \Pi^{\lambda^{\prime\prime}} \element^a , &
\lambda^{\prime\prime} & = -\eta_\grade^{}(a) \lambda .
\end{align}
Equating the selected chiralities gives the nonzero off-diagonal projector condition
\begin{align} \label{eq:Weyl pair constraint}
\lambda^\prime & = \lambda^{\prime\prime} , &
\eta_\grade^{}(a) & = -\begin{cases}
\eta_\grade^{}(t) & \text{for } \alpha = +,\\
\eta_\grade^{}(s) & \text{for } \alpha = - .
\end{cases}
\end{align}
The off-diagonal constraint therefore differs from the diagonal one by a single chirality sign.
The off-diagonal kinetic term has \(a = 1\), since \(\gamma_\mu\) has odd Clifford degree.
It survives the projector only if
\begin{align}
1 & = \begin{cases}
\eta_\grade^{}(t) & \text{for } \alpha = +,\\
\eta_\grade^{}(s) & \text{for } \alpha = - .
\end{cases}
\end{align}
The off-diagonal scalar Dirac mass survives the projector only if
\begin{align}
-1 & = \begin{cases}
\eta_\grade^{}(t) & \text{for } \alpha = +,\\
\eta_\grade^{}(s) & \text{for } \alpha = - .
\end{cases}
\end{align}
The off-diagonal kinetic term and scalar mass therefore require opposite adjoint chirality signs: a chirality-reversing adjoint retains the mass, whereas a chirality-preserving adjoint retains the kinetic term.
For every retained pair of off-diagonal terms, Hermiticity relates the two coefficients rather than constraining either term separately:
\begin{align} \label{eq:off-diagonal-Hermiticity}
\kappa^\mp & = -\xi_A^\alpha(s,t,1) (\kappa^\pm)^\ast , &
\mu^\mp & = \xi_A^\alpha(s,t,0) (\mu^\pm)^\ast , &
c^\mp_{i_1\cdots i_a} & = \xi_A^\alpha(s,t,a)(c^\pm_{i_1\cdots i_a})^\ast .
\end{align}
Together with the diagonal single-Weyl terms, the two-chirality projection supplies one candidate kinetic term and one candidate mass term for either adjoint chirality sign, subject to the remaining tests.

\subsubsection{\MWlong spinors}

\begin{table}
\begin{tabular}{ccccl} \toprule
\(d\bmod8\) & \(\eta_\rev^{}(d)\) & \multicolumn2c{non-derivative degree \(a\bmod4\)} & kinetic term \\ \cmidrule{3-4}
& & \MW & \sMW & \\ \midrule
\(0\) & \(+\) & \(2\) & \(0\) & none \\
\(2\) & \(-\) & \(3\) & \(1\) & \MW \\
\(4\) & \(+\) & \(0\) & \(2\) & none \\
\(6\) & \(-\) & \(1\) & \(3\) & \sMW \\
\bottomrule \end{tabular}
\caption[Projector and Grassmann tests for chiral Majorana bilinears]{
Projector and Grassmann nonvanishing tests for \MW and \sMW bilinears.
The Clifford degrees for which a non-derivative \MW or \sMW self-bilinear passes the chiral-projector condition \eqref{eq:MW constraint} and the respective Grassmann conditions in \cref{eq:Grassmann condition,eq:sM constraint} appear in the corresponding columns.
The kinetic term is displayed separately because the derivative and integration by parts reverse the Grassmann sign.
Only the projector and Grassmann nonvanishing tests are encoded in the table.
} \label{tab:majorana-weyl-lagrangian-terms}
\end{table}

\resetacronym{MW}

A \MW spinor obeys both the Weyl projection \eqref{eq:chiral projectors} and the real Majorana condition \eqref{eq:real structure}, which are compatible only for \(\delta = 0\bmod8\).
Although this compatibility establishes the field type, the projector and Grassmann tests still determine which self-bilinears survive.
The \MW adjoint is
\begin{align}
\psi^{\lambda\beta\ft} & = \psi^{\lambda\trans} \TransMat^\beta , &
\lambda & = \pm1 , &
\beta&\in\set{+,-} .
\end{align}
The candidate \MW terms are
\begin{align}
\mathcal L_\text{MW}^{\lambda\beta} & = \frac{\kappa^{\lambda\beta}}2
\psi^{\lambda\beta\ft}
\gamma_\mu \partial^\mu \psi^\lambda
-
\frac{\mu^{\lambda\beta}}2
\psi^{\lambda\beta\ft}\psi^\lambda
-
\frac12\sum_{a = 1}^{d}
c^{\lambda\beta}_{i_1\cdots i_a}
\psi^{\lambda\beta\ft}
\gamma^{i_1\cdots i_a}
\psi^\lambda .
\end{align}
The transpose pairing selected by the transposition intertwiner defines a chirality
\begin{align}
(\Pi^\lambda)^\trans \TransMat^\beta & = \TransMat^\beta \Pi^{\lambda^\prime} , &
\lambda^\prime & = \eta_\rev^{}(d) \lambda .
\end{align}
A homogeneous Clifford element of degree \(a\) maps chirality according to
\begin{align}
\element^a \Pi^\lambda & = \Pi^{\lambda^{\prime\prime}}\element^a , &
\lambda^{\prime\prime} & = \eta_\grade^{}(a) \lambda .
\end{align}
A same-chirality \MW bilinear therefore survives the projector only when the transpose pairing and the Clifford element select the same chirality:
\begin{align} \label{eq:MW constraint}
\lambda^\prime & = \lambda^{\prime\prime} , &
\eta_\rev^{}(d) & = \eta_\grade^{}(a) .
\end{align}
The corresponding periodicity is displayed in \cref{tab:majorana-weyl-lagrangian-terms}.
If this chiral projector condition holds, the term must also satisfy the Majorana Grassmann symmetry condition \eqref{eq:Grassmann condition}.
The \MW scalar mass survives the projector and Grassmann tests only if
\begin{align} \label{eq:MW condition}
\eta_\rev^{}(d) & = \eta_\grade^{}(0) = +1 , &
\xi_\TransMat^\beta(d,0) & = -1 .
\end{align}
The \MW kinetic term survives the projector and Grassmann tests only if
\begin{align}
\eta_\rev^{}(d) & = \eta_\grade^{}(1) = -1 , &
\xi_\TransMat^\beta(d,1) & = +1 .
\end{align}
The sign in the kinetic term is shifted relative to the non-derivative Grassmann rule by the derivative and integration by parts.
For each chirality \(\lambda\), the coefficients in \(\mathcal L_\text{MW}^{\lambda\beta}\) obey \eqref{eq:Majorana coefficient Hermiticity} with the corresponding \(\lambda\) index and \(\alpha = \beta\varsigma\), independently of the chiral-projector and Grassmann tests above.

\subsubsection{\sMWlong spinors}

A \sMWfirst spinor obeys both the Weyl projection \eqref{eq:chiral projectors} and the \sM condition \eqref{eq:quaternionic structure}, which are compatible only for \(\delta = 4\bmod8\).
As in the \MW case, the existence of the constrained field does not by itself ensure that its kinetic, mass, or higher-rank self-bilinears are nonzero.
The \sMW spinor and its adjoint are
\begin{align}
\psi^\lambda_A & = \Pi^\lambda\psi_A , &
\psi_A^{\lambda\beta\ft} & = \psi_A^{\lambda\trans} \TransMat^\beta , &
\lambda & = \pm1 , &
\beta&\in\set{+,-}
.
\end{align}
The candidate \sMW terms are
\begin{align}
\mathcal L_\text{\sMW}^{\lambda\beta} & = \frac{\kappa^{\lambda\beta}}2
\psi_A^{\lambda\beta\ft}
\gamma_\mu
\Omega_{AB}
\partial^\mu \psi^\lambda_B
-
\frac{\mu^{\lambda\beta}}2
\psi_A^{\lambda\beta\ft}
\Omega_{AB}
\psi^\lambda_B
-
\frac12\sum_{a = 1}^{d}
c^{\lambda\beta}_{i_1\cdots i_a}
\psi_A^{\lambda\beta\ft}
\gamma^{i_1\cdots i_a}
\Omega_{AB}
\psi^\lambda_B .
\end{align}
A same-chirality \sMW bilinear survives the projector only when the chirality selected by the transpose pairing defined by \(\TransMat^\beta\) agrees with the chirality produced by the Clifford element, as in \eqref{eq:MW constraint}.
If this chiral projector condition holds, the non-derivative terms must also satisfy the \sM Grassmann symmetry condition \eqref{eq:sM constraint}.
The \sMW scalar mass survives the projector and Grassmann tests only if
\begin{align} \label{eq:sM-Weyl condition}
\eta_\rev^{}(d) & = \eta_\grade^{}(0) = +1 , &
\xi_\TransMat^\beta(d,0) & = +1 .
\end{align}
The \sMW kinetic term survives the projector and Grassmann tests only if
\begin{align}
\eta_\rev^{}(d) & = \eta_\grade^{}(1) = -1 , &
\xi_\TransMat^\beta(d,1) & = -1 .
\end{align}
The sign in the kinetic term is shifted relative to the non-derivative Grassmann rule by the derivative and integration by parts, while the antisymmetric \(\Omega_{AB}\) reverses the \MW transpose condition.
For each chirality \(\lambda\), the coefficients in \(\mathcal L_\text{\sMW}^{\lambda\beta}\) obey \eqref{eq:Majorana coefficient Hermiticity} with the corresponding \(\lambda\) index and \(\alpha = \beta\varsigma\), independently of the symplectic, chiral-projector, and Grassmann tests above.

\section{Field theories} \label{sec:bott-class-field-content}

\begin{table}
\begin{panels}{.6}
\begin{tabular}{c*3{c@{ }c}} \toprule
\(\delta\bmod8\) & \multicolumn2c{Pin reality} & \multicolumn2c{Spin reality} & \multicolumn2c{Weyl structure} \\
\cmidrule(r){2-3} \cmidrule(lr){4-5} \cmidrule(l){6-7}
& \(\cl(s,t)\) & \((J_\Ch^+)^2\) & \(\cl(t,s)\) & \((J_\Ch^-)^2\) & \(\cl_\PT\) & \(J_\Ch^\pm(S_\PT^\lambda)\) \\ \midrule
\(0\) & \(\mathbb R\) & \(+\) & \(\mathbb R\) & \(+\) & \(2\mathbb R\) & \(S_\PT^\lambda\) \\
\(1\) & \(2\mathbb R\) & \(+\) & \(\mathbb C\) & \(0\) & \(\mathbb R\) & -- \\
\(2\) & \(\mathbb R\) & \(+\) & \(\mathbb H\) & \(-\) & \(\mathbb C\) & \(S_\PT^{-\lambda}\) \\
\(3\) & \(\mathbb C\) & \(0\) & \(2\mathbb H\) & \(-\) & \(\mathbb H\) & -- \\
\(4\) & \(\mathbb H\) & \(-\) & \(\mathbb H\) & \(-\) & \(2\mathbb H\) & \(S_\PT^\lambda\) \\
\(5\) & \(2\mathbb H\) & \(-\) & \(\mathbb C\) & \(0\) & \(\mathbb H\) & -- \\
\(6\) & \(\mathbb H\) & \(-\) & \(\mathbb R\) & \(+\) & \(\mathbb C\) & \(S_\PT^{-\lambda}\) \\
\(7\) & \(\mathbb C\) & \(0\) & \(2\mathbb R\) & \(+\) & \(\mathbb R\) & -- \\
\bottomrule \end{tabular}
\caption{Bott-class module types and conjugation structures} \label{tab:Bott-class module types}
\panel{.4}
\begin{tabular}{c*3c} \toprule
\(d\bmod8\) & \(\xi_\TransMat^+(d,a)\) & \multicolumn2c{\(\xi_\TransMat^-(d,a)\)} \\ \cmidrule(r){2-2} \cmidrule(l){3-4}
& \(a=0,1\) & \(a=0\) & \(a=1\) \\ \midrule
\(0\) & \(+\) & \(+\) & \(-\) \\
\(1\) & \(+\) & \(0\) & \(0\) \\
\(2\) & \(+\) & \(-\) & \(+\) \\
\(3\) & \(0\) & \(-\) & \(+\) \\
\(4\) & \(-\) & \(-\) & \(+\) \\
\(5\) & \(-\) & \(0\) & \(0\) \\
\(6\) & \(-\) & \(+\) & \(-\) \\
\(7\) & \(0\) & \(+\) & \(-\) \\
\bottomrule \end{tabular}
\caption{Transposition signatures} \label{tab:transposition signatures}
\end{panels}
\caption[Factorised classification of the thirty-two local Clifford periodicity sectors]{
Factorised Clifford-module types, conjugation structures, and transposition signatures for the \(32\) admissible periodicity sectors \((d,\delta)\), where both labels are understood modulo eight and have the same parity.
The matrix-size-independent types of the full, signature-reversed, and even real Clifford algebras, together with the squares of the induced conjugation maps, appear in panel \subref{tab:Bott-class module types}.
The symbols \(\mathbb R\), \(\mathbb C\), and \(\mathbb H\) denote real, complex, and quaternionic irreducible-module types, while a prefactor \(2\) denotes two simple summands and hence two inequivalent irreducible modules.
For even \(\delta\), the common image of a Weyl module under \(J_\Ch^+\) and \(J_\Ch^-\) appears in the final column; the dash marks the odd-dimensional classes, in which no Weyl decomposition exists.
The rank-zero and rank-one transposition signatures entering scalar and kinetic bilinears appear in panel \subref{tab:transposition signatures}, with zero denoting an absent intertwiner.
For a specified pair \((d,\delta)\), the congruence classes \(s = (d+\delta)/2\) and \(t = (d-\delta)/2\) modulo four determine the adjoint-intertwiner and collective-reflection signs through \cref{tab:symmetry A,eq:collective-reflection-signs}.
Their use in the Grassmann tests is explained in \cref{sec:bilinears-lagrangians}, and the \((d,\delta) = (4,2)\) sector is evaluated explicitly in \cref{sec:example-delta-two}.
} \label{tab:master-32}
\end{table}

\resetacronym{sM}
\resetacronym{MW}
\resetacronym{sMW}

Two periodic labels determine the local (s)pinor classification: the total dimension \(d = s+t\) and the Bott class \(\delta = s-t\), both taken modulo eight.
Since they have the same parity, real signatures realise only \(32\) of the \(64\) formal pairs.
The Bott class fixes the real Clifford-module types and conjugation signatures, while the total dimension fixes the independent transposition and chirality signs.
Combining the two entries in \cref{tab:master-32} with the corresponding adjoint and reflection signs specifies one matrix-size-independent periodicity sector.
In practice, the row labelled by \(\delta\) in \cref{tab:Bott-class module types} fixes the module types and available reality structures, while the row labelled by \(d\) in \cref{tab:transposition signatures} fixes the scalar and kinetic transpose signs.
Only the adjoint and collective-reflection signs require the separate congruence classes of \(s\) and \(t\), which follow from \(s = (d+\delta)/2\) and \(t = (d-\delta)/2\).
Representatives related by a change of spinor basis, conventional intertwiner phases, or Bott-periodic matrix-size factors belong to the same sector.
The resulting sectors classify local Clifford modules and intertwiners rather than complete field theories, which also depend on internal representations, flavour multiplicities, couplings, anomalies, dynamics, and global Spin or Pin structures.

The full Clifford algebra \(\cl(s,t)\) controls the Pin-level pinor module and its extension to spacetime reflections, whereas the even algebra \(\cl_\PT(s,t)\) controls the connected Spin-level spinor module and its Weyl decomposition in even dimension.
The labels real, complex, and quaternionic identify the division algebra over which the relevant matrix block is defined.
As defined in \cref{sec:spinor-pinor-classification}, a Majorana condition requires an equivariant real structure, while a \sM condition combines an equivariant quaternionic structure with the antisymmetric internal form \(\Omega_{AB}\).
In even dimension, preservation of each Weyl space permits a \MW or \sMW condition, whereas an antilinear map that exchanges the Weyl spaces imposes a condition only on their direct sum.
In odd dimension the central volume element labels the irreducible complex Clifford sectors, so a fixed sector contains only one member of each intertwiner pair.
When the real Clifford algebra splits into two simple blocks, the volume-element eigenvalue also distinguishes the two inequivalent real Pin modules.
In even dimension an odd Clifford lift of a reflection exchanges the Weyl spaces, while an even product of reflections preserves them, so a Pin action on the same chiral field space requires both Weyl modules.

Although the classification establishes whether the required field representation and invariant pairing can exist, an action term is allowed only after the compatibility, Grassmann, chiral-projector, covariance, internal-contraction, and Hermiticity requirements in \cref{sec:bilinears-lagrangians} have also been imposed.

\subsection{Explicit four-dimensional representatives} \label{sec:explicit-four-dimensional-representatives}

\begin{table}
\begin{tabular}{*{14}c} \toprule
\multicolumn2c{signature} & \multicolumn6c{Clifford intertwiners} & \multicolumn2c{reality structure} & \multicolumn4c{Pin lifts} \\ \cmidrule(r){1-2} \cmidrule(lr){3-8} \cmidrule(lr){9-10} \cmidrule(l){11-14}
& & \multicolumn2c{adjoint} & \multicolumn2c{transposition} & \multicolumn2c{conjugation} & Pin & Spin & \multicolumn3c{squares} & \(\comm\Plift\Tlift\) \\ \cmidrule(lr){3-4} \cmidrule(lr){5-6} \cmidrule(lr){7-8} \cmidrule(lr){9-9} \cmidrule(lr){10-10} \cmidrule(lr){11-13} \cmidrule(l){14-14}
\(s\) & \(t\) & \(A^+\) & \(A^-\) & \(\TransMat^+\) & \(\TransMat^-\) & \(\ConjMat^+\) & \(\ConjMat^-\) & \((J_\Ch^+)^2\) & \((J_\Ch^-)^2\) & \(\Plift^2\) & \(\Tlift^2\) & \((\PTlift)^2\) & \(\eta_\grade^{}(st)\) \\ \midrule
\(4\) & \(0\) & \(\matident\) & \(\omega\) & \(\gamma_{23}\) & \(\gamma_{14}\) & \(\gamma_{23}\) & \(\gamma_{14}\) & \(-\) & \(-\) & \(+\) & \(+\) & \(+\) & \(+\) \\
\(3\) & \(1\) & \(\gamma_{123}\) & \(\gamma_4\) & \(\gamma_{123}\) & \(\gamma_4\) & \(\matident\) & \(\omega\) & \(+\) & \(-\) & \(-\) & \(-\) & \(-\) & \(-\) \\
\(2\) & \(2\) & \(\gamma_{34}\) & \(\gamma_{12}\) & \(\gamma_{34}\) & \(\gamma_{12}\) & \(\matident\) & \(\omega\) & \(+\) & \(+\) & \(-\) & \(-\) & \(+\) & \(+\) \\
\(1\) & \(3\) & \(\gamma_1\) & \(\gamma_{234}\) & \(\gamma_{124}\) & \(\gamma_3\) & \(\gamma_{24}\) & \(\gamma_{13}\) & \(-\) & \(+\) & \(+\) & \(+\) & \(-\) & \(-\) \\
\(0\) & \(4\) & \(\omega\) & \(\matident\) & \(\gamma_{14}\) & \(\gamma_{23}\) & \(\gamma_{23}\) & \(\gamma_{14}\) & \(-\) & \(-\) & \(+\) & \(+\) & \(+\) & \(+\) \\
\bottomrule \end{tabular}
\caption[Four-dimensional intertwiner representatives and reflection signs]{
Explicit intertwiner representatives and collective-reflection signs for all five real signatures in four dimensions.
The \((3,1)\) and \((2,2)\) rows use four-by-four real matrices, whereas the \((4,0)\), \((1,3)\), and \((0,4)\) rows use two-by-two quaternionic matrices represented by four-by-four complex matrices as specified in \cref{fn:quaternions}.
The Clifford-generator representatives are defined in \cref{eq:gamma-40-quaternionic,eq:gamma-04-quaternionic,eq:gamma-31,eq:gamma-13-quaternionic,eq:gamma-22} for signatures \((4,0)\), \((0,4)\), \((3,1)\), \((1,3)\), and \((2,2)\), respectively.
Within each row, \(\gamma_{i_1\cdots i_a} = \gamma_{i_1}\cdots\gamma_{i_a}\) is the ordered product in that basis, and \(\omega = \gamma_1\gamma_2\gamma_3\gamma_4\).
For the quaternionic bases, complex conjugation, transposition, and Hermitian conjugation act on the complex representatives defined in \cref{fn:quaternions}.
The displayed intertwiners are the fixed unphased ordered Clifford products of \cref{eq:intertwiner candidates} with unitary normalisation.
Relative compatibility phases between the independently constructed representatives are retained explicitly in the corresponding relations.
The tuple \((\Plift^2,\Tlift^2,(\PTlift)^2,\eta_\grade^{}(st))\) occupies the final four columns, where the last entry is the relative commutation sign in \eqref{eq:collective-reflection-signs}.
In either definite-signature row, the collective lift associated with the absent coordinate sector is the identity, so its positive square is trivial rather than an additional reflection invariant.
} \label{tab:four-dimensional-intertwiner-representatives}
\end{table}

The five real signatures in four dimensions provide explicit Clifford representations for the four even Bott classes realised at \(d = 4\).
For the signatures \((4,0)\), \((1,3)\), and \((0,4)\), the ordered pairs \((\xi_A^\alpha(s,t,0),\xi_A^\alpha(s,t,1))\) are \((+,+)\) for \(A^+\) and \((+,-)\) for \(A^-\).
For \((2,2)\), they are \((-,-)\) for \(A^+\) and \((-,+)\) for \(A^-\), while the \((3,1)\) values are evaluated explicitly in the Lorentzian example in \cref{sec:example-delta-two}.
The two definite signatures \((4,0)\) and \((0,4)\) belong to the same periodicity sector because \(\delta = 4\bmod8\) in both cases, although the identity and volume element exchange their roles within the respective adjoint pairs.
The representatives in \cref{tab:four-dimensional-intertwiner-representatives} satisfy the basis-independent definitions \eqref{eq:intertwiner} and give the displayed \(\ConjMat\)-squares, while transposition gives \((\TransMat^+)^\trans = -\TransMat^+\) and \((\TransMat^-)^\trans = -\TransMat^-\) in every row.
The common rank-zero and rank-one transposition signatures are therefore
\begin{equation} \label{eq:four-dimensional-C-signatures}
\xi_\TransMat^+(4,0) = \xi_\TransMat^+(4,1) = \xi_\TransMat^-(4,0) = -\xi_\TransMat^-(4,1) = -1 .
\end{equation}
A scalar \sM self-bilinear fails this test for both intertwiners, while its kinetic counterpart requires \(\TransMat^+\).
For \((4,0)\), \((1,3)\), and \((0,4)\), Dirac-adjoint scalar coefficients are real for both fixed adjoints, while the \(A^+\) and \(A^-\) Dirac-adjoint kinetic coefficients are imaginary and real, respectively.
For \((2,2)\), Dirac-adjoint scalar coefficients are imaginary for both fixed adjoints, while the \(A^+\) and \(A^-\) Dirac-adjoint kinetic coefficients are real and imaginary, respectively.
These compensating phases arise from the fixed unphased representatives and do not change the underlying Dirac-adjoint bilinears or their physics.
For Majorana-type transpose bilinears, the corresponding coefficient phases additionally depend on the compatibility phase \(\zeta_M^{\alpha\varsigma}\) through \eqref{eq:Majorana coefficient Hermiticity}.
The signs in the \((J_\Ch^+)^2\) column identify the Pin-equivariant structure as real in signatures \((3,1)\) and \((2,2)\), but quaternionic in \((4,0)\), \((1,3)\), and \((0,4)\).
The signs in the \((J_\Ch^-)^2\) column identify the Spin-equivariant structure as real in \((2,2)\) and \((1,3)\), and quaternionic in the remaining signatures; in particular, the Majorana structure for the mostly-minus signature \((1,3)\) is defined by the Spin-equivariant map \(J_\Ch^-\).

\subsection{Lorentzian field theories} \label{sec:common-lorentzian-structure}

In the mostly-plus Lorentzian convention \((s,t) = (d-1,1)\), the relation \(\delta = d-2\bmod8\) selects eight of the \(32\) arbitrary-signature sectors \cite{VanProeyen:1999ni,DAuria:2000byu}.
As the spacetime dimension changes, this eightfold sequence fixes the real form of the spinor and pinor modules, the compatibility of chirality with Majorana-type conditions, and the signs entering fermion bilinears.
The examples below translate each algebra type in this sequence into familiar Lorentzian field terminology.

An equivariant real condition produces a Majorana field, whereas an equivariant quaternionic structure produces a \sM field only after it is combined with an even multiplet carrying \(\Omega_{AB}\).
An unconstrained complex field is called a Dirac spinor or pinor according to whether only the connected Spin action or a full Pin action is specified.
Local Lorentz covariance is controlled by \(\properSpin(d-1,1)\), whose complexified spinor representation splits into two Weyl representations for even \(d\), while no Weyl decomposition exists for odd \(d\).
The distinction between Dirac and Weyl spinors consequently concerns the connected Spin representation rather than the full Pin group.

Representing the disconnected components of the full Lorentz group requires Pin-level lifts of spatial and temporal reflections, which act on spinor indices as spacetime transformations rather than as additional internal symmetries.
Since the two lifts of a fixed reflection differ by the central sign, they have the same square and the same action on chirality, although inequivalent real Pin extensions can restrict to the same complexified Spin representation.

In even Lorentzian dimension the Clifford lifts \(\plift\) and \(\tlift\) of the elementary reflections have odd degree and therefore exchange the Weyl spaces, whereas their product has even degree and preserves them:
\begin{align}
\plift &: S_\PT^\lambda \to S_\PT^{-\lambda} , &
\tlift &: S_\PT^\lambda \to S_\PT^{-\lambda} , &
\ptlift &: S_\PT^\lambda \to S_\PT^{\lambda}.
\end{align}
A single Weyl module therefore does not carry an action of the full Pin group, which instead requires the direct sum of both chiralities on the same field space.
Alternatively, a reflection may map the chiral theory to a distinct theory of opposite chirality, in which case the original theory remains only Spin-covariant.

The Lorentzian projector conditions have the usual interpretation: for a single Weyl field the Dirac adjoint pairs with odd Clifford degree, so the chiral kinetic term survives while a Dirac-adjoint self-mass is projected out.
For a pair of opposite Weyl fields the off-diagonal condition is reversed, retaining the Dirac mass and projecting out an off-diagonal kinetic term.
In the Lorentzian classes admitting \MW or \sMW fields, the same-chirality projector and Grassmann tests retain the kinetic term and remove the scalar self-mass.
In every case, the coefficient phases must make the integrated action Hermitian and all free tensor indices must be contracted covariantly.

\subsubsection{\MWlong spinors in two and ten dimensions} \label{sec:example-delta-zero}

The physical reference points for the \(\delta = 0\bmod8\) class are two-dimensional worldsheet fermions and ten-dimensional \MW theories, including ten-dimensional super-\YM, the heterotic string, type-I supergravity, and type-IIB supergravity.
In this class the full Clifford algebra is real, while its even subalgebra splits into two real matrix blocks, so the real Majorana pinor restricts to two real Weyl spinor modules \(S_\PT^+\) and \(S_\PT^-\).
Both \(J_\Ch^+\) and \(J_\Ch^-\) are real structures that preserve each Weyl space.
A \MW condition can therefore be imposed on either irreducible chiral Spin module.
A Dirac pinor is an unconstrained field on the complexified real pinor module, a Majorana pinor satisfies the real condition on the full module, and a \MW spinor satisfies the compatible real condition on one chiral Spin module.
The Majorana Grassmann sign has the kinetic value \(\xi_\TransMat^\beta(d,1) = +1\) for the same-chirality \MW transpose bilinear in the Lorentzian representatives of this class.

\subsubsection{Majorana spinors in three and eleven dimensions} \label{sec:example-delta-one}

The \(\delta = 1\bmod8\) class contains the odd-dimensional real spinors of three-dimensional field theory and the Majorana spinor with thirty-two real components in eleven-dimensional supergravity.
The full Clifford algebra splits into two real matrix blocks, whereas the even Clifford algebra is a single real matrix algebra.
The connected Spin theory consequently has one real Majorana module, whereas extending it to Pin gives two inequivalent real Majorana modules.
The Pin modules are distinguished by the sign of the central odd-dimensional volume element and by the available intertwiners.
Complexifying the connected real spinor module gives a Dirac spinor, whereas a Pin action additionally requires a choice between the two real extensions.
The compatible transposition signs allow both a scalar Majorana mass and a Majorana kinetic term, subject to the remaining action-level requirements.

\subsubsection{Weyl spinors paired by Majorana reality in four dimensions} \label{sec:example-delta-two}

The \(\delta = 2\bmod8\) class contains the four-dimensional Lorentzian \SM, whose irreducible Spin fields are complex Weyl spinors.
A Dirac field combines two independent opposite-chirality Weyl fields, whereas a Majorana field combines one Weyl field with the opposite-chirality field fixed by the real structure.

The \((3,1)\) representative illustrates how to read the factorised classification: \((d,\delta) = (4,2)\) gives \(s = (d+\delta)/2 = 3\) and \(t = (d-\delta)/2 = 1\).
For \(\delta = 2\), the full Clifford algebra is real and its even subalgebra is complex, while both the Pin-equivariant real structure \(J_\Ch^+\) and the Spin-equivariant quaternionic structure \(J_\Ch^-\) exchange the two Weyl spaces.
The former defines a Majorana pinor and the latter a \sM spinor, each containing both chiralities.
Although the Dirac-adjoint self-mass of one Weyl field is absent, a Majorana mass is a same-chirality Weyl transpose contraction together with its complex conjugate, and the real structure relates the two terms.
The candidate \(\PC\) module map preserves chirality because both its candidate charge-conjugation factor and collective spatial inversion exchange it, whereas \(\PTC\) exchanges chirality because the full spacetime inversion preserves it.

At \(d = 4\), the transposition signatures imply that a Majorana scalar self-bilinear can use either \(\TransMat^+\) or \(\TransMat^-\), but a Majorana kinetic term requires \(\TransMat^-\).
The collective representatives in this signature satisfy
\begin{align}
\Plift^2 & = -\ident , &
\Tlift^2 & = -\ident , &
(\PTlift)^2 & = -\ident , &
\PTlift & = -\Tlift\Plift .
\end{align}
For the Pin-equivariant Majorana condition based on \(\ConjMat^+\), the choice \(\beta = -\) passes both the scalar and kinetic Grassmann tests, while \(\alpha = \beta\varsigma = -\) selects the compatible adjoint and the unphased representatives give
\begin{align}
\TransMat^- & = \gamma_4 = (\ConjMat^+)^\dagger A^- , &
\zeta_M^{-+} & = +1 .
\end{align}
Since \(\xi_A^-(3,1,0) = -1\) and \(\xi_A^-(3,1,1) = +1\), the coefficient conditions \eqref{eq:Majorana coefficient Hermiticity} reduce to
\begin{align}
(\kappa^-)^\ast & = -\kappa^- , &
(\mu^-)^\ast & = -\mu^- .
\end{align}
Under a rephasing \(\TransMat^-\to e^{\i\phi}\TransMat^-\), an unchanged bilinear density requires \(\zeta_M^{-+}\to e^{\i\phi}\zeta_M^{-+}\) and \(\kappa^-\to e^{-\i\phi}\kappa^-\), which explains why the rephasing-covariant coefficient relation contains \((\zeta_M^{-+})^2\).

\subsubsection{\sMlong spinors in five dimensions} \label{sec:example-delta-three}

Five-dimensional supersymmetric field theories and supergravity provide the Lorentzian representatives of the \(\delta = 3\bmod8\) class, for which the natural reality condition is \sM.
The full Clifford algebra is complex, whereas its even subalgebra is quaternionic, so the irreducible Pin module is of complex type but the connected spinor module admits a quaternionic structure.
Because this structure is only Spin-equivariant, the resulting \sM field is a spinor rather than a pinor.
The compatible transposition sign permits the \sM kinetic term, while the scalar self-mass vanishes by Grassmann symmetry and higher Clifford couplings remain rank dependent.

\subsubsection{\sMWlong spinors in six dimensions} \label{sec:example-delta-four}

Six-dimensional chiral supersymmetric theories, including \((1,0)\) and \((2,0)\) theories, lie in the \(\delta = 4\bmod8\) class and have \sMW fields as their minimal real chiral fermions.
The full Clifford algebra is quaternionic and its even subalgebra splits into two quaternionic matrix blocks, so the connected spinor module decomposes into two chiral quaternionic modules preserved by both \(J_\Ch^+\) and \(J_\Ch^-\).
Both maps are quaternionic structures, with \(J_\Ch^+\) Pin-equivariant and \(J_\Ch^-\) only automatically Spin-equivariant, so a \sMW condition can be imposed on either irreducible chiral Spin module.
A Dirac pinor is an unconstrained field in a complex realisation of the Pin module, whereas imposing the \sM condition at Pin level gives a \sM pinor whose chiral restrictions are \sMW spinors.
The \sM Grassmann sign has the kinetic value \(\xi_\TransMat^\beta(d,1) = -1\) for the same-chirality \sMW transpose bilinear in the Lorentzian representatives of this class.

\subsubsection{\sMlong spinors in seven dimensions} \label{sec:example-delta-five}

Seven-dimensional supersymmetric \YM theory and supergravity represent the \(\delta = 5\bmod8\) class, whose fermions satisfy an \sM condition.
The full Clifford algebra splits into two quaternionic matrix blocks and its even subalgebra is a single quaternionic matrix algebra, giving one connected quaternionic spinor module and two inequivalent quaternionic pinor modules.
The two Pin extensions are distinguished by the sign of the central odd-dimensional volume element and by the available intertwiners.
The quaternionic structure permits \sM spinors, while the two inequivalent Pin extensions of the connected Spin module each admit a \sM pinor.
Complexifying the connected module gives a Dirac spinor, whereas complexifying either full extension gives a Dirac pinor.
The compatible transposition signs permit both the scalar \sM mass and the \sM kinetic term, with the antisymmetry of \(\Omega_{AB}\) reversing the corresponding Majorana Grassmann conditions.
No chiral projector condition or off-diagonal Weyl mass interpretation is present.

\subsubsection{Majorana spinors and \sMlong pinors in eight dimensions} \label{sec:example-delta-six}

Eight-dimensional supersymmetric theories, often obtained by dimensional reduction from ten dimensions, lie in the \(\delta = 6\bmod8\) class.
The full Clifford algebra is quaternionic and its even subalgebra is complex, so the irreducible connected Spin fields are complex Weyl spinors in \(S_\PT^+\) or \(S_\PT^-\).
The Pin-equivariant quaternionic structure \(J_\Ch^+\) and the Spin-equivariant real structure \(J_\Ch^-\) both exchange these two Weyl spaces.
Hence \(J_\Ch^+\) defines a \sM pinor and \(J_\Ch^-\) defines a Majorana spinor on the paired Spin module, while neither defines a Majorana-type condition on one Weyl module alone.
The candidate \(\PC\) module map preserves chirality, whereas \(\PTC\) exchanges it.

\subsubsection{Majorana spinors in nine dimensions} \label{sec:example-delta-seven}

The nine-dimensional Majorana spinors appearing in circle reductions of ten-dimensional type-II theories and in nine-dimensional supergravity belong to the \(\delta = 7\bmod8\) class.
The full Clifford algebra is complex and its even subalgebra is real, so the connected spinor module admits a Majorana condition while the irreducible Pin module is of complex type.
The Majorana condition is Spin-equivariant but is not preserved by the odd Clifford generators.
Complexifying the real \(\properSpin(d-1,1)\)-module gives a Dirac spinor that admits the complex-type Pin extension, whereas the single real Majorana module does not carry a Pin action on the same real field space.
The compatible transposition sign permits the Majorana kinetic term, while the scalar self-mass vanishes by Grassmann symmetry and higher Clifford couplings remain rank dependent.

\subsection{Euclidean field theories} \label{sec:euclidean-field-theories}

For positive-definite Euclidean signature, \(t = 0\) implies \(d = s\) and \(\delta = d\bmod8\), so the Euclidean classes lie on the diagonal of the arbitrary-signature classification and inherit its Pin and Spin module types.
Depending on the application, these modules describe spin under ordinary spatial rotations, as for a Pauli spinor of \(\Spin(3)\cong\SU(2)\), internal symmetry representations, as for the \(\rep{16}\) of a \(\Spin(10)\) \GUT, or fermionic variables in a Wick-rotated functional integral.
The first two applications use only the representation theory, whereas the third also uses the transpose pairings and antilinear structures \cite{Wetterich:2010ni,Stone:2020vva}.
In a Euclidean functional integral, \(\psi\) and \(\widebar\psi\) are independent Grassmann variables, so a unitary Lorentzian interpretation is recovered through reflection positivity and analytic continuation rather than by imposing Lorentzian Hermiticity pointwise.

The Euclidean orthogonal group has the connected rotation component \(\SO(d)\) and one orientation-reversing component.
In even dimension an odd Clifford representative of the latter exchanges the Weyl spaces, so a single Weyl module carries a Spin representation but extends to a Pin representation only after the opposite chirality is included.
A reflection may instead map the theory to a distinct theory of opposite chirality, while in odd dimension the Pin-level distinction arises only when orientation-reversing transformations are included.

\subsubsection{Real chiral Euclidean spinors in eight dimensions} \label{sec:euclidean-delta-zero}

The \(d = 0\bmod8\) Euclidean class has real chiral Spin modules.
Algebraically, both \(J_\Ch^+\) and \(J_\Ch^-\) permit a \MW condition on either chiral module, while only \(J_\Ch^+\) extends this condition to a Majorana pinor on the full Pin module.
The physically central representative is \(\Spin(8)\), whose vector representation and two real chiral spinor representations all have real dimension eight and are permuted by triality.
\(\Spin(8)\) triality organises transverse bosonic and fermionic \DOFs in light-cone ten-dimensional string theory and supergravity, where \(\SO(8)\) is the transverse rotation group.
A nonzero chiral spinor in eight-dimensional Riemannian geometry has stabiliser \(\Spin(7)\), so a parallel chiral spinor reduces the holonomy to a subgroup of \(\Spin(7)\).

\subsubsection{Real Euclidean spinors in one and nine dimensions} \label{sec:euclidean-delta-one}

The \(d = 1\bmod8\) Euclidean class has real connected spinor modules and no chirality.
The connected real Spin module supports Majorana spinors and has two inequivalent real Pin extensions, each supporting Majorana pinors and distinguished by the sign of the central odd-dimensional volume element.
Since \(\SO(1)\) is trivial, the one-dimensional class has no nontrivial connected rotational action, whereas \(\Spin(9)\) has a physically important spinor representation of real dimension sixteen.
The sixteen-dimensional real \(\Spin(9)\) representation appears naturally as the transverse little-group spinor in eleven-dimensional supergravity and in related matrix-model descriptions.
The connected little-group representation itself only requires \(\Spin(9)\).

\subsubsection{Complex chiral Euclidean spinors in two and ten dimensions} \label{sec:euclidean-delta-two}

The \(d = 2\bmod8\) Euclidean class has complex chiral Spin modules.
The Pin-equivariant real structure \(J_\Ch^+\) pairs the two complex-conjugate Weyl modules into a Majorana pinor, whereas the Spin-equivariant quaternionic structure \(J_\Ch^-\) gives a \sM spinor on the pair.
Neither structure gives a Majorana-type condition on one Euclidean Weyl spinor alone.
The low-dimensional representative \(\Spin(2) \cong\U(1)\) has chiral spinor representations of complex dimension one with opposite half-integer charges, as used in two-dimensional Euclidean field theory, worldsheet fermions, and holomorphic spinor conventions.
The higher-dimensional representative \(\Spin(10)\) supplies the complex \(\rep{16}\) used to accommodate one chiral matter generation in \(\SO(10)\) \GUT models.
The two Weyl modules are \(\rep{16}\) and \(\widebar{\rep{16}}\), with their assignment to \(S_\PT^+\) and \(S_\PT^-\) fixed by the chirality convention.
A Pin-type extension would exchange these representations, but ordinary \(\Spin(10)\) model building uses only the connected group and therefore requires no pinor extension.

\subsubsection{Quaternionic Euclidean spinors in three dimensions} \label{sec:euclidean-delta-three}

The \(d = 3\bmod8\) Euclidean class has quaternionic connected spinor modules.
The Spin-equivariant quaternionic structure permits a \sM spinor, whereas the complex-type Pin module has no Pin-equivariant Majorana-type condition.
The basic physical example is ordinary non-relativistic spin, whose rotation group satisfies \(\Spin(3) \cong\SU(2)\).
The Pauli spinor is the fundamental spin representation of three-dimensional spatial rotations and has complex dimension two, equivalently quaternionic dimension one.
Its quaternionic character is the group-theoretic origin of symplectic reality conditions for three-dimensional Euclidean spinors.
Pinors become relevant only when the rotational symmetry is enlarged from \(\SO(3)\) to \(\O(3)\).

\subsubsection{Quaternionic chiral Euclidean spinors in four dimensions} \label{sec:euclidean-delta-four}

The \(d = 4\bmod8\) Euclidean class has quaternionic chiral Spin modules.
Both \(J_\Ch^+\) and \(J_\Ch^-\) permit a \sMW condition on either chiral module, while only \(J_\Ch^+\) extends this condition to a \sM pinor on the full Pin module.
In four dimensions the connected spin group factorises as \(\Spin(4) \cong \SU(2)^+\times \SU(2)^-\), and each chiral module is the fundamental representation of one \(\SU(2)\) factor, with complex dimension two or quaternionic dimension one.
The Euclidean Weyl modules are therefore quaternionic rather than real or genuinely complex.
Instantons, self-duality, and the decomposition of two-forms into self-dual and anti-self-dual parts are naturally expressed using the two \(\SU(2)\) factors.

\subsubsection{Quaternionic Euclidean spinors in five dimensions} \label{sec:euclidean-delta-five}

The \(d = 5\bmod8\) Euclidean class has quaternionic connected spinor modules and no chirality.
The connected quaternionic Spin module supports \sM spinors and has two inequivalent quaternionic Pin extensions, each supporting \sM pinors and distinguished by the sign of the central odd-dimensional volume element.
The basic group-theoretic representative is \(\Spin(5) \cong \Sp(2)\).
Its fundamental spinor representation has complex dimension four, equivalently quaternionic dimension two.
Besides double covering the five-dimensional Euclidean rotation group, \(\Spin(5)\) appears as an internal symmetry group in supersymmetric theories, for example as the \(R\)-symmetry group of six-dimensional \((2,0)\) theories.

\subsubsection{Complex chiral Euclidean spinors in six dimensions} \label{sec:euclidean-delta-six}

The \(d = 6\bmod8\) Euclidean class has complex chiral Spin modules.
The Pin-equivariant quaternionic structure \(J_\Ch^+\) pairs the two complex-conjugate Weyl modules into a \sM pinor, whereas the Spin-equivariant real structure \(J_\Ch^-\) gives a Majorana spinor on the pair.
Neither structure acts within one Weyl module.
The main representative is \(\Spin(6) \cong \SU(4)\).
Its two chiral spinor representations are the \(\rep 4\) and \(\widebar{\rep 4}\), each of complex dimension four.
The group \(\Spin(6)\) governs six-dimensional Euclidean rotations, occurs in compactification and internal-rotation problems, and is also the \(R\)-symmetry group of four-dimensional \(\mathcal N = 4\) super-\YM theory.
In the latter application the Euclidean spinor representations are internal symmetry representations.
If only the connected \(\Spin(6)\) symmetry is used, the two chiral representations can be treated as independent complex conjugate modules.

\subsubsection{Real Euclidean spinors in seven dimensions} \label{sec:euclidean-delta-seven}

The \(d = 7\bmod8\) Euclidean class has real connected spinor modules and no chirality.
The real Spin module admits a Majorana condition, whereas its complex-type Pin extension requires complexification and does not preserve the real Majorana subspace.
The group \(\Spin(7)\) has a spinor representation of real dimension eight.
Seven-dimensional Riemannian spinors are important in special geometry because the stabiliser of a nonzero real spinor in \(\Spin(7)\) is \(\G_2\), which is the group underlying \(\G_2\)-structures and \(\G_2\)-holonomy compactifications.
This class is therefore directly relevant to supersymmetric compactification geometry and real Killing-spinor constructions.

\subsection{Conformal extension}

The Clifford-algebraic classification applies directly to finite-dimensional spinor modules in conformal embedding space.
For a flat spacetime of signature \((s,t)\), the global conformal Lie algebra is \(\so(s+1,t+1)\), and the connected orthogonal action lifts to \(\properSpin(s+1,t+1)\) \cite{Costa:2011mg,Isono:2017grm}.
Finite-dimensional conformal-spinor modules therefore follow from the signature shift \((s,t)\to(s+1,t+1)\), which is precisely the split recursive construction \eqref{eq:recursive construction 2}:
\begin{equation}
 \cl(s+1,t+1) \cong \cl(s,t)\otimes\cl(1,1) \cong M^2\bigl(\cl(s,t)\bigr).
\end{equation}
The conformal Clifford algebra consequently has the same underlying real, complex, quaternionic, or split type as the spacetime Clifford algebra, while the split factor doubles the matrix size.
Equivalently, the two classification parameters transform as
\begin{align}
 d_\text{conf} &= d+2 , &
 \delta_\text{conf} &= (s+1)-(t+1)=\delta .
\end{align}
The conformal extension therefore increases the dimension by two while remaining in the same Bott class \(\delta\bmod8\).
All classification properties depending only on \(\delta\bmod8\) are inherited unchanged, whereas those depending explicitly on \(d\bmod8\) must be evaluated at \(d+2\).
The corresponding change in the reversion sign is
\begin{equation}
 \eta_\rev^{}(d+2)=-\eta_\rev^{}(d).
\end{equation}
Chirality signatures controlled by \(\eta_\rev(d)\) consequently change sign under the conformal extension, while those controlled only by \(\eta_\rev(\delta)\) remain unchanged.

The origin of the enlarged signature is transparent in the embedding-space formulation of conformal field theory \cite{Costa:2011mg,Isono:2017grm}.
The conformal compactification of spacetime is represented by null rays in a space of signature \((s+1,t+1)\), on which the conformal group acts linearly.
A point in an affine spacetime patch is recovered by choosing a section of the projective null cone, while the conformal dimension of a field is represented by its homogeneity in the embedding coordinates.
The nonlinear conformal transformations of the physical coordinates are thereby realised as ordinary orthogonal transformations in two additional dimensions.

At the Lie-algebra level, the enlargement is expressed by the graded decomposition
\begin{align}
 \so(s+1,t+1)
 &=
 \mathfrak g_{-1}\oplus\mathfrak g_0\oplus\mathfrak g_{+1}, &
 \mathfrak g_0&=\so(s,t)\oplus\mathbb R D .
\end{align}
Here \(D\) generates dilatations, while \(\mathfrak g_{+1}\) and \(\mathfrak g_{-1}\) are generated by translations and special conformal transformations, respectively, using \([D,X]=nX\) for \(X\in\mathfrak g_n\) \cite{Simmons-Duffin:2016gjk,Poland:2018epd}.
A local conformal primary at the origin is therefore characterised by a representation of the Lorentz spin group \(\properSpin(s,t)\) together with a conformal dimension, while the special conformal generators annihilate the primary and translations generate its descendants.
The resulting conformal multiplet should consequently be distinguished from a finite-dimensional spinor module of \(\properSpin(s+1,t+1)\).

The two Clifford algebras therefore describe related but distinct representation spaces: \(\cl(s,t)\) governs the local Lorentz spin carried by a field on physical spacetime, whereas \(\cl(s+1,t+1)\) governs finite-dimensional spinorial indices transforming linearly under the conformal spin group.
In the embedding-space formulation the latter modules are used directly, with homogeneity and projection or equivalence conditions reducing the embedding-space spinor to the physical conformal primary \cite{Isono:2017grm}.

For Euclidean signature the conformal extension reads \(\cl(s,0)\to\cl(s+1,1)\), whereas for Lorentzian signature with one time direction it reads \(\cl(s,1)\to\cl(s+1,2)\).

\section{Conclusion}

A convention-explicit treatment of fermions in arbitrary real signature has been developed that determines their local Clifford-theoretic field content and bilinear constraints.
Separating the full real Clifford algebra from its even subalgebra distinguishes pinors from spinors and prevents an extension to disconnected spacetime reflections from being inferred from covariance under the connected Spin group.
It also determines whether a real or quaternionic structure is Pin-equivariant or only Spin-equivariant and whether it preserves or exchanges the Weyl modules.

Combining the Bott class \(\delta = s-t\bmod8\) with the total dimension \(d = s+t\), considered modulo eight, organises the resulting module and intertwiner structures into thirty-two admissible local periodicity sectors.
The factorisation in \cref{tab:master-32} retains the Pin and Spin module types, conjugation structures, chirality properties, and transposition signatures needed for field-theory applications, while keeping the signature dependence of the adjoint and reflection signs explicit.
Fixed unphased ordered products connect the basis-independent intertwiner definitions to calculations in a chosen Clifford representation, with compatibility phases kept explicit and conventional map phases treated separately.
This translation is illustrated for all five four-dimensional real signatures, clarifying which familiar Majorana structures extend from the connected Spin group to a Pin action.

For local operators, the representation type supplies only the first of several independent requirements, since an invariant pairing, compatible intertwiners, the Grassmann and chiral selection rules, covariant contraction of the remaining indices, and Hermiticity of the Lorentzian action must all be established before a candidate term is allowed.
The resulting criteria provide the complete local Clifford-algebraic input, while gauge and flavour representations, anomalies, global Spin or Pin structures, and dynamics remain properties of the specific theory.

Once their action on the Hamiltonian is specified, the basis-independent maps also acquire a one-particle \QM interpretation.
For free-fermion systems, adjoining a normalised mass matrix to the Clifford algebra generated by kinetic and symmetry directions produces the extension problem whose real and complex Bott sequences reproduce the \AZlong tenfold way.

Together, these results provide a direct route from real Clifford algebras to fermion field content and candidate local operators in higher-dimensional theories, dimensional reductions, and Lorentzian or Euclidean formulations.

\subsection*{Acknowledgements}

This work is supported by the \FCT project \textnumero\ \href{https://doi.org/10.54499/2020.03969.CEECIND/CP1587/CT0014}{2020.\allowbreak 03969.\allowbreak CEECIND/\allowbreak CP1587/\allowbreak CT0014}.
The author thanks the Particles and Cosmology group at the University of Basel for their hospitality.

\appendix

\section{\QMlong} \label{sec:ABC-euclidean-QM}

In \QM the same Clifford intertwiners act on one-particle spin spaces, although their physical interpretation changes because time is an external evolution parameter rather than a coordinate of the Euclidean Clifford space.
For a system in \(s\) spatial dimensions, \(\cl(s,0)\) describes the spinorial action of spatial rotations, whereas a physical symmetry must also act consistently on the Hamiltonian, internal \DOFs, and external backgrounds.

\subsection{One-particle interpretation}

The one-particle states belong to the Hilbert space
\begin{equation} \label{eq:QM-Hilbert-space}
\mathscr H = L^2(\mathbb R^s)\otimes S^{\mathbb C}\otimes V,
\end{equation}
where \(S^{\mathbb C}\) is the complexified pinor module defined in \eqref{eq:pinor} and \(V\) contains any flavour, charge, orbital, band, or Nambu \DOFs.
The three tensor factors describe the spatial dependence of the wavefunction, its spinorial index, and its model-dependent internal \DOFs, respectively.
Spatial rotations require only the restriction to \(\Spin(s)\), whereas a spatial reflection requires the corresponding extension to \(\Pin(s)\).
If neither an odd Clifford operator nor a reflection occurs, the spin factor may instead be restricted to an irreducible Spin constituent of \(S^{\mathbb C}\).

The Euclidean specialisation has \(d = s\), \(t = 0\), and \(\delta = s\bmod8\).
Consequently, the Clifford generators are Hermitian by \eqref{eq:gamma hermiticity}, and the adjoint, conjugation, and transposition relations are the \(t = 0\) cases of \eqref{eq:intertwiner}.

A spatial orthogonal transformation acts on both the position and spin factors.
For \(M\in\O(s)\) and a lift \(g\in\Pin(s)\), the resulting action is
\begin{equation} \label{eq:QM-Pin-reflection}
(U_{g}\Psi)(\vec x) = g\Psi(M^{-1}\vec x),
\end{equation}
where the lift is understood in the chosen Clifford representation and belongs to \(\Spin(s)\) when \(M\) lies in the identity component.
A matrix acting only on \(S^{\mathbb C}\) therefore represents only the spinorial part of a spatial transformation.

The \(t = 0\) specialisation of \eqref{eq:preferred Dirac adjoint} selects the \(A^+\)-adjoint, which is the positive-definite Hilbert-space adjoint by the empty-product convention in \eqref{eq:intertwiner candidates}.
In even dimension, the complementary \(A^-\) is proportional to \(\chi\) by \eqref{eq:intertwiner negation} and can provide a unitary grading of a Hamiltonian, but only if the complete Hamiltonian anticommutes with it.
On a fixed irreducible odd-dimensional Pin module, \(A^-\) is absent because no complex-linear endomorphism negates all Clifford generators.

\subsection{Antilinear Clifford-module maps}

The complex-conjugation intertwiners define the basis-independent antilinear maps \(J_\Ch^\pm\) in \eqref{eq:Majorana structure}, and the same maps can be applied to one-particle states.
Their pointwise extension to the Hilbert space is antiunitary when the intertwiners and the internal action on \(V\) are unitary.
For each existing map on the Euclidean diagonal, its invariant square is
\begin{equation} \label{eq:QM-J-square}
(J_\Ch^\pm)^2 = \xi_\ConjMat^\pm(s)\ident.
\end{equation}
A positive square defines a real structure on an invariant state space, whereas a negative square defines a quaternionic structure and admits no nonzero fixed state.
Unlike a Majorana condition on a Grassmann-valued field, this statement concerns the scalar structure of a one-particle Hilbert space or one of its invariant subspaces.

The Clifford action of \(J_\Ch^\pm\) alone does not fix the physical interpretation of the map.
Depending on the internal factor \(V\) and the Hamiltonian relation, the map may define a real or quaternionic structure, exchange conjugate charge sectors, or provide the spinor part of particle-hole symmetry or time reversal.

The transposition intertwiner \(\TransMat^\pm\) represents an invariant complex-bilinear pairing rather than a linear Hilbert-space operator.
Using the positive-definite inner product to identify this pairing with an antilinear map gives the basis-covariant map \(K_\alT^\pm\) introduced in the main discussion of antilinear transformations.
In a unitary Clifford basis its coordinate expression and Clifford action are
\begin{align} \label{eq:QM-C-induced}
K_\alT^\pm\Psi& = (\TransMat^\pm)^{-1}\Psi^\ast, &
K_\alT^\pm\gamma_i(K_\alT^\pm)^{-1}& = \pm\gamma_i.
\end{align}
The transpose signature fixes its square without a basis choice:
\begin{equation} \label{eq:QM-T-square}
(K_\alT^\pm)^2 = \xi_\TransMat^\pm(s)\ident.
\end{equation}
Multiplication by a conventional phase leaves the square of any antilinear map unchanged.

Once the positive Hilbert-space adjoint is used, substitution of \eqref{eq:preferred Dirac adjoint,eq:intertwiner candidates} into \eqref{eq:compatible-ABC} shows that the two antilinear constructions coincide up to phase:
\begin{equation} \label{eq:QM-B-C-equivalence}
K_\alT^\varsigma = (\zeta_M^{+\varsigma})^\ast J_\Ch^\varsigma.
\end{equation}
Consequently, the Euclidean diagonal has
\begin{equation} \label{eq:QM-B-C-sign-equivalence}
\xi_\ConjMat^\varsigma(s) = \xi_\TransMat^\varsigma(s),
\end{equation}
as is also visible directly in \cref{tab:symmetry trans conj}.
The \(\ConjMat^\varsigma\)-construction displays the module structure, while the \(\TransMat^\varsigma\)-construction obtains the same antilinear map from the invariant transpose pairing.

In even dimension, using \(\alpha = -\) in \eqref{eq:compatible-ABC} gives the second candidate
\begin{equation} \label{eq:QM-two-time-reversals}
K_\alT^{-\varsigma}\propto(A^-)^\dagger J_\Ch^\varsigma\propto\chi J_\Ch^\varsigma.
\end{equation}
Thus the two maps differ by the chirality operator and have the square pattern already listed in \cref{tab:symmetry trans conj}.
In odd dimension only one member of each pair acts on a fixed irreducible Pin module, while the missing member can act only after both inequivalent Pin sectors are included and the map is allowed to exchange them.

\subsection{Hamiltonian symmetries and spectral flattening}

For a translation-invariant Hermitian Hamiltonian \(\mathcal H(\vec k)\), time reversal, particle-hole symmetry, and spectral symmetry are defined respectively by
\begin{align}  \label{eq:QM-Hamiltonian-symmetries}
\mathcal T\mathcal H(\vec k)\mathcal T^{-1}& = \mathcal H(-\vec k),  &
\mathcal C\mathcal H(\vec k)\mathcal C^{-1}& = -\mathcal H(-\vec k),  &
\mathcal S\mathcal H(\vec k)\mathcal S^{-1}& = -\mathcal H(\vec k).
\end{align}
The operations \(\mathcal T\) and \(\mathcal C\) are antiunitary, while \(\mathcal S\) is unitary and can be chosen proportional to \(\mathcal T\mathcal C\) when both antilinear symmetries are present.
A map induced by \(J_\Ch^\pm\) or \(K_\alT^\pm\) acquires one of these physical names only after the corresponding relation in \eqref{eq:QM-Hamiltonian-symmetries} has been verified for the complete system.
The same module map can therefore have different physical roles after different actions on \(V\) are supplied.

In particular, a conjugation-intertwiner map need not be the one-particle analogue of second-quantised charge conjugation, because a Hilbert space containing a particle of fixed charge need not contain its conjugate-charge sector.
Similarly, particle-hole symmetry of a Nambu-doubled Hamiltonian is a first-quantised energy-reflection constraint.

Realification means retaining the vectors of \(\mathscr H\) while restricting scalar multiplication from \(\mathbb C\) to \(\mathbb R\).
On the resulting real vector space \(\mathscr H_{\mathbb R}\), multiplication by \(\i\) becomes a real-linear operator that squares to minus the identity.
Every antiunitary operation \(\mathcal A\) reverses this structure:
\begin{align} \label{eq:QM-realification}
\i^2& = -\ident, & \acomm{\mathcal A}{\i}& = 0.
\end{align}
On an invariant subspace where \(\mathcal A^2 = \epsilon\ident\) with \(\epsilon = \pm1\), every state \(\Psi\) has a partner satisfying
\begin{align} \label{eq:QM-antilinear-pair}
\Psi^\prime& = \mathcal A\Psi, & \mathcal A\Psi^\prime& = \epsilon\Psi.
\end{align}
For \(\epsilon = +1\), the invariant subspace has an adapted real basis and the fixed vectors form a real subspace.
For \(\epsilon = -1\), no nonzero vector is fixed, and antiunitarity gives
\begin{equation} \label{eq:QM-Kramers-orthogonality}
\langle\mathcal A\Psi,\Psi\rangle = \langle\mathcal A\Psi,\mathcal A^2\Psi\rangle = -\langle\mathcal A\Psi,\Psi\rangle.
\end{equation}
Hence \(\langle\mathcal A\Psi,\Psi\rangle = 0\).
The pair \((\Psi,\mathcal A\Psi)\) is therefore a quaternionic doublet.

For time reversal, the two states have the same energy, but they lie in the same momentum fibre only at a time-reversal-invariant momentum satisfying \(\vec k\equiv-\vec k\).
Kramers degeneracy consequently follows from \(\mathcal T^2 = -\ident\) only at such a momentum or in a zero-dimensional system.
Particle-hole symmetry instead sends energy \(E\) to \(-E\), so it acts within one eigenspace only at zero energy.
These conclusions depend only on antiunitarity and the invariant square, not on a chosen matrix basis.

For a Hamiltonian gapped at zero, spectral flattening and the occupied-state projector are defined by
\begin{align} \label{eq:flattened Hamiltonian}
Q(\vec k)& = \frac{\mathcal H(\vec k)}{\abs{\mathcal H(\vec k)}}, &
Q^\dagger& = Q, &
Q^2& = \ident, &
P& = \frac{\ident-Q}2.
\end{align}
Because flattening replaces each nonzero energy eigenvalue by its sign, it discards its magnitude while preserving the eigenspaces, the gap, and every symmetry relation in \eqref{eq:QM-Hamiltonian-symmetries}.
Stable equivalence further permits the addition of decoupled flat bands, which changes matrix size without changing the stable phase.
The flattened operator need not anticommute with the kinetic matrices of a Dirac representative.

\section{Condensed matter} \label{sec:condensed-matter-bott}

The basis-independent \QM framework applies directly to stable gapped free-fermion Hamiltonians and reproduces the condensed-matter tenfold way \cite{Schnyder:2008tya,Kitaev:2009mg,Ryu:2010zza,Kennedy:2014cia}.
The same real and complex Bott patterns recur, although the quadratic space is now generated by symmetry, momentum, and mass directions rather than by spacetime vectors, so its signature serves only as an auxiliary classification device.
Encoding the symmetry and momentum constraints as Clifford generators reduces the classification to determining which normalised mass matrices can be added without closing the gap.

\subsection{Extended coefficient space} \label{sec:cm-az-generator-space}

The auxiliary quadratic space separates the physical coefficients of a Dirac Hamiltonian from formal directions that encode its symmetry constraints.
For an \(s\)-dimensional Dirac Hamiltonian, the coefficients of the mass and momentum matrices span the physical coefficient space
\begin{align} \label{eq:cm-hamiltonian-subspace}
V_{\text{phys}}^s& = V_{\text{mass}}\oplus V_{\text{mom}}^s, &
V_{\text{mass}}& = \operatorname{span}_{\mathbb R}\{e_m\}, &
V_{\text{mom}}^s& = \operatorname{span}_{\mathbb R}\{e_1,\ldots,e_s\}.
\end{align}
Further directions encode the symmetry constraints:
\begin{equation}
V_{\text{sym}}^{p,q} = \operatorname{span}_{\mathbb R}\{e_1^+,\ldots,e_p^+,e_1^-,\ldots,e_q^-\},
\end{equation}
where the superscripts record the signs of the corresponding unit directions.
The extended coefficient space is therefore
\begin{equation} \label{eq:cm-extended-space}
V_{\text{ext}}^{p,q;s} = V_{\text{phys}}^s\oplus V_{\text{sym}}^{p,q}.
\end{equation}

A coefficient vector can be written as
\begin{equation} \label{eq:cm-coefficient-vector}
x = me_m+\sum_{i = 1}^s k_i e_i+\sum_{a = 1}^p\sigma_a^+e_a^++\sum_{b = 1}^q\sigma_b^-e_b^-.
\end{equation}
The quadratic form is chosen as
\begin{equation} \label{eq:cm-coefficient-form}
Q_{\text{ext}}(x) = m^2-\sum_{i = 1}^s k_i^2+\sum_{a = 1}^p(\sigma_a^+)^2-\sum_{b = 1}^q(\sigma_b^-)^2,
\end{equation}
so \(V_{\text{phys}}^s\), \(V_{\text{sym}}^{p,q}\), and \(V_{\text{ext}}^{p,q;s}\) have signatures \((1,s)\), \((p,q)\), and \((p+1,q+s)\), respectively.
At this stage the symbols \(e\) denote basis vectors of a real quadratic space rather than operators: the coefficients appearing in the physical Hamiltonian occupy \(V_{\text{phys}}^s\), whereas the auxiliary coordinates \(\sigma_a^\pm\) are not additional couplings.

Removing the mass axis gives the source space
\begin{align} \label{eq:cm-source-hyperplane}
V_{\text{src}}^{p,q;s}& = e_m^\perp = V_{\text{mom}}^s\oplus V_{\text{sym}}^{p,q}, &
\operatorname{sig}(V_{\text{src}}^{p,q;s})& = (p,q+s).
\end{align}
Adjoining the mass direction, and hence opening the Dirac gap, corresponds geometrically to
\begin{equation} \label{eq:cm-space-extension}
V_{\text{src}}^{p,q;s} \to V_{\text{src}}^{p,q;s}\oplus V_{\text{mass}} = V_{\text{ext}}^{p,q;s}.
\end{equation}

\subsection{Orthogonal transformations and double covers} \label{sec:cm-orthogonal-space}

The source and extended quadratic spaces carry the orthogonal frame groups
\begin{align}
\O(V_{\text{src}}^{p,q;s})&\cong\O(p,q+s), &
\O(V_{\text{ext}}^{p,q;s})&\cong\O(p+1,q+s).
\end{align}
Although these transformations preserve the quadratic forms, a general frame transformation need not preserve the distinguished mass, momentum, and symmetry subspaces and is therefore not automatically a physical symmetry of the Hamiltonian.

Applying the Clifford construction \eqref{eq:Clifford algebra} gives
\begin{align}
\cl(V_{\text{src}}^{p,q;s})&\cong\cl(p,q+s), &
\cl(V_{\text{ext}}^{p,q;s})&\cong\cl(p+1,q+s).
\end{align}
For \(g\in\Pin(p+1,q+s)\), its projected frame transformation \(\Mmat\) is defined by
\begin{align} \label{eq:cm-lifted-frame-action}
\eval{\Ad^\PT_g}_{V_{\text{ext}}^{p,q;s}}& = \Mmat, &
Q_{\text{ext}}(\Mmat x)& = Q_{\text{ext}}(x).
\end{align}
As established in \cref{sec:pin-spin-groups}, products of unit vectors lift orthogonal frame transformations to the corresponding Pin group, whereas even products lift the identity component to the Spin group.
For the extended space, the double covers are
\begin{align}
&\shortexact{\C^2_\cover}{\Pin(p+1,q+s)}{\O(p+1,q+s)}, &
&\shortexact{\C^2_\cover}{\properSpin(p+1,q+s)}{\properSO(p+1,q+s)}.
\end{align}
The inclusion \eqref{eq:cm-space-extension} therefore induces the Clifford extension \(\cl(p,q+s) \to\cl(p+1,q+s)\).

Under the simultaneous stabilisation \((p,q)\to(p+1,q+1)\), the split recursion \eqref{eq:recursive construction 2} gives
\begin{equation}
\cl(p+1,q+1)\cong\cl(p,q)\otimes\cl(1,1)\cong M^2\bigl(\cl(p,q)\bigr).
\end{equation}
This tensor product enlarges the matrix representation without changing the underlying real, complex, quaternionic, or split matrix-algebra type.
In the condensed-matter setting, this algebraic stabilisation corresponds physically to adding symmetry-compatible trivial bands.
Consequently, the stable real symmetry label is
\begin{equation} \label{eq:cm-bott-label}
r=p-q\bmod8,
\end{equation}
while \(p+q\) and the individual values of \(p\) and \(q\) specify a particular unstabilised representative.

\subsection{Clifford representation and mass extension} \label{sec:cm-clifford-extension}

The geometric extension becomes a Hamiltonian problem after the extended Clifford algebra is represented on the realified one-particle Hilbert space \(\mathscr H_{\mathbb R}\).
The orthonormal frame is represented by operators according to
\begin{align} \label{eq:cm-basis-representation}
e_m&\mapsto\gamma_0, &
e_i&\mapsto\gamma_i, &
e_a^+&\mapsto E_a^+, &
e_b^-&\mapsto E_b^-.
\end{align}
Their squares reproduce the quadratic form,
\begin{align} \label{eq:cm-represented-squares}
\gamma_0^2& = +\ident, &
\gamma_i^2& = -\ident, &
(E_a^+)^2& = +\ident, &
(E_b^-)^2& = -\ident,
\end{align}
and operators assigned to distinct directions anticommute.
The symmetry generators \(E_a^+\) and \(E_b^-\) are real-linear combinations and products of the complex structure \(\i\), the operations \(\mathcal T\), \(\mathcal C\), and \(\mathcal S\), and any other internal symmetry structures required by the class.
Their squares and mutual commutation relations are fixed using the basis-independent operations of \cref{sec:ABC-euclidean-QM} together with their action on the internal factor \(V\).
Because one named symmetry can generate several real-linear Clifford directions, \(p+q\) is not the number of physical symmetries.

A low-energy Dirac representative of the physical coefficient space is
\begin{equation} \label{eq:cm-dirac-hamiltonian}
\mathcal H_D(\vec k) = m\gamma_0+\i\sum_{i = 1}^s k_i\gamma_i,
\end{equation}
where \(\gamma_0\) is Hermitian and the \(\gamma_i\) are anti-Hermitian, so every term in \(\mathcal H_D\) is Hermitian.
Here \(\i\) is the real-linear complex structure in \eqref{eq:QM-realification}, and multiplication by it converts each represented negative Clifford generator into a Hermitian kinetic matrix without adding a direction to the coefficient space.
The matrices \(\gamma_0\) and \(\gamma_i\) commute with the real-linear complex structure \(\i\), as required for \(\mathcal H_D\) to be complex-linear on \(\mathscr H\).
The auxiliary coordinates do not occur in \(\mathcal H_D\), but their represented generators constrain the admissible kinetic and mass matrices.

The source generators are collectively denoted by
\begin{equation}
E_A\in\{\gamma_1,\ldots,\gamma_s,E_1^+,\ldots,E_p^+,E_1^-,\ldots,E_q^-\}.
\end{equation}
A normalised symmetry-compatible mass matrix is a positive Clifford generator satisfying
\begin{align} \label{eq:cm-mass-generator}
\gamma_0^2& = \ident, & \acomm{E_A}{\gamma_0}& = 0.
\end{align}
Classifying symmetry-compatible masses is therefore the represented form of the extension
\begin{equation} \label{eq:cm-real-extension}
\cl(p,q+s) \to\cl(p+1,q+s).
\end{equation}
For the Dirac representative, flattening at \(\vec k = 0\) gives \(Q(0) = \sgn(m)\gamma_0\), so the added generator is the normalised mass direction.

With the sign convention of \eqref{eq:cm-coefficient-form}, the stable space of positive-generator extensions is
\begin{equation}
R_{p-(q+s)} = R_{r-s},
\end{equation}
and its connected components give the strong stable classification
\begin{equation} \label{eq:cm-real-classification}
\pi_0(R_{r-s}).
\end{equation}
Here \(R_{r-s}\) is the stabilised space of admissible normalised mass matrices, so two masses lie in the same component when they can be continuously deformed into one another without violating the symmetry constraints or closing the gap.
Each momentum direction contributes one negative generator and therefore shifts the zero-dimensional symmetry index by \(-1\), while the real classifying spaces satisfy \(R_{n+8}\simeq R_n\) \cite{Kitaev:2009mg}.
The group \(\pi_0(R_{r-s})\) describes the strong continuum classification obtained after compactifying momentum space to a sphere.
For a lattice Hamiltonian over the Brillouin torus, the corresponding real K-theory group can additionally contain weak invariants and is not generally exhausted by \(\pi_0(R_{r-s})\).

When no antiunitary structure is imposed, the extension is complex and depends only on the parity \(n\bmod2\) of the complex Clifford class.
The zero-dimensional extension \(\cl(n,\mathbb C)\to\cl(n+1,\mathbb C)\) has classifying space \(C_n\), while \(s\) momentum directions give \(C_{n-s}\).
The complex strong classification is consequently \(\pi_0(C_{n-s})\), with \(C_{n+2}\simeq C_n\).
Adding decoupled flat bands changes only the matrix-size representative in either sequence.

\subsection{\BdGlong Hamiltonians and Nambu doubling} \label{sec:cm-bdg}

For superconducting systems, the \BdG construction gives a direct realisation of the realified one-particle module \cite{Alicea:2012ux, Beenakker:2011np}.
Nambu doubling places particles and holes in a single one-particle space, on which the Fermi constraint becomes an antilinear operation.
In terms of fermionic creation and annihilation operators \(c_i^\dagger\) and \(c_i\), a general quadratic Hamiltonian is
\begin{align}
 H
 & =
 \sum_{i,j}c_i^\dagger h_{ij}c_j
 +
 \frac12\sum_{i,j}
 \left(
 \Delta_{ij}c_i^\dagger c_j^\dagger
 +
 \Delta_{ij}^\ast c_jc_i
 \right),
 &
 h^\dagger& = h,
 &
 \Delta^\trans& = -\Delta.
\end{align}
The Nambu spinor packages the annihilation and creation operators into
\begin{align}
 H
 & =
 \frac12\Psi^\dagger\mathcal H_{\BdG}\Psi+E_0,
 &
 \Psi
 & =
 \begin{pmatrix}
 c\\
 c^\dagger
 \end{pmatrix},
 &
 \mathcal H_{\BdG}
 & =
 \begin{pmatrix}
 h & \Delta\\
 -\Delta^\ast & -h^\ast
 \end{pmatrix}.
\end{align}
The particle-hole operation acts antilinearly on the Nambu module and satisfies
\begin{equation} \label{eq:particle-hole symmetry}
\mathcal C\mathcal H_{\BdG}(\vec k)\mathcal C^{-1} = -\mathcal H_{\BdG}(-\vec k).
\end{equation}
This relation is the Nambu-space Fermi constraint conventionally called particle-hole symmetry, rather than an additional spacetime symmetry.
It sends \(\Phi_E(\vec k)\) to a state of energy \(-E\) at \(-\vec k\), as follows directly from \eqref{eq:QM-Hamiltonian-symmetries}.
At zero energy, \(\mathcal C^2 = +\ident\) permits a particle-hole-fixed vector
\begin{equation} \label{eq:Majorana Bogoliubov}
\mathcal C\Phi_0 = \Phi_0,
\end{equation}
whose Bogoliubov quasiparticle operator is self-adjoint.
Thus a Majorana zero mode is a real one-particle Nambu state.
For \(\mathcal C^2 = -\ident\), no nonzero zero mode is fixed and the zero modes instead form the quaternionic pairs of \eqref{eq:QM-antilinear-pair}.
The full Nambu representation has a canonical positive-square Fermi constraint, while internal symmetries or an irreducible block reduction can lead to the effective negative-square particle-hole structure.
Spectral flattening is performed by \eqref{eq:flattened Hamiltonian} and introduces no additional condensed-matter-specific calculation.
Near a Dirac gap closing, its zero-momentum mass block gives the positive generator classified in \cref{sec:cm-clifford-extension}.

\subsection{The \AZlong symmetry classes} \label{sec:cm-az-classes}

\begin{table}
\begin{panels}[t]{.35}
\begin{tabular}{c*2c} \toprule
& \multicolumn2c{\(n\bmod2\)} \\ \cmidrule{2-3}
& \(0\) & \(1\) \\ \midrule
class & A & AIII \\ \midrule
\(\mathcal T^2\) & \(0\) & \(0\) \\
\(\mathcal C^2\) & \(0\) & \(0\) \\
\(\mathcal S\) & \(0\) & \(+\) \\ \midrule
\(\cl(n,\mathbb C)\) & \(\mathbb C\) & \(2\mathbb C\) \\ \midrule
\(\pi_0(C_n)\) & \(\mathbb Z\) & \(0\) \\
\bottomrule \end{tabular}
\caption{Complex classes}
\panel{.65}
\begin{tabular}{c*8c} \toprule
& \multicolumn8c{\(r \bmod 8\)} \\ \cmidrule{2-9}
& \(0\) & \(1\) & \(2\) & \(3\) & \(4\) & \(5\) & \(6\) & \(7\) \\ \midrule
class & AI & BDI & D & DIII & AII & CII & C & CI \\ \midrule
\(\mathcal T^2\) & \(+\) & \(+\) & \(0\) & \(-\) & \(-\) & \(-\) & \(0\) & \(+\) \\
\(\mathcal C^2\) & \(0\) & \(+\) & \(+\) & \(+\) & \(0\) & \(-\) & \(-\) & \(-\) \\
\(\mathcal S\) & \(0\) & \(+\) & \(0\) & \(+\) & \(0\) & \(+\) & \(0\) & \(+\) \\ \midrule
\(\xi_\bott(r)\) & \(+\) & \(+\) & \(+\) & \(0\) & \(-\) & \(-\) & \(-\) & \(0\) \\
\(\cl(r,0)\) & \(\mathbb R\) & \(2\mathbb R\) & \(\mathbb R\) & \(\mathbb C\) & \(\mathbb H\) & \(2\mathbb H\) & \(\mathbb H\) & \(\mathbb C\) \\ \midrule
\(\pi_0(R_r)\) & \(\mathbb Z\) & \(\mathbb Z_2\) & \(\mathbb Z_2\) & \(0\) & \(\mathbb Z\) & \(0\) & \(0\) & \(0\) \\
\bottomrule \end{tabular}
\caption{Real classes}
\end{panels}
\caption[Zero-dimensional \AZlong classes, Clifford algebras, and stable groups]{
Zero-dimensional \AZshort symmetry classes obtained from the complex label \(n\bmod2\) and the stable real symmetry label \(r = p-q\bmod8\).
A zero in a symmetry row denotes the absence of that operation, while the signs distinguish real and quaternionic antilinear structures.
The Clifford rows suppress matrix-size factors but retain both simple summands in the split classes.
The stable zero-dimensional classification groups appear in the final rows.
In \(s\) spatial dimensions, the strong classification follows by replacing \(n\) with \(n-s\) or \(r\) with \(r-s\).
} \label{tab:tenfold-way}
\end{table}

The Hamiltonian relations \eqref{eq:QM-Hamiltonian-symmetries} and the invariant squares of their antilinear operations determine the conventional \AZfirst labels \cite{Altland:1997zz}.
In zero spatial dimensions, the flattened Hamiltonian \(Q\) is the positive mass generator adjoined to \(\cl(p,q)\).
The time-reversal and particle-hole squares follow by adjoining the positive or negative generator preserved by the corresponding operation.
Time reversal preserves \(Q\), so its real or quaternionic square is read from the extended algebra \(\cl(p+1,q)\), whose Bott class is \(r+1\).
Particle-hole symmetry reverses both \(Q\) and the complex structure \(\i\), so it preserves the negative generator \(\i Q\), which satisfies \((\i Q)^2 = -\ident\) and extends the source algebra to \(\cl(p,q+1)\) of Bott class \(r-1\).
The Bott signature therefore gives
\begin{align} \label{eq:cm-AZ-signs}
\mathcal T^2& = \xi_\bott(r+1)\ident, &
\mathcal C^2& = \xi_\bott(r-1)\ident,
\end{align}
where a zero value means that the corresponding operation is absent.
The grading \(\mathcal S\) is present precisely for odd \(r\), when both antilinear operations occur and their product is linear.
These three statements generate the eight real columns of \cref{tab:tenfold-way} directly from the Bott sequence in \cref{tab:bott signature}.
The two complex classes retain only \(n\bmod2\): class A has no spectral grading and class AIII has one.

The two complex and eight real Clifford classes therefore reproduce the ten symmetry classes and their zero-dimensional stable groups, while the spatial shifts derived in \cref{sec:cm-clifford-extension} generate the corresponding strong periodic table.
Although the invariant squares in \cref{tab:tenfold-way} arise from the same real and quaternionic module structures as those of \(J_\Ch^\pm\), their physical roles are fixed by the Hamiltonian relations.



\end{document}